\documentclass[runningheads]{llncs}

\usepackage[usenames,dvipsnames]{xcolor}
\usepackage{tikz}
\usepackage{wrapfig}
\usepackage{subfigure}
\usepackage{fixltx2e}
\usepackage{amsmath}
\usepackage{amssymb}
\usepackage{amsfonts}
\usepackage{latexsym}
\usepackage{xspace}
\usepackage{rotating}
\usepackage{enumerate}
\usepackage{paralist}
\usepackage{pgfplots}
\usepackage{booktabs}
\usepackage{enumitem}
\usepackage{etoolbox}
\usepackage{cite}
\usepackage{multirow}

\usepackage{float}
\usepackage{pifont}

\usetikzlibrary{patterns,arrows,calc,shapes,shadows,decorations.pathmorphing,decorations.pathreplacing,automata,shapes.multipart,positioning,shapes.geometric,fit,circuits,trees,shapes.gates.logic.US,fit, automata ,petri}

\tikzstyle{every picture}=[thick]
\tikzstyle{every loop}=[->]
\tikzstyle{every scope}=[>=latex]
\tikzstyle{dot}=[circle,thick,minimum size=0.5mm,fill=black, inner sep=1pt]
\tikzstyle{every state}=[draw=black,line width=.5pt,fill=white,minimum size=10pt,initial text=]

\pgfplotscreateplotcyclelist{mycolor}{%
blue,every mark/.append style={fill=blue!80!black},mark=*, densely dashed\\%
violet,every mark/.append style={fill=violet!80!black},mark=square*, densely dashed\\%
BurntOrange!60!black,every mark/.append style={fill=BurntOrange!80!black},mark=otimes*\\%
red,mark=diamond*,every mark/.append style={fill=red!80!black}\\%
}

\input{dft-tikz}

\tikzset{
  place/.style={
    circle,
    thick,
    draw=black,
    fill=white,
    minimum size=6mm
  },
  iplace/.style={
    circle,
    thick,
    draw=blue!80!black!100,
    fill=blue!15,
    minimum size=6mm
  },
  itransition/.style={
    rectangle,
    thick,
    fill=black,
    minimum height=8mm,
    inner xsep=2pt
  },
  ttransition/.style={
    rectangle,
    thick,
    draw,
    minimum height=8mm,
    inner xsep=2pt
  },
  Itransition/.style={
    rectangle,
    thick,
    fill=black,
    minimum width=8mm,
    inner ysep=2pt
  },
  Ttransition/.style={
    rectangle,
    thick,
    draw,
    minimum width=8mm,
    inner ysep=2pt
  }
}

\tikzstyle{dummy}=[draw=none, fill=none]

\newcommand{\dftcalc}{DFTCalc}
\newcommand{\storm}{Storm}
\newcommand{\galileo}{Galileo}

\newcommand{\cmark}{\ding{51}}%
\newcommand{\xmark}{\ding{55}}%\
\renewcommand{\checkmark}{\cmark}
\newcommand{\cross}{\xmark}

\newcommand{\sep}{\ensuremath{~|~}}

\newcommand{\ie}{i.e., }
\newcommand{\eg}{e.g., }
\newcommand{\wrt}{w.r.t.\ }
\newcommand{\cf}{cf.\ }

\newcommand{\interfaces}{\ensuremath{\mathcal{I}}}

\newcommand{\AND}{\textsf{AND}\xspace}
\newcommand{\OR}{\textsf{OR}\xspace}
\newcommand{\PAND}{\textsf{PAND}\xspace}
\newcommand{\PANDincl}{\ensuremath{\PAND_\leq}}
\newcommand{\PANDexcl}{\ensuremath{\PAND_<}}
\newcommand{\POR}{\textsf{POR}\xspace}
\newcommand{\PORincl}{\ensuremath{\POR_\leq}}
\newcommand{\PORexcl}{\ensuremath{\POR_<}}
\newcommand{\VOT}[1]{\textsf{VOT}\textsubscript{#1}\xspace}
\newcommand{\SEQ}{\textsf{SEQ}\xspace}
\newcommand{\SPARE}{\textsf{SPARE}\xspace}
\newcommand{\FDEP}{\textsf{FDEP}\xspace}
\newcommand{\PDEP}{\textsf{PDEP}\xspace}
\newcommand{\BE}{\textsf{BE}\xspace}

\newcommand{\MUTEX}{\textsf{MUTEX}\xspace}

\newcommand{\Tp}{\ensuremath{\mathit{Tp}}}

\newcommand{\Top}{\ensuremath{\mathit{top}}}
\newcommand{\DFT}{\ensuremath{\mathcal{F}}\xspace}
\newcommand{\children}{\ensuremath{\sigma}}
\newcommand{\child}[2]{\ensuremath{#1_{#2}}}

\newcommand{\DFTtuple}{\ensuremath{(V,\children,\Tp,\Top)}\xspace}

\newcommand{\Failed}{\ensuremath{\textsf{Failed}}}

\newcommand{\Claimed}{\ensuremath{\textsf{Claimed}}}

\newcommand{\Unavailable}{\ensuremath{\textsf{Unavail}}}
\newcommand{\Consider}{\ensuremath{\textsf{Next}}}
\newcommand{\Active}{\ensuremath{\textsf{Active}}}
\newcommand{\Failsafe}{\ensuremath{\textsf{FailSafe}}}
\newcommand{\Disabled}{\ensuremath{\textsf{Disabled}}}
\newcommand{\Init}{\ensuremath{\textsf{Init}}}
\newcommand{\Evidence}{\ensuremath{\textsf{Evidence}}}
\newcommand{\Collect}{\ensuremath{\textsf{Collect}}}
\newcommand{\Coin}{\ensuremath{\textsf{Coin}}}
\newcommand{\Flip}{\ensuremath{\textsf{Flip}}}
\newcommand{\Forward}{\ensuremath{\textsf{Forward}}}
\newcommand{\DFTnodes}{\ensuremath{V}}
\newcommand{\tunavailable}{\ensuremath{\textsf{unavailable}}}
\newcommand{\tclaim}{\ensuremath{\textsf{claim}}}
\newcommand{\tchildfail}{\ensuremath{\textsf{child-fail}}}

\newcommand{\prioVar}{\ensuremath{\vec{\pi}}}
\newcommand{\prioConc}{\ensuremath{\vec{c}}}

\newcommand{\gspn}{\ensuremath{G}}
\newcommand{\petriNet}{\ensuremath{\mathcal{N}}}
\newcommand{\petriTemplate}{\ensuremath{\mathcal{T}}}
\newcommand{\setOfTemplates}{\ensuremath{\mathbb{T}}}
\newcommand{\petriPlaces}{\ensuremath{P}}
\newcommand{\petriTransitions}{\ensuremath{T}}
\newcommand{\petriTimed}{\ensuremath{\petriTransitions_t}}
\newcommand{\petriImmediate}{\ensuremath{\petriTransitions_i}}
\newcommand{\petriWeight}{\ensuremath{W}}
\newcommand{\petriPriority}{\ensuremath{\Pi}}
\newcommand{\petriPD}{\ensuremath{\petriPriority\textsf{Dom}}}
\newcommand{\petriPartition}{\ensuremath{\mathcal{D}}}
\newcommand{\marking}{\ensuremath{M}}
\newcommand{\conc}[1]{\ensuremath{\textsf{conc}(#1)}}

\newcommand{\enabled}[1]{\ensuremath{\textsf{enabled}(#1)}}
\newcommand{\fire}{\ensuremath{\textsf{fire}}}

\newcommand{\merge}{\ensuremath{\textsf{merge}}}
\newcommand{\templ}[2]{\ensuremath{\textsf{templ}\,_{#1}(#2)}}
\newcommand{\templDef}[1]{\ensuremath{\textsf{templ}\,_{#1}}}
\newcommand{\initMarkTempl}{\ensuremath{\textsf{templ}\,_{\textsf{init}}}}

\newcommand{\mapping}{\ensuremath{\mathcal{M}}\xspace}
\newcommand{\maxchildren}{\ensuremath{\children_\textsf{max}}}
\newcommand{\sizechildren}{\ensuremath{|\children_v|}}

\newcommand{\NN}{\ensuremath{\mathbb{N}}}

\newcommand{\RRplus}{\ensuremath{\mathbb{R}_{>0}}}

\definecolor{lightblue}{RGB}{224,224,255}
\definecolor{lightred}{RGB}{255,224,224}
\definecolor{lightgreen}{RGB}{224,255,224}
\definecolor{lightyellow}{RGB}{255,255,224}
\definecolor{lightpurple}{RGB}{255,224,255}
\definecolor{darkerred}{RGB}{64,0,0}
\definecolor{darkred}{RGB}{128,0,0}
\definecolor{darkblue}{RGB}{0,0,128}
\definecolor{darkgreen}{RGB}{0,128,0}
\definecolor{darkpurple}{RGB}{128,0,128}

\makeatletter
\renewcommand{\paragraph}{%
  \@startsection{paragraph}{4}%
  {\z@ }{-5\p@ \@plus -4\p@ \@minus -4\p@ }{-0.5em \@plus -0.22em \@minus -0.1em}%
  {\normalfont\normalsize\itshape}%
}
\makeatother

\title{One Net Fits All%
\thanks{This work is supported by the CDZ project CAP, the DFG RTG 2236 ``UnRAVeL'', the STW project 154747 SEQUOIA, and the EU project SUCCESS.}
}
\subtitle{A unifying semantics of Dynamic Fault Trees using GSPNs}

\author{Sebastian Junges\inst{1}, Joost-Pieter Katoen\inst{1,2},\\ Mari\"{e}lle Stoelinga\inst{2,3}, Matthias Volk\inst{1}}
\institute{Software Modeling and Verification, RWTH Aachen University, Germany \and Formal Methods and Tools, University of Twente, Netherlands \and Department of Software Science, Radboud University Nijmegen, Netherlands}

\newcommand{\dftscale}{0.6}

\begin{document}
\maketitle

%========================================================================================
% Sections
%========================================================================================
\begin{abstract}
Dynamic Fault Trees (DFTs) are a prominent model in reliability engineering.
They are strictly more expressive than static fault trees, but this comes at a price:
their interpretation is non-trivial and leaves quite some freedom.
This paper presents a GSPN semantics for DFTs.
This semantics is rather simple and compositional.
The key feature is that this GSPN semantics unifies \emph{all} existing DFT semantics from the literature.
All semantic variants can be obtained by choosing appropriate priorities and treatment of non-determinism.
\end{abstract}

\section{Introduction}
\label{sec:intro}
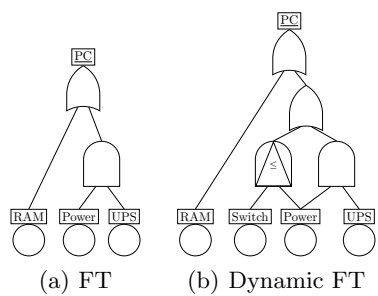
\begin{wrapfigure}[12]{R}{0.43\textwidth}
\centering
\subfigure[FT]{
    \scalebox{\dftscale}{
    \begin{tikzpicture}[scale=.6,text=black]
    \node[or2] (or) {};
    \node[labelbox] (or_label) at (or.east) {\underline{$\text{PC}$}};
    \node[and2,below=1.3cm of or.west] (and) {};
    \node[be,below=1.3cm of and.center, xshift=-0.5cm] (power) {};
    \node[labelbox] (power_label) at (power.north) {Power};
    \node[be,below=1.3cm of and.center, xshift=0.5cm] (ups) {};
    \node[labelbox] (ups_label) at (ups.north) {UPS};
    \node[be,left=0.4cm of power] (ram) {};
    \node[labelbox] (ram_label) at (ram.north) {RAM};

    \draw[-] (or.input 1) -- (ram_label.north);
    \draw[-] (or.input 2) -- (and.output);
    \draw[-] (and.input 1) -- (power_label.north);
    \draw[-] (and.input 2) -- (ups_label.north);
\end{tikzpicture}%
    }
    \label{fig:toy_sft}
}
\subfigure[Dynamic FT]{
    \scalebox{\dftscale}{
    \begin{tikzpicture}[scale=.6,text=black]
    \node[or2] (or) {};
    \node[labelbox] (or_label) at (or.east) {\underline{$\text{PC}$}};
    \node[or2, below=0.9cm of or.west] (or2) {};
    \node[and2,below=1.2cm of or2,yshift=0.1cm] (and) {};
    \node[and2, left=1.4cm of and.east] (pand) {{\tiny\rotatebox{270}{$\leq$}}};
    \node[triangle,scale=1.62,yshift=-3.5,xscale=0.80] (triangle) at (pand) {};

    \node[be,below=1.3cm of and.center, xshift=-0.8cm] (power) {};
    \node[labelbox] (power_label) at (power.north) {Power};
    \node[be,below=1.3cm of and.center, xshift=0.5cm] (ups) {};
    \node[labelbox] (ups_label) at (ups.north) {UPS};
    \node[be,left=0.4cm of power] (switch) {};
    \node[labelbox] (switch_label) at (switch.north) {Switch};
    \node[be,left=0.5cm of switch] (ram) {};
    \node[labelbox] (ram_label) at (ram.north) {RAM};

    \draw[-] (or.input 1) -- (ram_label.north);
    \draw[-] (or.input 2) -- (or2.output);
    \draw[-] (or2.input 1) -- (pand.output);
    \draw[-] (or2.input 2) -- (and.output);
    \draw[-] (pand.input 1) -- (switch_label.north);
    \draw[-] (pand.input 2) -- (power_label.north);
    \draw[-] (and.input 1) -- (power_label.north);
    \draw[-] (and.input 2) -- (ups_label.north);
\end{tikzpicture}%
    }
    \label{fig:toy_dft}
}
\caption{Fault tree examples}
\label{fig:toy_fault_trees}
\end{wrapfigure}%
Fault trees (FTs)~\cite{trivedi_bobbio_2017} are a popular model in reliability engineering.
They are used by engineers on a daily basis, are recommended by standards in e.g., the automotive, aerospace and nuclear power industry.
Various commercial and academic tools support FTs; see \cite{RS15} for a survey.
FTs visualise how combinations of components faults (their leaves, called basic events) lead to a system failure.
Inner tree nodes (called gates) are like logical gates in circuits such as \AND and \OR.
The simple FT in Fig.~\ref{fig:toy_sft} models that a PC fails if either the RAM, or both power and  UPS fails.

\begin{figure}[tb]
\centering
\subfigure[DFT]{
    \scalebox{0.8}{
    \begin{tikzpicture}[scale=.6,text=black]
    \node[and2] (pand) {{\tiny\rotatebox{270}{$\leq$}}};
    \node[triangle,scale=1.62,yshift=-3.5,xscale=0.80] (triangle_b) at (pand) {};
    \node[labelbox] (pand_label) at (pand.east) {$\underline{Z}$};

    \node[or2,below=1.8cm of pand.center, yshift=0.7cm] (or) {};
    \node[labelbox] (or_label) at (or.east) {$X$};

    \node[be,below=1.0cm of or.center, xshift=-0.4cm] (A) {};
    \node[labelbox] (a_label) at (A.north) {$A$};
    \node[be,below=1.0cm of or.center, xshift=0.7cm] (B) {};
    \node[labelbox] (b_label) at (B.north) {$B$};

    \draw[-] (pand.input 2) -- (b_label.north);
    \draw[-] (pand.input 1) -- (or_label.north);
    \draw[-] (or.input 1) -- (a_label.north);
    \draw[-] (or.input 2) -- (b_label.north);
\end{tikzpicture}%
    }
    \label{fig:idea_dft}
}
\hspace{0.5cm}
\subfigure[Basic scheme]{
    \scalebox{0.7}{
    \begin{tikzpicture}[every label/.append style={font=\scriptsize}]
    \node[iplace, label=0:{$\Failed_Z$}] (zfailed) {};
    \node[scale=0.8, cloud, draw,cloud puffs=10,cloud puff arc=120, aspect=2, inner ysep=0.5em, below=0.2cm of zfailed] (zcloud) {$Z$};
    \node[iplace, below=1.0cm of zfailed, xshift=-0.7cm, label=180:{$\Failed_X$}] (xfailed) {};
    \node[scale=0.8,cloud, draw,cloud puffs=10,cloud puff arc=120, aspect=2, inner ysep=0.5em, below=0.2cm of xfailed] (xcloud) {$X$};
    \node[iplace, below=1.2cm of xfailed, xshift=-0.6cm, label=180:{$\Failed_A$}] (afailed) {};
    \node[scale=0.8,cloud, draw,cloud puffs=10,cloud puff arc=120, aspect=2, inner ysep=0.5em, below=0.2cm of afailed] (acloud) {$A$};
    \node[iplace, below=1.2cm of xfailed, xshift=0.9cm, label=0:{$\Failed_B$}] (bfailed)  {};
    \node[scale=0.8,cloud, draw,cloud puffs=10,cloud puff arc=120, aspect=2, inner ysep=0.5em, below=0.2cm of bfailed] (bcloud) {$B$};

    \draw[->] (xcloud) -- (xfailed);
    \draw[->] (zcloud) -- (zfailed);
    \draw[->] (acloud) -- (afailed);
    \draw[->] (bcloud) -- (bfailed);
    \draw[->] (bfailed) -- (zcloud);
    \draw[->] (afailed) -- (xcloud);
    \draw[->] (bfailed) -- (xcloud);
    \draw[->] (xfailed) -- (zcloud);
\end{tikzpicture}%
    }
    \label{fig:idea_translation}
}
\hspace{0.5cm}
\subfigure[Simplified resulting GSPN]{
    \scalebox{0.8}{
    \begin{tikzpicture}[every label/.append style={font=\scriptsize}]
  	\node[Ttransition, label={[name=label_lama] left:$\lambda_A$}] (Af) {};
  	\node[Ttransition, right=1.8cm of Af, label={[name=label_lamb] right:$\lambda_B$}] (Bf) {};
  	\node[iplace, above=0.4cm of Af, label={[name=label_a] left:$\Failed_A$}] (AF) {};
  	\node[iplace, above=0.4cm of Bf, label={[name=label_b] right:$\Failed_B$}] (BF) {};

  	\draw[o->] (Af) -- (AF);
  	\draw[o->] (Bf) -- (BF);

  	\node[iplace, above=1.0cm of AF, label={[name=label_x] left:$\Failed_X$}] (XF) {};
    \node[Itransition, above=0.4cm of AF, label={[name=label_f1] left:$t_1$@$3$}] (Xf1) {};
    \node[Itransition, right=0.3cm of Xf1, label={[name=label_f2] below:$t_2$@$3$}] (Xf2) {};

	\draw[<->] (AF) -- (Xf1);
  	\draw[<->] (BF) -- (Xf2);

  	\draw[o->] (Xf1) -- (XF);
  	\draw[o->] (Xf2) -- (XF);

    \node[iplace, above=2.4cm of BF, label={[name=label_z] above:$\Failed_Z$}] (v_failed) {};
    \node[place, left=0.9cm of v_failed, label={[name=label_fs] above:$\Failsafe_Z$},] (v_failsafe) {};
    \node[Itransition, below=0.6cm of v_failsafe,label={[name=label_t1] 135:$t_3$@$2$}] (Zt1) {};
    \node[Itransition, below=0.6cm of v_failed, label={[name=label_t2] right:$t_4$@$2$}] (Zt2) {};

    \draw[<->] (XF) -- (Zt2);
  	\draw[<->] (BF) -- (Zt2);
  	\draw[-o] (v_failsafe) -- (Zt2);
  	\draw[<-o] (v_failed) -- (Zt2);
  	\draw[<->] (BF) -- (Zt1);
  	\draw[-o] (XF) -- (Zt1);
  	\draw[<-o] (v_failsafe) -- (Zt1);

    \node[draw,dotted,fit=(label_lama) (Af)] {};
    \node[draw,dotted,fit=(label_lamb) (Bf)] {};
    \draw[dotted] (label_f1.south west) |- (label_f1.north west) -| (label_f2.south east) -| (label_f1.south west);
    \draw[dotted] (label_t1.south west) |- (label_fs.north west) -- (label_fs.north east) |- (label_t2.north west) -- (label_t2.north east) -| (label_t2.south east) -| (label_t1.south west);
\end{tikzpicture}%
    }
    \label{fig:idea_gspn}
}
\caption{Compositional semantics of DFTs using GSPNs}
\label{fig:dft_to_gspn_scheme}
\end{figure}
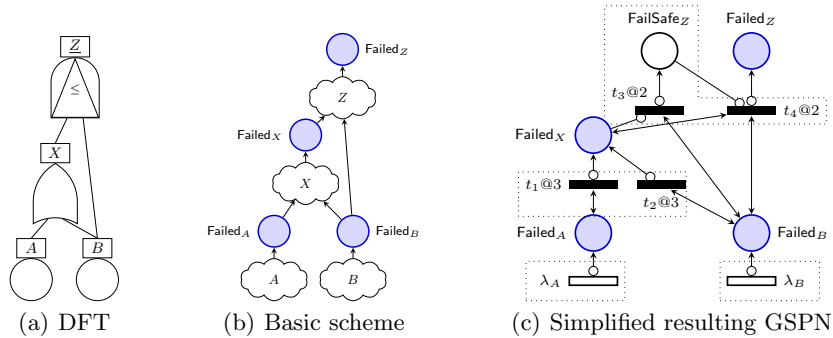

Standard FTs appeal due to their simplicity.
However, they lack expressive power to faithfully model many aspects of realistic systems such as spare components, redundancies, etc.
This deficiency is remedied by \emph{Dynamic Fault Trees} (DFTs, for short)~\cite{DBB90}.
They involve a variety of new gates such as spares and functional dependencies.
These gates are dynamic as their behaviour depends on the failure history.
For instance, the DFT in Fig.~\ref{fig:toy_dft} extends our sample FT.
If the power fails while the switch is operational, the system can switch to the UPS.
However, if the power fails after the switch failed, their parent \PAND-gate causes the system to immediately fail\footnote{A \PAND-gate fails if all its children fail in a left-to-right order.}.
The expressive power of DFTs allows for modelling complex failure combinations succinctly.
This power comes at a price:
the interpretation of DFTs leaves quite some freedom and the complex interplay between the gates easily leads to misinterpretations~\cite{JGKS16}.
The DFT in Fig.~\ref{fig:idea_dft} raises the question whether $B$'s failure first causes $X$ to fail which in turn causes $Z$ to fail, or whether $B$'s failure is first propagated to $Z$ making it impossible for $Z$ to fail any more?
These issues are not just of theoretical interest.
Slightly different interpretations may lead to significantly divergent reliability measures and give rise to distinct underlying stochastic (decision) processes.

This paper defines a \emph{unifying} semantics of DFTs using generalised stochastic Petri nets (GSPNs)~\cite{DBLP:journals/tocs/MarsanCB84,Mar95}.
The use of GSPNs to give a meaning to DFTs is not new; GSPN semantics of (dynamic) fault trees have received quite some attention in the literature~\cite{DBLP:journals/entcs/Raiteri05,bobbio2004parametric,BFGP03,KWP15}.
Many DFT features are naturally captured by GSPN concepts, e.g., the failure of a basic event can be modelled by a timed transition, the instantaneous failure of a gate by an immediate transition, and places can be exploited to pass on failures.
This work builds upon the GSPN-based semantics in~\cite{DBLP:journals/entcs/Raiteri05}.
The appealing feature of our GSPN semantics is that it \emph{unifies} various existing DFT semantics, in particular various state-space based meanings using Markov models~\cite{CSD00,Boudali2010,VJK17}, such as continous-time Markov Chains (CTMC), Markov automata (MA)~\cite{EHZ10b}, a form of continous-time Markov decision process, or I/O interactive Markov chain (IOIMC)~\cite{DBLP:books/sp/Hermanns02}.
The key is that we capture all these distinct interpretations by a \emph{single} GSPN.
The structure of the net is the same for all possible meanings.
Only two net features vary: the transition priorities and the partitioning of immediate transitions.
The former steer the ordering of how failures propagate through a DFT, while the latter control the possible ways in which to resolve conflicts (and confusion)~\cite{EHKZ13}.

\begin{table}[tb]
\scriptsize{
\centering
\caption{Semantic differences between supported semantics}
\label{tab:semantic_differences}
\begin{tabular}{@{}p{2.4cm}|p{1.8cm}p{1.7cm}p{1.8cm}p{1.9cm}p{1.9cm}@{}}
\toprule
                                                                     & Monolithic CTMC~\cite{CSD00}  & IOIMC~\cite{Boudali2010}            & Monolithic MA~\cite{VJK17}          & Orig. GSPN~\cite{DBLP:journals/entcs/Raiteri05}              & New GSPN \\
\midrule
Tool support                                                         & \galileo~\cite{SDC99}         & \dftcalc~\cite{ABBGS13}             & \storm~\cite{DehnertJKV17}     & |                               & | \\
Underlying model                                                     & CTMC                          & IMC~\cite{DBLP:books/sp/Hermanns02} & MA~\cite{EHZ10b}               & GSPN/CTMC\newline\cite{DBLP:journals/tocs/MarsanCB84,Mar95}               & GSPN/MA~\cite{EHKZ13} \\
\midrule
Priority gates                                                       & $\leq$                        & $<$                                 & $\leq$                         & $<$                             & $\leq$ and $<$ \\
Nested spares                                                        & not supported                 & late claiming                       & early claiming                 & not supported                             & early claiming \\
Failure propagation                                                  & bottom-up                     & arbitrary                           & bottom-up                      & arbitrary                       & bottom-up \\
\FDEP forwarding                                                     & first                  & interleaved & last                    & interleaved & first \\
Non-determinism                                                      & uniform & true \newline (everywhere)                  & true \newline \FDEP       & uniform  & true \newline (\PAND, \SPARE) \\
\bottomrule
\end{tabular}
}
\end{table}

The benefits of a unifying GSPN are manifold.
First and foremost, it gives insights in the choices that DFT semantics from the literature --- and the tools realising these semantics --- make.
We show that already three DFT aspects distinguish them all: failure propagation, forwarding in functional dependencies, and non-determinism, see the last three rows in Table~\ref{tab:semantic_differences}.
Mature tool-support for GSPNs such as SHARPE \cite{TS09}, SMART~\cite{DBLP:journals/sigmetrics/CiardoMW09}, GreatSPN~\cite{DBLP:journals/sigmetrics/BaarirBCPDF09} and its editor~\cite{Amparore14} can be exploited for all covered DFT semantics.
Thirdly, our \emph{compositional} approach, with simple GPSNs for each DFT gate, is easy to extend with more gates.
The compositional nature is illustrated in Fig.~\ref{fig:dft_to_gspn_scheme}.
The occurrence of an event like the failure of a DFT node is reflected by a dedicated (blue) place.
The behaviour of a gate is represented by immediate transitions (solid bars) and auxiliary (white) places.
Failing \BE{}s are triggered by timed transitions (open bars).

Our framework allows for expressing different semantics by a mild variation of the GSPN; e.g., whether $B$'s failure is first propagated to $X$ or to $Z$ can be accommodated by imposing different transition priorities.
The paper supports a rich class of DFTs as indicated in Table~\ref{tab:syntax_overview}.
The first column refers to the framework, the next four columns to existing semantics from the literature, and the last column to a new instantiation with mild restrictions, but presumably more intuitive semantics.
The meaning of the rows is clarified in Sect.~\ref{sec:dfts}.

\begin{table}[tb]
\centering
\scriptsize{
\caption{Syntax supported by different semantics}
\label{tab:syntax_overview}
\begin{tabular}{@{}l|cccccc@{}}
\toprule
DFT Feature                & Framework  & Monolithic CTMC & IOIMC       & Monolithic MA & Orig. GSPN & New GSPN \\
\midrule
Share \SPARE{}s            & \checkmark & \checkmark      & \checkmark  & \checkmark    & \cross     & \checkmark \\
\SPARE w/ subtree      & \checkmark & \cross          & \checkmark  & \checkmark    & \cross     & \checkmark \\
Shared primary             & \checkmark          & \cross               & \checkmark  & \cross        & \cross          & \checkmark \\
Priority gates             & \PAND/\POR & \PAND           & \PAND       & \PAND/\POR    & \PAND      & \PAND/\POR \\
Downward \FDEP{}s          & \checkmark & \cross          & \checkmark  & \checkmark    & \cross     & \cross \\
\SEQ{}s on gates           & \cross     & \checkmark      & \cross            & \checkmark    & \cross     & \cross \\
\PDEP                      & \checkmark & \cross          & \cross      & \checkmark    & \cross     & \checkmark \\
\bottomrule
\end{tabular}
}
\end{table}

\paragraph{Related work.}
The semantics of DFTs is naturally expressed by a state-transition diagram such as a Markov model~\cite{CSD00,Boudali2010,VJK17}.
Support of nested dynamic gates is an intricate issue, and the resulting Markov model is often complex.
To overcome these drawbacks, semantics using higher-order formalisms such as Bayesian Networks \cite{DBLP:journals/ress/MontaniPBC08,DBLP:journals/tr/BoudaliD06}, Boolean logic driven Markov processes~\cite{Bouissou03,Rauzy15} or GSPNs~\cite{BFGP03,DBLP:journals/entcs/Raiteri05} have been proposed.
DFT semantics without an underlying state-space have also been investigated, \cf \eg \cite{MRL14,Walker2009}.
These semantics often consider restricted classes of DFTs, but can circumvent the state-space explosion.
Fault trees have been expressed or extracted from domain specific languages for reliability analysis such as Hip-HOPS, which internally may use Petri net semantics \cite{CHEN201391}.
For a preliminary comparison, we refer to \cite{JGKS16, trivedi_bobbio_2017}.
Semantics for DFTs with repairs~\cite{bobbio2004parametric}, or maintenance~\cite{DBLP:conf/icfem/GuckSS15} are more involved~\cite{DBLP:journals/ress/Raiteri11}, and not considered in this paper.

\paragraph{Organisation of the paper.}
Sect.~\ref{sec:preliminaries} introduces the main concepts of GSPNs and DFTs.
Sect.~\ref{sec:translation} presents our compositional translation from DFTs to GSPNs for the most common DFT gate types.
It includes some elementary properties of the obtained GSPNs and reports on prototypical tool-support.
Sect.~\ref{sec:semantics} discusses DFT semantics from the literature based on  the unifying GSPN semantics.
Sect.~\ref{sec:conclusion} concludes and gives a short outlook into future work.
App.~\ref{sec:extensions} shows how other DFT gates can be captured in our framework while proofs are provided in App.~\ref{sec:proofs}.

\section{Preliminaries}
\label{sec:preliminaries}
\subsection{Generalised Stochastic Petri Nets}
This section summarises the semantics of GSPNs as given in~\cite{EHKZ13}.
The GSPNs are (as usual) Petri nets with timed and immediate transitions.
The former model the failure of basic events in DFTs, while the latter represent the instantaneous behaviour of DFT gates.
Inhibitor arcs ensure that transitions do not fire repeatedly, to naturally model that components do not fail repeatedly.
Transition weights allow to resolve possible non-determinism.
Priorities will (as explained later) be the key to distinguish the different DFT semantics; they control the order of transition firings for, e.g., the failure propagation in DFTs.
Finally, partitions of immediate transitions allow for a flexible treatment of non-determinism.

\begin{definition}[GSPN]
	A \emph{generalised stochastic Petri net (GSPN)} $\gspn$ is a tuple $(\petriPlaces, \petriTransitions, I, O, H, m_0, \petriWeight, \petriPD, \petriPriority, \petriPartition)$ where
    \begin{compactitem}
        \item $\petriPlaces$ is a finite set of \emph{places}.
        \item $\petriTransitions = \petriImmediate \cup \petriTimed$ is a finite set of \emph{transitions}, partitioned into the set $\petriImmediate$ of \emph{immediate transitions} and the set $\petriTimed$ of \emph{timed transitions}.
        \item $I, O, H\colon \petriTransitions \rightarrow (\petriPlaces \rightarrow \NN)$, the \emph{input-}, \emph{output-} and \emph{inhibition-multiplicities} of each transition, respectively.
        \item $m_0 \in \marking$ is the \emph{initial marking} with $\marking = \petriPlaces \rightarrow \NN$ the set of \emph{markings}.
        \item $\petriWeight \colon \petriTransitions \rightarrow \RRplus$ are the \emph{transition-weights}.
        \item $\petriPD$ is the \emph{priority domain} and
                 $\petriPriority \colon \petriTransitions \rightarrow \petriPD$ the \emph{transition-priorities}.
        \item $\petriPartition \in 2^{\petriImmediate}$, a \emph{partition} of the immediate transitions.
    \end{compactitem}
\end{definition}
For convenience, we write $\gspn = (\petriNet, \petriWeight, \petriPD, \petriPriority, \petriPartition)$ and $\petriNet = (\petriPlaces, \petriTransitions, I, O, H, m_0)$.
The definition is as in \cite{EHKZ13} extended by priorities and with a mildly restricted (i.e., marking-independent) notion of partitions.
An example GSPN is given in Fig.~\ref{fig:idea_gspn} on page~\pageref{fig:idea_gspn}.
Places are depicted by circles, transitions by open (solid) bars for timed (immediate) transitions.
If $I(t,p) > 0$, we draw a directed arc from place $p$ to transition $t$.
If $O(t,p) > 0$, we draw a directed arc from $t$ to $p$.
If $H(t,p) > 0$, we draw a directed arc from $p$ to $t$ with a small circle at the end.
The arcs are labelled with the multiplicities.
For all gates in the main text, all multiplicities are one (and are omitted).
Some gates in App.~\ref{sec:extensions} require a larger multiplicity.
Transition weights are prefixed with a $w$, transition priorities with an $@$, and may be omitted to avoid clutter.

We describe the GSPN semantics for $\petriPD = \NN$, and assume in accordance with \cite{Mar95} that for all $t \in \petriTimed: \petriPriority(t)= 0$ and for all $t \in \petriImmediate: \petriPriority(t) = c > 0$.
Other priority domains are used in Sect.~\ref{sec:semantics}.
The semantics of a GSPN are defined by its \emph{marking graph} which constitutes the state space of a MA.
In each marking, a set of transitions are enabled.
\begin{definition}[Concession, enabled transitions, firing]
	The set \conc{m} of \emph{conceded transitions} in $m \in \marking$ is:
	\[ \conc{m} = \{ t \in T \sep \forall p \in \petriPlaces: m(p) \geq I(t)(p) \land m(p) < H(t)(p) \} \]
	The set $\enabled{m}$ of \emph{enabled transitions in $m$} is:
	\[ \enabled{m} = \conc{m} \cap \{ t \in \petriTransitions \sep \petriPriority(t) = \max_{t \in \conc{m}} \petriPriority(t)  \} \]
The \emph{effect of firing $t \in \enabled{m}$ on $m \in \marking$} is a marking $\fire(m, t)$ such that:
\[\forall p \in \petriPlaces: \fire(m,t)(p) = m(p) - I(t)(p) + O(t)(p). \]
\end{definition}
\begin{example}
    Consider again the GSPN in Fig.~\ref{fig:idea_gspn}.
    Let $m \in \marking$ be a marking with $m(\Failed_B)=1$ and $m(p)=0$ for all $p \in P \setminus \{\Failed_B\}$.
    Then the transitions $t_2$ and $t_3$ have concession, but only $t_2$ is enabled.
    Firing $t_2$ on $m$ leads to the marking $m'$ with $m'(\Failed_B)=1=m'(\Failed_X)$, and $m'(p)=0$ for $p \in \{\Failed_A, \Failed_Z, \Failsafe_Z\}$.
\end{example}
If multiple transitions are enabled in a marking $m$, there is a \emph{conflict} which transition fires next.
For transitions in different partitions, this conflict is resolved non-deterministically (as in non-stochastic Petri nets).
For transitions in the same partition the conflict is resolved probabilistically (as in the GSPN semantics of \cite{Mar95}).
Let $C = \enabled{m} \cap D$ be the set of enabled transitions in $D \in \petriPartition$.
Then transition $t \in C$ fires next with probability $\frac{\petriWeight(t)}{\sum_{t' \in C} \petriWeight(t')}$.
If in a marking only timed transitions are enabled, in the corresponding state, the sojourn time is exponentially distributed with exit rate $\sum_{t' \in C} \petriWeight(t')$.
If a marking enables both timed and immediate transitions, the latter prevail as the probability to fire a timed transition immediately is zero.

A Petri net is \emph{$k$-bounded} for $k \in \NN$ if for every place $p \in \petriPlaces$ and for every reachable marking $m(p) \leq k$.
Boundedness of a GSPN is a sufficient criterion for the finiteness of the marking graph.
A $k$-bounded GSPN has a \emph{time-trap} if its marking graph contains a cycle $m \xrightarrow{t_1} m_1 \xrightarrow{t_2} \hdots \xrightarrow{t_n} m$ such that for all $1 \leq i \leq n$, $t_i \in \petriImmediate$.
The absence of time-traps is important for analysis purposes.

\label{sec:gspns}
\subsection{Dynamic Fault Trees}
\label{sec:dfts}
This section, based on~\cite{VJK17}, introduces DFTs 
and their nodes, and gives some formal definitions for concise notation in the remainder of the paper.
The DFT semantics are clarified in depth in the main part of the paper.

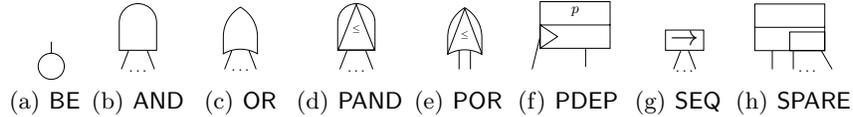
\begin{figure}[t]
\centering
\subfigure[\BE]{
 \centering
%  \tikzsetnextfilename{dftelements_be}
\makebox[0.065\linewidth]{
\scalebox{0.62}{
 \begin{tikzpicture}[  scale=.8,font=\LARGE,text=black, every node/.style={transform shape}, node distance=0.3cm]
	\node[be] (pand) {};
	\node[above=of pand] (output) {};
	\draw[-] (pand) -- (output);
\end{tikzpicture}}}
 \label{fig:BE}
}
%\subfigure[\small$\VOT k$]{
% \centering
%%  \tikzsetnextfilename{dftelements_vot}
%\makebox[0.095\linewidth]{
%\scalebox{0.62}{
% \begin{tikzpicture}
%    \node[and3] (and) {\rotatebox{270}{$k$}};
%    \node[below=0.4 cm of and.input 1, xshift=-0.2cm]  (i1) {};
%    \node[below=0.3 cm of and.input 2]  (dots) {$\hdots$};
%    
%    \node[below=0.4 cm of and.input 3, xshift=0.2cm] (i2) {};
%    
%    \draw[-] (and.input 1) -- (i1);
%    \draw[-] (and.input 3) -- (i2);
%  \end{tikzpicture}}}
% \label{fig:VOT}
%}
\subfigure[$\AND$]{
  \centering
%   \tikzsetnextfilename{dftelements_and}
\makebox[0.085\linewidth]{
\scalebox{0.62}{
 \begin{tikzpicture}
    \node[and3] (and) {};
    \node[below=0.4 cm of and.input 1, xshift=-0.2cm]  (i1) {};
    \node[below=0.3 cm of and.input 2]  (dots) {$\hdots$};
    
    \node[below=0.4 cm of and.input 3, xshift=0.2cm] (i2) {};
    
    \draw[-] (and.input 1) -- (i1);
    \draw[-] (and.input 3) -- (i2);
  \end{tikzpicture}
  }
  }
 \label{fig:AND} 
}
\subfigure[$\OR$]{
  \centering
%   \tikzsetnextfilename{dftelements_or}
\makebox[0.085\linewidth]{
\scalebox{0.62}{
 \begin{tikzpicture}
    \node[or3] (and) {};
    \node[below=0.4 cm of and.input 1, xshift=-0.2cm]  (i1) {};
    \node[below=0.3 cm of and.input 2]  (dots) {$\hdots$};
    
    \node[below=0.4 cm of and.input 3, xshift=0.2cm] (i2) {};
    
    \draw[-] (and.input 1) -- (i1);
    \draw[-] (and.input 3) -- (i2);
  \end{tikzpicture}
  }
  }
 \label{fig:OR}
}
 \subfigure[\PAND]{
 \centering
\makebox[0.1\linewidth]{
 \scalebox{0.62}{
   \begin{tikzpicture}   
    \node[and3] (and) {{\tiny\rotatebox{270}{$\leq$}}};
    \node[triangle,scale=1.62,yshift=-3.5,xscale=0.80] (triangle_a) at (and) {};
    \node[below=0.4 cm of and.input 1, xshift=-0.2cm]  (i1) {};
    \node[below=0.3 cm of and.input 2]  (dots) {$\hdots$};
    
    \node[below=0.4 cm of and.input 3, xshift=0.2cm] (i2) {};
    
    \draw[-] (and.input 1) -- (i1);
    \draw[-] (and.input 3) -- (i2);
  \end{tikzpicture}}}
  \label{fig:DFTElements_PAND}
  
 }
\subfigure[\POR]{
  \centering
\makebox[0.083\linewidth]{
  \scalebox{0.62}{
    \begin{tikzpicture}   
    \node[or2] (and) {{\tiny\rotatebox{270}{$\leq$}}};
    \node[btriangle,scale=1.61,yscale=0.915, xshift=-0.113cm] (triangle_b) at (and) {};
    \node[below=0.4 cm of and.input 1]  (i1) {};
    
    \node[below=0.4 cm of and.input 2] (i2) {};
    
    \draw[-] (and.input 1) -- (i1);
    \draw[-] (and.input 2) -- (i2);
  \end{tikzpicture}}}
  \label{fig:DFTElements_POR}
 }
 \subfigure[\PDEP]{
   \centering
   \scalebox{0.62}{
      \begin{tikzpicture}   
    
    \node[fdep] (and) {};
    \node[above=0.07cm of and.center] (x) {$p$};
    \node[below=0.7 cm of and.T, xshift=-0.2cm]  (i1) {};
    
    \node[below=0.4 cm of and.EB] (i2) {};
    \draw[-] (and.T) -- (i1);
    \draw[-] (and.EB) -- (i2);
    
  \end{tikzpicture}
  }
    \label{fig:DFTElements_PDEP}
 }
 \subfigure[\SEQ]{
  \centering
  \makebox[0.083\linewidth]{
  \scalebox{0.62}{
    \begin{tikzpicture}   
    \node[seq] (and) {$\rightarrow$};
    \node[below=0.4 cm of and.250, xshift=-0.2cm]  (i1) {};
    \node[below=0.3 cm of and.270]  (dots) {$\hdots$};
    
    \node[below=0.4 cm of and.290, xshift=0.2cm] (i2) {};
    
    \draw[-] (and.250) -- (i1);
    \draw[-] (and.290) -- (i2);
  \end{tikzpicture}}}
  \label{fig:DFTElements_SEQ}
 }
 \subfigure[\SPARE]{
\centering
\makebox[0.115\linewidth]{
\scalebox{0.62}{
  \begin{tikzpicture}   
    \node[spare] (and) {};
    \node[below=0.4 cm of and.P]  (i1) {};

    \node[below=0.4 cm of and.SA] (i2) {};
    \node[below=0.3 cm of and.SC] (i3) {$\hdots$};
    
    \node[below=0.4 cm of and.SE, xshift=0.3cm] (i4) {};
    
    \draw[-] (and.P) -- (i1);
    \draw[-] (and.SA) -- (i2);
    \draw[-] (and.SE) -- (i4);
  \end{tikzpicture} }}
  \label{fig:DFTElements_SPARE}
 }
 \caption{Node types in ((a)-(c)) static and (all) dynamic fault trees.} % hard-coded refs
 \label{fig:DFTElements}
\end{figure}

Fault trees (FTs) are directed acyclic graphs with typed nodes.
Nodes without successors (or: \emph{children}), are \emph{basic events} (\BE{}s).
All other nodes are \emph{gates}.
\BE{}s represent system components that can fail.
Initially, a \BE is \emph{operational};  it \emph{fails} according to a negative exponential distribution.
A gate fails if its \emph{failure condition} over its children is fulfilled.
The key gates for static fault trees (SFTs) are typed \AND and \OR, shown in Fig.~\ref{fig:DFTElements}(b,c). 
These gates fail if all (\AND) or at least one (\OR) children have failed, respectively.
Typically, FTs express for which occurrences of \BE failures, a specifically marked node (\emph{top-event}) fails.

SFTs lack an internal state --- the failure condition is independent of the history.
Therefore, SFTs lack expressiveness~\cite{JGKS16,RS15}.
Several extensions commonly referred to as \emph{Dynamic Fault Trees} (DFTs) have been introduced to increase the expressiveness.
The extensions introduce new node types, shown in Fig.~\ref{fig:DFTElements}(d-h);
we categorise them as \emph{priority gates}, \emph{dependencies}, \emph{restrictors}, and \emph{spare gates}.
\subsubsection{Priority gates.}
These gates extend static gates by imposing a condition on the ordering of failing children and allow for order-dependent failure propagation.
A \emph{priority-and} (\PAND) fails if all its children have failed in order from left to right.
Fig.~\ref{fig:PandVsSeq_A} depicts a \PAND with two children.
It fails if $A$ fails before $B$ fails.
The \emph{priority-or} (\POR)~\cite{Walker2009} only fails if the leftmost child fails before any of its siblings do.
The semantics for simultaneous failures is discussed in Sect.~\ref{sec:templates}.
If a gate cannot fail any more, e.g., when $B$ fails before $A$ in Fig.~\ref{fig:PandVsSeq_A}, it is \emph{fail-safe}.
\subsubsection{Dependencies.}
Dependencies do not propagate a failure to their parents, instead, when their \emph{trigger} (first child) fails, they update their \emph{dependent events} (remaining children).
We consider \emph{probabilistic dependencies} (\PDEP{}s) \cite{DBLP:journals/ress/MontaniPBC08}.
Once the trigger of a \PDEP fails, its dependent events fail with probability $p$.
Fig.~\ref{fig:PDEPexample} shows a \PDEP where the failure of trigger $A$ causes a failure of \BE $B$ with probability $0.8$ (provided it has not failed before).
\emph{Functional dependencies} (\FDEP{}s) are \PDEP{} with probability one (we omit the $p$ then).
\subsubsection{Restrictors.}
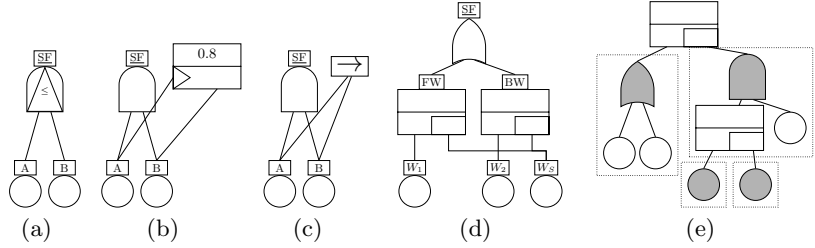
\begin{figure}[tb]
\centering
\subfigure[]{
  \makebox[.1\linewidth]{ \scalebox{\dftscale}{
    \begin{tikzpicture}[scale=.6,text=black]
        \node[and2] (pand_a)  {{\tiny\rotatebox{270}{$\leq$}}};
        \node[triangle,scale=1.62,yshift=-3.5,xscale=0.80] (triangle_a) at (pand_a) {};
        \node[labelbox] (panda_label) at (pand_a.east){\underline{SF}};

        \node[be, below=1.4cm of pand_a.input 1, xshift=-0.3cm] (c1) {};
        \node[labelbox] (c1_label) at (c1.north){A};

        \node[be, below=1.4cm of pand_a.input 2, xshift=0.3cm] (cn) {};
        \node[labelbox] (cn_label) at (cn.north){B};

        \draw[-] (pand_a.input 1)   -- (c1_label.north);
        \draw[-] (pand_a.input 2)   -- (cn_label.north);
    \end{tikzpicture}
  }}
%   \caption{Regarding sequences via \PAND{}s.}
   \label{fig:PandVsSeq_A}
 }%\hfill%
 \subfigure[]{ \centering
  \makebox[.14\linewidth]{ \scalebox{\dftscale}{
    \begin{tikzpicture}[scale=.6,text=black]
        \node[and2] (pand_a) {};
        \node[labelbox] (panda_label) at (pand_a.east){\underline{SF}};

        \node[fdep, right=0.8cm of pand_a] (fdep) {};
        \node[above=0.07cm of fdep.center] (x) {$0.8$};
        \node[be, below=1.4cm of pand_a.input 1, xshift=-0.3cm] (c1) {};
        \node[labelbox] (c1_label) at (c1.north){A};

        \node[be, below=1.4cm of pand_a.input 2, xshift=0.3cm] (cn) {};
        \node[labelbox] (cn_label) at (cn.north){B};

        \draw[-] (pand_a.input 1)   -- (c1_label.north);
        \draw[-] (pand_a.input 2)   -- (cn_label.north);
        \draw[-] (fdep.T) -- (c1_label.north);
        \draw[-] (fdep.EB) -- (cn_label.north);
    \end{tikzpicture}
  }}
%   \caption{Restricting sequences via \SEQ{}s.}
 \label{fig:PDEPexample}
}%\hfill%
\subfigure[]{ \centering
  \makebox[.14\linewidth]{ \scalebox{\dftscale}{
    \begin{tikzpicture}[scale=.6,text=black]
        \node[and2] (pand_a) {};
        \node[labelbox] (panda_label) at (pand_a.east){\underline{SF}};

        \node[seq,right=0.7cm of pand_a] (seq) {$\rightarrow$};

        \node[be, below=1.4cm of pand_a.input 1, xshift=-0.3cm] (c1) {};
        \node[labelbox] (c1_label) at (c1.north){A};

        \node[be, below=1.4cm of pand_a.input 2, xshift=0.3cm] (cn) {};
        \node[labelbox] (cn_label) at (cn.north){B};

        \draw[-] (pand_a.input 1)   -- (c1_label.north);
        \draw[-] (pand_a.input 2)   -- (cn_label.north);
        \draw[-] (seq.260) -- (c1_label.north);
        \draw[-] (seq.280) -- (cn_label.north);
    \end{tikzpicture}
  }}
%   \caption{Restricting sequences via \SEQ{}s.}
 \label{fig:PandVsSeq_B}
}%\hfill%
\subfigure[]{
  \scalebox{\dftscale}{
    \begin{tikzpicture}[scale=.6,text=black, node distance=1cm]
        \node[or2] (and) {};
        \node[labelbox] (andl) at (and.east) {\underline{SF}};
        \node[spare,below=of and,fill=white!100!black, xshift=-1.2cm] (spare1) {};

        \node[spare,right=0.35cm of spare1,fill=white!100!black] (spare2) {};
        \node[labelbox] (rl1) at (spare1.north) {FW};
        \node[labelbox] (rl2) at (spare2.north) {BW};

        \node[be,node distance=0.9cm,  below =of spare1.P] (p1) {};
        \node[be,node distance=0.9cm,  below =of spare2.P] (p2) {};
        \node[be,right =0.35cm of p2] (S) {};
        \node[labelbox] (pl1) at (p1.north) {$W_1$};
        \node[labelbox] (pl2) at (p2.north) {$W_2$};
        \node[labelbox] (sl) at (S.north) {$W_S$};

        \draw[-] (rl1.north) -- (and.input 1);
        \draw[-] 	(rl2.north) -- (and.input 2);
        \draw[-] 	(spare1.P) -- (pl1.north);
        \draw[-] 	(spare2.P) -- (pl2.north);
        \draw[-] (spare1.SC) -| +(0,-0.6)  -| (sl.north);
        \draw[-] (spare2.SC) -| +(0,-0.6)  -| (sl.north);
    \end{tikzpicture}
  }
  \label{fig:SpareExample}
}
\subfigure[]{
  \scalebox{\dftscale}{
    \begin{tikzpicture}[scale=.6,text=black, node distance=1.1cm, box/.style={draw, densely dotted, inner sep=4pt}]
        \node[spare,fill=white!100!black, xshift=-3.8cm] (spare1) {};
        \node[or2,fill=black!30!white, below=1.4cm of spare1.center, yshift=1.4cm] (a1) {};

        \node[be, below=of a1.center, xshift=-0.4cm] (a2) {};
        \node[be, below=of a1.center, xshift=0.4cm] (a3) {};

        \node[and2,fill=black!30!white,below=1.2cm of spare1.center, yshift=-1cm] (b1) {};

        \node[spare,below=0.6cm of b1, xshift=-0.8cm] (spare2) {};
        \node[be, below=of b1, anchor=east, xshift=0.9cm] (b2) {};

        \node[be,fill=black!30!white, below=0.4cm of spare2.P, xshift=-0.2cm] (c1) {};
        \node[be,fill=black!30!white, below=0.4cm of spare2.SC, xshift=0.2cm] (d1) {}; 

        \node[box, fit=(a1) (a2) (a3)] {};
        \node[box, fit=(b1) (spare2) (b2)] {};
        \node[box, fit=(c1)] {};
        \node[box, fit=(d1)] {};

        \draw[-] (spare1.P) -- (a1.output);
        \draw[-] (spare1.SD) -- (b1.output);
        \draw[-] (a1.input 1) -- (a2.north);
        \draw[-] (a1.input 2) -- (a3.north);
        \draw[-] (b1.input 1) -- (spare2);
        \draw[-] (b1.input 2) -- (b2.north);
        \draw[-] (spare2.P) -- (c1.north);
        \draw[-] (spare2.SD) -- (d1.north);
    \end{tikzpicture}
  }
  \label{fig:sparemodules}
}
\caption{Simple examples of dynamic nodes \cite{VJK17}.}
\label{fig:examples}
\end{figure}

Restrictors limit possible failure propagations.
\emph{Sequence enforcers} (\SEQ{}s) enforce that their children only fail from left to right.
This differs from priority-gates which do not prevent certain orderings, but only propagate if an ordering is met.
The \AND $SF$ in Fig.~\ref{fig:PandVsSeq_B} fails if $A$ and $B$ have failed (in any order), but the \SEQ enforces that $A$ fails prior to $B$.
In contrast to Fig.~\ref{fig:PandVsSeq_A}, $SF$ is never fail-safe.
Another restrictor is the \MUTEX (not depicted) which ensures that exactly one of its children fails.

% Figure with usual examples of DFTs

\subsubsection{Spare gates.}
Consider the DFT in Fig.~\ref{fig:SpareExample} modelling (part of) a motor bike with a spare wheel.
A bike needs two wheels to be operational.
Either wheel can be replaced by the spare wheel, but not both.
The spare wheel is less likely to fail until it is in use.
Assume the front wheel fails.
The spare wheel is available and used, but from now on, it is more likely to fail.
If any other wheel fails, no spare wheel is available any more, and the parent \SPARE fails.

\SPARE{}s involve two mechanisms: \emph{claiming} and \emph{activation}.
Claiming works as follows.
\SPARE{}s \emph{use} one of their children.
If this child fails, the \SPARE tries to \emph{claim} another child (from left to right).
Only operational children that have not been claimed by another \SPARE can be claimed.
If claiming fails --- modelling that all spare components have failed --- the \SPARE fails.
Let us now consider activation.
\SPARE{}s may have (independent, i.e., disjoint) sub-DFTs as children.
This includes nested \SPARE{}s, \SPARE{}s having \SPARE{}s as children.
A \emph{spare module} is a set of nodes linked to each child of the \SPARE.
This child is the \emph{module representative}.
Fig.~\ref{fig:sparemodules} gives an example of spare modules (depicted by boxes) and the representatives (shaded nodes).
Here, a spare module contains all nodes which have a path to the representative without an intermediate \SPARE.
Every leaf of a spare module is either a \BE or a \SPARE.
Nodes outside of spare modules are \emph{active}.
For each active \SPARE and used child $v$, the nodes in $v$'s spare module are activated.
Active \BE{}s fail with their active failure rate, all other \BE{}s with their passive failure rate.

\subsubsection{DFTs formally.}
We now give the formal definition of DFTs.
\begin{definition}[DFT]
	A \emph{Dynamic Fault Tree} $\DFT$ (DFT) is a tuple $(\DFTnodes, \children, \Tp, \Top)$:
	\begin{compactitem}
		\item $\DFTnodes$ is a finite set of \emph{nodes}.
		\item $\children\colon V \rightarrow V^{*}$ defines the (ordered) \emph{children} of a node.
		\item $\Tp\colon V \rightarrow \{ \BE \} \cup \{\AND, \OR, \PAND, \dots\}$ defines the \emph{node-type}.
		\item $\Top \in \DFTnodes$ is the \emph{top event}.
	\end{compactitem}
\end{definition}
For node $v \in \DFTnodes$, we also write $v \in \DFT$.
If $\Tp(v) = K$ for some $K \in \{ \BE, \AND, \dots \}$, we write $v \in \DFT_K$.
We use $\children(v)_i$ to denote the $i$-th child of $v$ and $\child{v}{i}$ as shorthand.

We assume (as all known literature) that DFTs are \emph{well-formed}, \ie
(1) The directed graph induced by $\DFTnodes$ and $\children$ is acyclic, i.e., the transitive closure of the parent-child order
is irreflexive, and
(2) Only the leaves have no children.

For presentation purposes, for the main body we restrict the DFTs to \emph{conventional DFTs}, and discuss how to lift the restrictions in App.~\ref{sec:extensions}.
\begin{definition}[Conventional DFT]
	A DFT is \emph{conventional} if
    \begin{enumerate}
	    \item Spare modules are only shared via their (unique) representative. In particular, they are disjoint. \label{restrpt:sharingviarepr}
		\item All children of a \SEQ{} are \BE{s}. \label{restrpt:seqenfoverbe}
		\item All children of an \FDEP are \BE{}s. \label{restrpt:fdepoverbe}
	\end{enumerate}
\end{definition}
Restriction~\ref{restrpt:sharingviarepr} restricts the DFTs syntactically and in particular ensures that spare modules can be seen as a single entity \wrt claiming and activation. Lifting this restriction to allow for non-disjoint spare modules raises new semantic issues \cite{JGKS16}.
Restriction~\ref{restrpt:seqenfoverbe} ensures that the fallible \BE{s} are immediately deducible.
Restriction~\ref{restrpt:fdepoverbe} simplifies the presentation, in Sect.~\ref{sec:semantics_further_issues} we relax this restriction.

\section{Generic Translation of DFTs to GSPNs}
\label{sec:translation}
The goal of this section is to define the semantics of a DFT $\DFT$ as a GSPN $\petriTemplate_\DFT$.
We first introduce the notion of GSPN templates, and present templates for the common DFT node types such as \BE, \AND, \OR, \PAND, \SPARE, and \FDEP in Sect.~\ref{sec:templates}.
(Other node types such as \PDEP, \SEQ, \POR, and so forth are treated in App.~\ref{sec:extensions}.)
Sect.~\ref{sec:formal_translation} presents how to combine the templates so as to obtain a template for an entire DFT.
Some properties of the resulting GSPNs are described in Sect.~\ref{sec:properties} while tool-support is shortly presented in Sect.~\ref{sec:tool_support}.

\subsection{GSPN templates and interface places}
Recall the idea of the translation as outlined in Fig.~\ref{fig:dft_to_gspn_scheme}.
We start by introducing the set $\interfaces_\DFT$ of \emph{interface places}:
\[
  \interfaces_\DFT = \{ \Failed_v, \Unavailable_v, \Active_v \sep v \in \DFT \} \cup \{ \Disabled_v \sep v \in \DFT_\BE\}
\]
The places $\interfaces_\DFT$ manage the communication for the different mechanisms in a DFT.
A token is placed in $\Failed_v$ once the corresponding DFT gate $v$ fails.
On the failure of a gate, the tokens in the failed places of its children are not removed as a child may have multiple parents.
Inhibitor arcs connected to $\Failed_v$ prevent the repeated failure of an already failed gate.
The $\Unavailable_v$ places are used for the claiming mechanism of \SPARE{}s, $\Active_v$ manages the activation of spare components, while $\Disabled_v$ is used for \SEQ{}s.

Every DFT node is translated into some \emph{auxiliary places}, transitions, and arcs.
The arcs either connect interface or auxiliary places with the transitions.
For each node-type, we define a template that describes how a node of this type is translated into a GSPN (fragment).

To translate contextual behaviour of the node, we use \emph{priority variables} $\prioVar = \{ \prioVar_v \sep v \in \DFT \}$.
Transition priorities are functions over the priority variables $\prioVar$, i.e., $\petriPriority\colon T \rightarrow \NN[\prioVar]$.
These variables are instantiated with concrete values in Sect.~\ref{sec:semantics}, yielding priorities in $\NN$.
This section does not exploit the partitioning of the immediate transitions; the usage of this GSPN ingredient is deferred to Sect.~\ref{sec:semantics}.
Put differently, for the moment it suffices to let each immediate transition constitute its (singleton) partition.

\begin{definition}[GSPN-Template]
The GSPN $\petriTemplate = (\petriNet, \petriWeight, \NN[\prioVar], \petriPriority, \petriPartition)$ is a \emph{($\prioVar$-parameterised) \emph{template} over $\interfaces \subseteq \petriPlaces$}.
The \emph{instantiation} of $\petriTemplate$ with $\prioConc \in \NN^n$ is the GSPN $\petriTemplate[\vec{c}] = (\petriNet, \petriWeight, \NN, \petriPriority', \petriPartition)$ with $\petriPriority'(t) = \petriPriority(t)(\prioConc)$ for all $t \in \petriTransitions$.
\end{definition}
The instantiation replaces the $n$ priority variables by their concrete values.

\subsection{Templates for common gate types}
\label{sec:templates}

We use the following notational conventions.
Gates have $n$ children.
Interface places $\interfaces$ are depicted using a blue shade; their initial marking is defined by the initialisation template, \cf Sect.~\ref{sec:init_template}. %capturing initially failed \BE{}s, active or unavailable spare components, and disabled \BE{}s.
Other places have an initial token if it is drawn in the template.
Transition priorities are indicated by $@$ and the priority function, e.g., $@\prioVar_v$.
The role of the priorities is discussed in detail in Sect.~\ref{sec:semantics}.

\subsubsection{Basic events.}
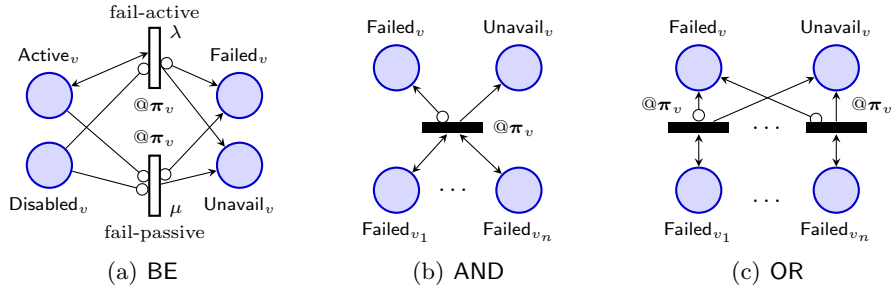
\begin{figure}[tb]
\centering
\subfigure[\BE]{
    \begin{tikzpicture}[every label/.append style={font=\scriptsize}]
    \node[iplace,label=above:{$\Active_v$}] (v_active) {};
    \node[iplace,label=below:{$\Disabled_v$}, below=0.3cm of v_active] (v_enabled) {};
    \node[iplace,label=90:{$\Failed_v$}, right=1.9cm of v_active] (v_failed) {};
    \node[iplace,label=below:{$\Unavailable_v$}, below=0.3cm of v_failed] (v_unavailable) {};
    \node[ttransition, label=above:{fail-active}, label=below:{@$\prioVar_v$}, right=1cm of v_active, yshift=0.5cm, label=60:{$\lambda$}] (v_tactive) {};
    \node[ttransition, label=below:{fail-passive}, label=above:{@$\prioVar_v$}, right=1cm of v_active, yshift=-1.2cm, label=300:{$\mu$}] (v_tpassive) {};

    \draw[-o] (v_active) -- (v_tpassive);
    \draw[<->] (v_active) -- ([yshift=2pt] v_tactive.west);
    \draw[o->] (v_tactive) -- (v_failed);
    \draw[o->] (v_tpassive) -- (v_failed);
    \draw[->] (v_tactive) -- (v_unavailable);
    \draw[->] (v_tpassive) -- (v_unavailable);

    \draw[-o] (v_enabled) edge (v_tactive);
    \draw[-o] (v_enabled) edge ([yshift=-2pt] v_tpassive.west);
\end{tikzpicture}%
    \label{fig:be_template}
}
\hspace{0.5cm}
\subfigure[$\AND$]{
    \begin{tikzpicture}[every label/.append style={font=\scriptsize}]
    \node[iplace,label=above:{$\Failed_v$}] (v_failed) {};
    \node[iplace,label=above:{$\Unavailable_v$}, right=1cm of v_failed] (v_unavailable) {};

    \node[Itransition, below=0.4cm of v_failed, xshift=0.75cm, label=right:{@$\prioVar_v$}] (v_tfail) {};
    \node[place,dummy, below=0.4cm of v_tfail] (dummy) {$\hdots$};
    \node[iplace, below=1cm of v_failed, label=below:{$\Failed_{\child{v}{1}}$}] (sv1_failed) {};
    \node[iplace, below=1cm of v_unavailable, label=below:{$\Failed_{\child{v}{n}}$}] (svn_failed) {};

    \draw[<->] (sv1_failed) -- (v_tfail);
    \draw[<->] (svn_failed) -- (v_tfail);
    \draw[o->] (v_tfail) -- (v_failed);
    \draw[->] (v_tfail) -- (v_unavailable);
\end{tikzpicture}%
    \label{fig:and_template}
}
\hspace{0.5cm}
\subfigure[$\OR$]{
    \begin{tikzpicture}[every label/.append style={font=\scriptsize}]
    \node[iplace,label=above:{$\Failed_v$}] (v_failed) {};
    \node[iplace,label=above:{$\Unavailable_v$}, right=1.2cm of v_failed] (v_unavailable) {};

    \node[Itransition, below=0.4cm of v_failed, label=135:{@$\prioVar_v$}] (v_tfail1) {};
    \node[Itransition, below=0.4cm of v_unavailable, label=45:{@$\prioVar_v$}] (v_tfailn) {};

    \node[iplace, below=1cm of v_failed, label=below:{$\Failed_{\child{v}{1}}$}] (sv1_failed) {};
    \node[iplace, below=1cm of v_unavailable, label=below:{$\Failed_{\child{v}{n}}$}] (svn_failed) {};

    \node[place,dummy, right=0.15cm of v_tfail1] (tdummy) {$\hdots$};
    \node[place,dummy, below=0.3cm of tdummy] (dummy) {$\hdots$};

    \draw[<->] (sv1_failed) -- (v_tfail1);
    \draw[<->] (svn_failed) -- (v_tfailn);
    \draw[o->] (v_tfail1) -- (v_failed);
    \draw[o->] (v_tfailn) -- (v_failed);
    \draw[->] (v_tfail1) -- (v_unavailable);
    \draw[->] (v_tfailn) -- (v_unavailable);
\end{tikzpicture}%
    \label{fig:or_template}
}
\caption{GSPN templates for basic events and static gates}
\label{fig:staticgate_templates}
\end{figure}

Fig.~\ref{fig:be_template} depicts the template $\templ{\BE}{v}$ of BE $v$.
It consists of two timed transitions, one for active failure and one for passive failure.
Place $\Failed_v$ contains a token if $v$ has failed.
The inhibitor arcs emanating $\Failed_v$ prevent both transitions to fire once the \BE has failed.
A token in $\Unavailable_v$ indicates that $v$ is unavailable for claiming by a \SPARE.
If $\Active_v$ holds a token, the node fails with the active failure rate $\lambda$, otherwise it fails with the passive failure rate $\mu$ which typically is $c{\cdot}\lambda$ with $0 < c \leq 1$.
The place $\Disabled_v$ contains a token if the \BE is not supposed to fail.
It is used in the description of the semantics of, e.g., \SEQ in App.~\ref{sec:seq_template}.

\subsubsection{\AND and \OR.}
Fig.~\ref{fig:and_template} shows the template $\templ{\AND}{v}$ for the \AND gate $v$.
A token is put in $\Failed_v$ as soon as the places $\Failed_{\child{v}{i}}$ for all children $\child{v}{i}$ contain a token.
Place $\Failed_v$ is thus marked if $v$ has failed.
Firing the (only) immediate transition puts tokens in $\Failed_v$ and $\Unavailable_v$, and returns the tokens taken from $\Failed_{\child{v}{i}}$.
Similar to the \BE template, an inhibitor arc prevents the multiple execution of the failed-transition once $v$ failed.
The template for an \OR gate is constructed analogously, see Fig.~\ref{fig:or_template}.
The failure of one child suffices for $v$ to fail; thus each child has a transition to propagate  its failure to $\Failed_v$.

\subsubsection{\PAND.}
\label{sec:templ_pand}
\begin{figure}[tb]
\centering
\subfigure[Inclusive $\PANDincl$]{
    \begin{tikzpicture}[every label/.append style={font=\scriptsize}]
    \node[iplace,label=above:{$\Failed_v$}] (v_failed) {};
    \node[iplace,label=above:{$\Unavailable_v$}, right=0.6cm of v_failed] (v_unavailable) {};
    \node[place,label=above:{$\Failsafe$}, left=1.1cm of v_failed] (v_failsafe) {};

    \node[iplace,below=1.4cm of v_failsafe,label=below:{$\Failed_{\child{v}{1}}$}] (sv1_failed) {};
    \node[iplace,right=0.4cm of sv1_failed,label=below:{$\Failed_{\child{v}{2}}$}] (sv2_failed) {};
    \node[place,dummy, right=0.3cm of sv2_failed] (dummy) {$\hdots$};
    \node[iplace,below=1.4cm of v_unavailable, label=270:{$\Failed_{\child{v}{n}}$}] (svn_failed) {};

    \node[Itransition, above=0.7cm of sv1_failed, label=120:{@$\prioVar_v$}] (t_vfailsafe1) {};
    \node[Itransition, above=0.7cm of sv2_failed, label=70:{@$\prioVar_v$}] (t_vfailsafe2) {};
    \node[Itransition, above=0.7cm of svn_failed, label=60:{@$\prioVar_v$}] (t_vfailed) {};
    \node[Itransition,dummy, above=0.7cm of dummy] (tdummy) {$\hdots$};

    \draw[-o] (sv1_failed) -- (t_vfailsafe1.south);
    \draw[<->] (sv2_failed) -- ([xshift=5pt] t_vfailsafe1.south);
    \draw[o->] (t_vfailsafe1) -- (v_failsafe);

    \draw[-o] (sv2_failed) -- (t_vfailsafe2.south);
    \draw[<->] (dummy) -- ([xshift=5pt] t_vfailsafe2.south);
    \draw[o->] (t_vfailsafe2) -- (v_failsafe);

    \draw[<->] (sv1_failed.60) -- ([xshift=-7pt] t_vfailed.south);
    \draw[<->] (sv2_failed.60) -- ([xshift=-2pt] t_vfailed.south);
    \draw[<->] (svn_failed) -- (t_vfailed.south);

    \draw[o->] ([xshift=-2pt] t_vfailed.north) -- (v_failed);
    \draw[->] (t_vfailed.north) -- (v_unavailable);
    \draw[o-]  ([xshift=-7pt] t_vfailed.north) -- (v_failsafe.10);
\end{tikzpicture}%
    \label{fig:pandincl_template}
}
\hspace{0.5cm}
\subfigure[Exclusive $\PANDexcl$]{
    \begin{tikzpicture}[every label/.append style={font=\scriptsize}]
    \node[iplace,label=above:{$\Failed_v$}] (v_failed) {};
    \node[iplace,label=above:{$\Unavailable_v$}, right=0.6cm of v_failed] (v_unavailable) {};
    \node[place,label=above:{$X_1$}, left=2.3cm of v_failed] (v_failsafe) {};
    \node[place,label=above:{$X_2$}, right=0.4cm of v_failsafe] (v_failsafe2) {};
    \node[place,dummy, right=0.4cm of v_failsafe2] (xdummy) {$\hdots$};

    \node[iplace,below=1.4cm of v_failsafe, label=below:{$\Failed_{\child{v}{1}}$}] (sv1_failed) {};
    \node[iplace,below=1.4cm of v_failsafe2, label=below:{$\Failed_{\child{v}{2}}$}] (sv2_failed) {};
    \node[place,dummy, below=1.4cm of xdummy] (dummy) {$\hdots$};
    \node[iplace,below=1.4cm of v_failed,label=below:{$\Failed_{\child{v}{n}}$}] (svn_failed) {};

    \node[Itransition, above=0.7cm of sv1_failed, label=120:{@$\prioVar_v$}] (t_vfailsafe1) {};
    \node[Itransition, above=0.7cm of sv2_failed, label=70:{@$\prioVar_v$}] (t_vfailsafe2) {};
    \node[Itransition,dummy, above=0.7cm of dummy] (tdummy) {};
    \node[Itransition, below=0.6cm of v_failed, label=right:{@$\prioVar_v$}] (t_vfailed) {};

    \draw[<->] (sv1_failed) -- (t_vfailsafe1.south);
    \draw[-o] (sv2_failed) -- ([xshift=5pt] t_vfailsafe1.south);
    \draw[o->] (t_vfailsafe1) -- (v_failsafe);

    \draw[<->] (sv2_failed) -- (t_vfailsafe2);
    \draw[-o] (dummy) -- ([xshift=5pt] t_vfailsafe2.south);
    \draw[o->] (t_vfailsafe2) -- (v_failsafe2);
    \draw[->] (v_failsafe) -- (t_vfailsafe2);

    \draw[->] (v_failsafe2) -- (tdummy);

    \draw[->] (xdummy) -- ([xshift=-3pt] t_vfailed.north);
    \draw[o->] (t_vfailed.north) -- (v_failed);
    \draw[->] ([xshift=3pt] t_vfailed.north) -- (v_unavailable);
    \draw[<->] (svn_failed) -- (t_vfailed);
\end{tikzpicture}%
    \label{fig:pandexcl_template}
}
\caption{GSPN templates for inclusive and exclusive \PAND}
\label{fig:pand_templates}
\end{figure}
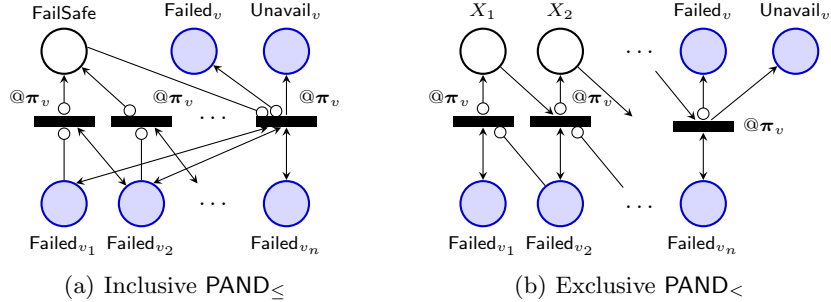

We distinguish two versions \cite{CSD00} of the priority gate \PAND: \emph{inclusive} (denoted $\leq$) and \emph{exclusive} (denoted $<$).

The \emph{inclusive} $\PANDincl$ $v$ fails if all its children failed in order from left to right while \emph{including simultaneous failures} of children.
Fig.~\ref{fig:pandincl_template} depicts its template.
If child $\child{v}{i}$ failed but its left sibling $\child{v}{i-1}$ is still operational, the $\PANDincl$ becomes fail-safe, as reflected by placing a token in $\Failsafe$.
The inhibitor arc of $\Failsafe$ now prevents the rightmost transition to fire, so no token can be put in $\Failed_v$ any more.
If all children failed from left-to-right and $\PANDincl$ is not fail-safe, the rightmost transition can fire modelling the failure of the $\PANDincl$.

The \emph{exclusive} $\PANDexcl$ $v$ is similar but \emph{excludes the simultaneous failure} of children.
Its template is shown in Fig.~\ref{fig:pandexcl_template} and uses the auxiliary places $X_1, \dots, X_{n-1}$ which indicate if the previous child failures agree with the strict failure order.
A token is placed in $X_i$ if a token is in $X_{i-1}$ and the child $\child{v}{i}$ has just failed but its right sibling $\child{v}{i+1}$ is still operational.
A token can only be put in $\Failed_v$ if the rightmost child fails and $X_{n-1}$ contains a token.
If the child $\child{v}{i}$ violates the order, the inhibitor arc from its corresponding transition prevents to put a token in $X_{i-1}$.
This models that $\PANDexcl$ becomes fail-safe.

The behaviour of both \PAND variants crucially depends on whether children fail simultaneously or strictly ordered.
The moment children fail depends on the order in which failures propagate, and is discussed in detail in Sect.~\ref{sec:semantics_failure_propagation}.

\subsubsection{\SPARE.}
\begin{figure}[tb]
\centering
\begin{tikzpicture}[every label/.append style={font=\scriptsize}]
    % Child 1
	\node[place, label=above:{$\Consider_1$}, tokens=1] (consider1) {};
	\node[iplace, right=0.1cm of consider1, yshift=-0.7cm, label=right:{$\Unavailable_{\child{v}{1}}$}] (unavailable1) {};
	\node[place, below=1cm of unavailable1, label=below:{$\Claimed_1$}] (claimed1) {};
	\node[iplace, below=0.5cm of claimed1, label=below:{$\Failed_{\child{v}{1}}$}] (failed1) {};

	\node[itransition, right=0.8cm of claimed1, label=below:{$\tchildfail$}, label=right:{@$\prioVar_v$}] (tfailedclaimed1) {};
	\node[Itransition, below=0.4cm of unavailable1, label=left:{$\tclaim$}, label=right:{@$\prioVar_v$}] (tclaimsuccess1) {};
	\node[itransition, right=1.5cm of consider1, label=above:{$\tunavailable$}, label=35:{@$\prioVar_v$}] (tclaimfail1) {};

	\node[place, label=above:{$\Consider_{2}$}, right=2.9cm of consider1] (consider2) {};

	\draw[->] (consider1) -- (tclaimfail1.west);
	\draw[<->] (unavailable1) -- ([yshift=-3pt] tclaimfail1.west);
	\draw[->] (tclaimfail1) -- (consider2);

	\draw[->] (consider1) -- ([xshift=-3pt] tclaimsuccess1.north);
	\draw[<-o] (unavailable1) -- (tclaimsuccess1.north);
	\draw[->] (tclaimsuccess1) -- (claimed1);

	\draw[->] (claimed1) -- (tfailedclaimed1.west);
	\draw[<->] (failed1.35) -- ([yshift=-3pt] tfailedclaimed1.west);
	\draw[->] (tfailedclaimed1) -- (consider2);

    % Child 2
	\node[iplace, right=0.1cm of consider2, yshift=-0.7cm, label=right:{$\Unavailable_{\child{v}{2}}$}] (unavailable2) {};
	\node[place, below=1cm of unavailable2, label=below:{$\Claimed_2$}] (claimed2) {};
	\node[iplace, below=0.5cm of claimed2, label=below:{$\Failed_{\child{v}{2}}$}] (failed2) {};

	\node[itransition, right=0.8cm of claimed2, label=below:{$\tchildfail$}, label=right:{@$\prioVar_v$}] (tfailedclaimed2) {};
	\node[Itransition, below=0.4cm of unavailable2, label=left:{$\tclaim$}, label=right:{@$\prioVar_v$}] (tclaimsuccess2) {};
	\node[itransition, right=1.5cm of consider2, label=above:{$\tunavailable$}, label=35:{@$\prioVar_v$}] (tclaimfail2) {};

	\node[place,dummy, right=2.9cm of consider2] (considerdummy) {$\hdots$};

	\draw[->] (consider2) -- (tclaimfail2.west);
	\draw[<->] (unavailable2) -- ([yshift=-3pt] tclaimfail2.west);
	\draw[->] (tclaimfail2) -- (considerdummy);

	\draw[->] (consider2) -- ([xshift=-3pt] tclaimsuccess2.north);
	\draw[<-o] (unavailable2) -- (tclaimsuccess2.north);
	\draw[->] (tclaimsuccess2) -- (claimed2);

	\draw[->] (claimed2) -- (tfailedclaimed2.west);
	\draw[<->] (failed2.35) -- ([yshift=-3pt] tfailedclaimed2.west);
	\draw[->] (tfailedclaimed2) -- (considerdummy);

    % Child dummy
	\node[place,dummy, below=1.7cm of considerdummy] (claimeddummy) {$\hdots$};
	\node[place,dummy, below=0.5cm of claimeddummy] (faileddummy) {$\hdots$};

    % Child n
	\node[iplace, right=0.6cm of considerdummy, yshift=-0.7cm, label=right:{$\Unavailable_{\child{v}{n}}$}] (unavailablen) {};
	\node[place, below=1cm of unavailablen, label=below:{$\Claimed_n$}] (claimedn) {};
	\node[iplace, below=0.5cm of claimedn, label=below:{$\Failed_{\child{v}{n}}$}] (failedn) {};

	\node[itransition, right=0.8cm of claimedn, label=below:{$\tchildfail$}, label=right:{@$\prioVar_v$}] (tfailedclaimedn) {};
	\node[Itransition, below=0.4cm of unavailablen, label=left:{$\tclaim$}, label=right:{@$\prioVar_v$}] (tclaimsuccessn) {};
	\node[itransition, right=2.0cm of considerdummy, label=above:{$\tunavailable$}, label=35:{@$\prioVar_v$}] (tclaimfailn) {};

	\node[iplace, label=above:{$\Failed_v$}, right=3.3cm of considerdummy] (v_failed) {};

	\draw[->] (considerdummy) -- (tclaimfailn.west);
	\draw[<->] (unavailablen) -- ([yshift=-3pt] tclaimfailn.west);
	\draw[->] (tclaimfailn) -- (v_failed);

	\draw[->] (considerdummy) -- ([xshift=-3pt] tclaimsuccessn.north);
	\draw[<-o] (unavailablen) -- (tclaimsuccessn.north);
	\draw[->] (tclaimsuccessn) -- (claimedn);

	\draw[->] (claimedn) -- (tfailedclaimedn.west);
	\draw[<->] (failedn.35) -- ([yshift=-3pt] tfailedclaimedn.west);
	\draw[->] (tfailedclaimedn) -- (v_failed);

	\node[iplace,label=below:{$\Unavailable_v$}, below=0.7cm of v_failed] (v_unavailable) {};

	\draw[->] (tfailedclaimedn) -- (v_unavailable);
	\draw[->] (tclaimfailn) -- (v_unavailable);
\end{tikzpicture}%
\caption{GSPN template for \SPARE, the claiming mechanism}
\label{fig:spare_template}
\end{figure}
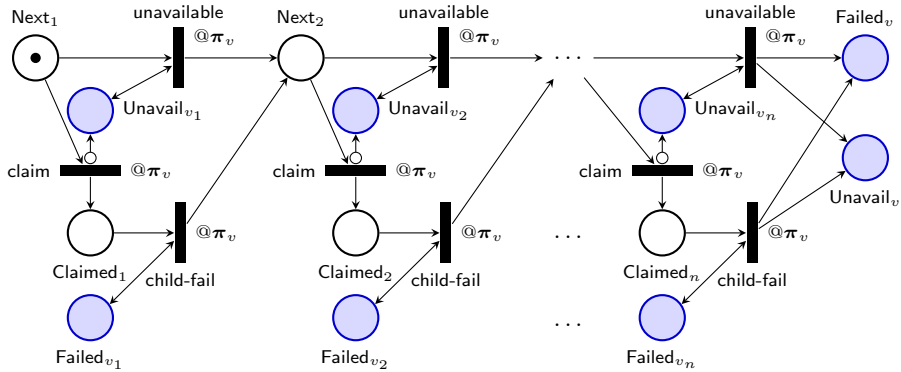

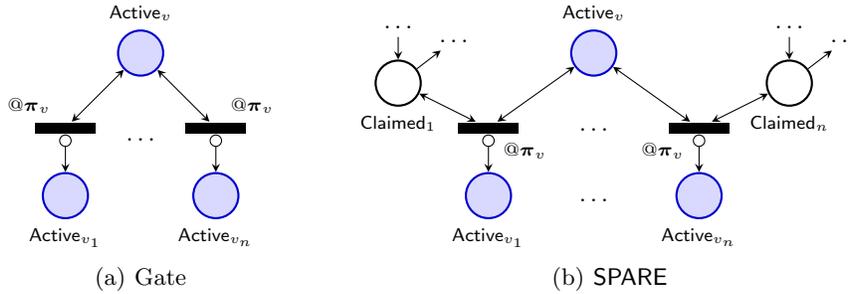
\begin{figure}[tb]
\centering
\subfigure[Gate]{
    \begin{tikzpicture}[every label/.append style={font=\scriptsize}]
	\node[iplace, label=above:{$\Active_{v}$}] (active) {};

	\node[Itransition, below=0.6cm of active, xshift=-1cm, label=135:{@$\prioVar_v$}] (tactive1) {};
	\node[iplace, below=0.5cm of tactive1, label=below:{$\Active_{\child{v}{1}}$}] (active1) {};
	\node[below=0.7cm of active] (activedots) {$\hdots$};
	\node[Itransition, below=0.6cm of active, xshift=1cm, label=45:{@$\prioVar_v$}] (tactiven) {};
	\node[iplace, below=0.5cm of tactiven, label=below:{$\Active_{\child{v}{n}}$}] (activen) {};

    \draw[<->] (active) -- (tactive1);
    \draw[o->] (tactive1) -- (active1);
    \draw[<->] (active) -- (tactiven);
    \draw[o->] (tactiven) -- (activen);
\end{tikzpicture}%
    \label{fig:gate_activation}
}
\hspace{0.5cm}
\subfigure[\SPARE]{
    \begin{tikzpicture}[every label/.append style={font=\scriptsize}]
	\node[iplace, label=above:{$\Active_{v}$}] (active) {};

	\node[Itransition, below=0.6cm of active, xshift=-1.4cm, label=315:{@$\prioVar_v$}] (tactive1) {};
	\node[iplace, below=0.5cm of tactive1, label=below:{$\Active_{\child{v}{1}}$}] (active1) {};
	\node[place, above=0.2cm of tactive1, xshift=-1.2cm, label=below:{$\Claimed_1$}] (claimed1) {};
	\node[above=0.3cm of claimed1] (dots1) {$\hdots$};
	\node[above right=0.3cm of claimed1] (dots12) {$\hdots$};

	\node[Itransition,dummy, below=0.6cm of active] (tdummy) {$\hdots$};
	\node[place,dummy, below=0.5cm of tdummy] (dummy) {$\hdots$};

	\node[Itransition, below=0.6cm of active, xshift=1.4cm, label=225:{@$\prioVar_v$}] (tactiven) {};
	\node[iplace, below=0.5cm of tactiven, label=below:{$\Active_{\child{v}{n}}$}] (activen) {};
	\node[place, above=0.2cm of tactiven, xshift=1.2cm, label=below:{$\Claimed_n$}] (claimedn) {};
	\node[above=0.3cm of claimedn] (dotsn) {$\hdots$};
	\node[above right=0.3cm of claimedn] (dotsn2) {$\hdots$};

	\draw[->] (dots1) -- (claimed1);
	\draw[->] (claimed1) -- (dots12);
    \draw[<->] (active) -- (tactive1);
    \draw[<->] (claimed1) -- (tactive1);
    \draw[o->] (tactive1) -- (active1);
	\draw[->] (dotsn) -- (claimedn);
	\draw[->] (claimedn) -- (dotsn2);
    \draw[<->] (active) -- (tactiven);
    \draw[<->] (claimedn) -- (tactiven);
    \draw[o->] (tactiven) -- (activen);
\end{tikzpicture}%
    \label{fig:spare_activation}
}
\caption{GSPN template extensions for the activation mechanism of DFT elements}
\label{fig:activation}
\end{figure}

We depict the template $\templ{\SPARE}{v}$ for \SPARE in two parts: Claiming%
\footnote{We consider \emph{early} claiming; the concept of \emph{late} claiming is described in App.~\ref{sec:nested_spares}.}
is depicted in Fig.~\ref{fig:spare_template}, activation is shown in Fig.~\ref{fig:activation}.
\paragraph{Claiming.}
$\templ{\SPARE}{v}$ has two sorts of auxiliary places for each child $i$: $\Consider_i$ and $\Claimed_i$.
A token in $\Consider_i$ indicates that the spare component $\child{v}{i}$ is the next in line to be considered for claiming.
Initially, only $\Consider_1$ is marked as the primary child is to be claimed first.
A token in $\Claimed_i$ indicates that \SPARE $v$ has currently claimed the spare component $\child{v}{i}$.
This token moves (possibly via $\Claimed_i$) through places $\Consider_i$ and ends in $\Failed_v$ if all children are unavailable or already claimed.
The claiming mechanism considers the $\Unavailable$ places of the children.
If $\Unavailable_i$ is marked, the $i$-th spare component cannot be claimed as either the $i$-th child has failed or it has been claimed by another \SPARE.
In this case, the transition $\tunavailable$ fires and the token is moved to $\Consider_{i+1}$.
Then, spare component $i{+}1$ has to be considered next.

An empty place $\Unavailable_i$ indicates that the $i$-th spare component is available.
The \SPARE can claim it by firing the $\tclaim$ transition.
This results in tokens in $\Claimed_i$ and $\Unavailable_i$, marking the spare component unavailable for other \SPARE{}s.
If a spare component is claimed (token in $\Claimed_i$) and it fails, the transition $\tchildfail$ fires, and the next child is considered for claiming.

\paragraph{Activation.}
When an active \SPARE claims a spare component $c$, all nodes in the spare module (the subtree) $M_c$ become active,
i.e.,  \BE{}s in $M_c$ now fail with their active (rather than passive) failure rate, and \SPARE{}s in $M_c$ propagate the activation downwards.
The GSPN extensions for the activation mechanism are given in Fig.~\ref{fig:activation}.
The activation in \SPARE{}s is depicted in Fig.~\ref{fig:spare_activation}.
If a token is in $\Claimed_i$ indicating that the \SPARE claimed the $i$th-child, and the \SPARE itself is active, the transition can fire and places a token in $\Active_{\child{v}{i}}$ indicating that the $i$th-child has become active.
Other gates simply propagate the activation to their children as depicted in Fig.~\ref{fig:gate_activation}.

\subsubsection{\FDEP.}
\begin{figure}[tb]
\centering
\begin{tikzpicture}[every label/.append style={font=\scriptsize}]
	\node[iplace, label=above:{$\Failed_{\child{v}{1}}$}] (sv1_failed) {};
	\node[iplace,label=above:{$\Failed_v$}, right=1.8cm of sv1_failed] (v_failed) {};
	\node[iplace,label=above:{$\Unavailable_v$}, right=0.7cm of v_failed] (v_unavailable) {};

	\node[Itransition, below=0.4cm of sv1_failed, xshift=-2.4cm, label=left:{@$\prioVar_{v}$}] (tf1) {};
	\node[Itransition,dummy, below=0.4cm of sv1_failed] (tdummy) {$\hdots$};
	\node[Itransition, below=0.4cm of sv1_failed, xshift=2.4cm, label=right:{@$\prioVar_{v}$}] (tfn) {};

	\node[iplace, below=0.5cm of tf1, label=below:{$\Failed_{\child{v}{2}}$}] (sv2_failed) {};
	\node[iplace, below=0.5cm of tf1, xshift=-1.4cm, label=below:{$\Unavailable_{\child{v}{2}}$}] (sv2_unavail) {};
	\node[iplace, below=0.5cm of tf1, xshift=1.4cm, label=below:{$\Disabled_{\child{v}{2}}$}] (sv2_enabled) {};
	\node[iplace, below=0.5cm of tfn, label=below:{$\Failed_{\child{v}{n}}$}] (svn_failed) {};
	\node[iplace, below=0.5cm of tfn, xshift=-1.4cm, label=below:{$\Unavailable_{\child{v}{n}}$}] (svn_unavail) {};
	\node[iplace, below=0.5cm of tfn, xshift=1.4cm, label=below:{$\Disabled_{\child{v}{n}}$}] (svn_enabled) {};
	\node[place,dummy, below=0.5cm of tdummy] (dummy) {$\hdots$};

	\draw[<->] (sv1_failed) -- (tf1.north);
	\draw[<->] (sv1_failed) -- (tfn.north);
	\draw[-o] (sv2_enabled) -- (tf1);
	\draw[-o] (svn_enabled) -- (tfn);
	\draw[<-o] (sv2_failed) -- (tf1);
	\draw[<-o] (svn_failed) -- (tfn);
	\draw[<-] (sv2_unavail) -- (tf1);
	\draw[<-] (svn_unavail) -- (tfn);
\end{tikzpicture}%
\caption{GSPN template for \FDEP}
\label{fig:fdep_template}
\end{figure}
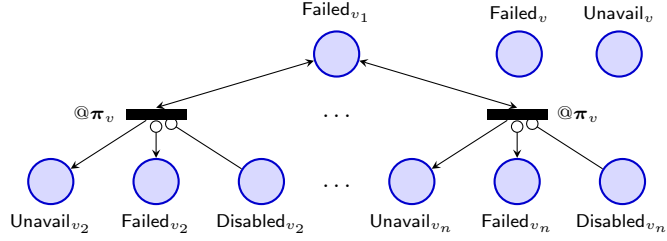

Fig.~\ref{fig:fdep_template} depicts the template $\templ{\FDEP}{v}$ for \FDEP $v$; the generalized \PDEP is discussed in App.~\ref{sec:pdep_template}.
If the first child of the \FDEP fails, the dependent children fail too.
Thus, if $\Failed_{\child{v}{1}}$ is marked, then all transitions can fire and place tokens in the $\Failed$ places of the children indicating the failure propagation to dependent nodes.
There is no arc to $\Failed_v$ as the \FDEP itself cannot fail.

\FDEP{}s introduce several semantic problems for DFTs, \cf \cite{JGKS16}.
This leads to different semantic interpretations which can be captured in our GSPN translation by different values for the priority variables $\prioVar_v$; as elaborated in Sect.~\ref{sec:semantics}.

\subsection{Gluing templates}
\label{sec:formal_translation}

It remains to describe how the GSPN templates for the DFT elements are combined.
We define the merging of templates.
A more general setting is provided via graph-rewriting, \cf \cite{DBLP:journals/entcs/Raiteri05}.
\begin{definition}[Merging Templates]
Let $\petriTemplate_i = (\petriNet_i, \petriWeight_i, \NN[\prioVar], \petriPriority_i, \petriPartition_i)$ for $i=1,2$ be $\prioVar$-parameterised templates over $\petriPlaces_1 \cap \petriPlaces_2 = \interfaces$.
    The \emph{merge} of $\petriTemplate_1$ and $\petriTemplate_2$ is the $\prioVar$-parameterised template over $\interfaces$, $\merge(\petriTemplate_1, \petriTemplate_2) = (\petriNet, \petriWeight, \NN[\prioVar], \petriPriority, \petriPartition)$ with
\begin{itemize}
    \item $\petriPlaces = \petriPlaces_1 \cup \petriPlaces_2$
    \item $\petriTransitions = \petriTransitions_1 \uplus \petriTransitions_2$, $I = I_1 \uplus I_2$, $O = O_1 \uplus O_2$, $H = H_1 \uplus H_2$
    \item $m_0 = m_{0,1} + m_{0,2}$
    \item $\petriWeight = \petriWeight_1 \uplus \petriWeight_2$,
    $\petriPriority = \petriPriority_1 \uplus \petriPriority_2$,
    $\petriPartition = \petriPartition_1 \uplus \petriPartition_2$.
\end{itemize}
\end{definition}
An $n$-ary merge of templates over $\interfaces_\DFT$ is obtained by concatenation of the binary merge.
As the (disjoint) union on sets is associative and commutative, so is the merging of templates.
Let $\merge(\setOfTemplates \cup \petriTemplate)$, where $\setOfTemplates$ is a finite non-empty set of templates over some $\interfaces$ and $\petriTemplate$ is a template over $\interfaces$, denote $\merge(\petriTemplate, \merge(\setOfTemplates))$.

The GSPN translation converts each DFT node $v$ into the corresponding GSPN using its type-dependent template $\templDef{\Tp(v)}$.

\begin{definition}[Template for a DFT]
Let DFT $\DFT = \DFTtuple$ and $\{\templ{\Tp(v)}{v} \sep v \in \DFT\}$ be the set of templates over $\interfaces_\DFT$ each with priority-variable $\prioVar_v$.
The \emph{GSPN template} $\petriTemplate_\DFT$ for DFT $\DFT$ with places $\petriPlaces \supset \interfaces_\DFT$ is defined by $\petriTemplate_\DFT = \merge \left(\{\templ{\Tp(v)}{v} \sep v \in \DFT\} \cup \{\initMarkTempl\} \right)$.
\end{definition}
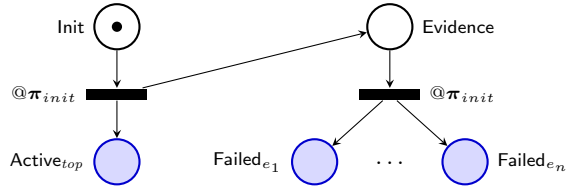
\begin{figure}[tb]
\centering
\begin{tikzpicture}[every label/.append style={font=\scriptsize}]

	\node[place, label=left:{$\Init$}, tokens=1] (init) {};
	\node[Itransition, below=0.5cm of init, label=left:{@$\prioVar_{init}$}] (t1) {};
	\node[iplace, below=0.5cm of t1, label=left:{$\Active_{\Top}$}] (active_top) {};

	\node[place, right=3cm of init, label=right:{$\Evidence$}] (evidence) {};
	\node[Itransition, below=0.5cm of evidence, label=right:{@$\prioVar_{init}$}] (t2) {};
	\node[iplace, below=0.5cm of t2, xshift=-1cm, label=left:{$\Failed_{e_1}$}] (evidence1) {};
	\node[circle, below=0.5cm of t2] (evidencedots) {$\hdots$};
	\node[iplace, below=0.5cm of t2, xshift=1cm, label=right:{$\Failed_{e_n}$}] (evidencen) {};

	\draw[->] (init) -- (t1);
	\draw[->] (t1) -- (active_top);
	\draw[->] (t1) -- (evidence);
	\draw[->] (evidence) -- (t2);
	\draw[->] (t2) -- (evidence1);
	\draw[->] (t2) -- (evidencen);
\end{tikzpicture}%
\caption{GSPN template for initialisation}
\label{fig:init_template}
\end{figure}

\subsubsection{Initialisation template.}
\label{sec:init_template}
The initialisation template $\initMarkTempl$, see Fig.~\ref{fig:init_template}, is ensured to fire once and first, and allows to change the initial marking, \eg already initially failed DFT nodes.
This construct allows to fit the initial marking to the requested semantics without modifying  the overall translation.
The leftmost transition fires initially, and places a token in $\Active_\Top$.
The transition models starting the top-down activation propagation from the top-level node.
Furthermore, a token is placed in the place \textsf{Evidence}, enabling the setting of \emph{evidence}, \ie already failed DFT nodes.
If $\{e_1, \dots, e_n\} \subseteq \DFT_\BE$ is the set of already failed \BE{}s, firing the rightmost transition puts a token in each $\Failed_{e_i}$ for all already failed \BE $e_i$.

\subsection{Properties}
\label{sec:properties}
We discuss some properties of the obtained GSPN $\petriTemplate_\DFT$ for a DFT $\DFT$.
Details can be found in App.~\ref{sec:proofs}.

\emph{The size of $\petriTemplate_\DFT$ is linear in the size of $\DFT$.}
Let $\maxchildren = \max_{v \in \DFT} |\children(v)|$ be the maximal number of children in $\DFT$.
The GSPN $\petriTemplate_\DFT$ has no more than $6 \cdot |V| \cdot \maxchildren + 2$  places and immediate transitions, and  $2 \cdot |\DFT_{\BE}|$ timed transitions.

\emph{Transitions in $\petriTemplate_\DFT$ fire at most once}.
Therefore, $\petriTemplate_\DFT$ does not contain time-traps.
Tokens in the interface places $\Failed_v$, $\Active_v$ and $\Unavailable_v$ are never removed.
For such a place $p$ and any transition $t$, $O(p)(t) \leq I(p)(t)$.
Typically, the inhibitor arcs of interface places prevent a re-firing of a transition.
In $\templ{\PANDexcl}{v}$, $\templ{\SPARE}{v}$ and $\initMarkTempl$ tokens move from left to right, and no transition is ever enabled after it has fired.

\emph{The GSPN $\petriTemplate_\DFT$ is two-bounded, all places except $\Unavailable_v$ are one-bounded}.
Typically, either the inhibitor arcs prevent adding tokens to places that contain a token, or a token moves throughout the (cycle-free) template.
However, two tokens can be placed in $\Unavailable_v$:
One token is placed in $\Unavailable_v$ if $v$ is claimed by a \SPARE.
Another token is placed in $\Unavailable_v$ if $v$ failed.
The GSPN templates can be easily extended to  ensure $1$-boundedness of $\Unavailable_v$ as well, \cf App.~\ref{sec:extension_adaptions}.

\subsection{Tool support}
\label{sec:tool_support}
We realised the GSPN translation of DFTs within the model checker \textsc{\storm}~\cite{DehnertJKV17}, version 1.2.1\footnote{\url{http://www.stormchecker.org/publications/gspn-semantics-for-dfts.html}}.
Storm can export the obtained  GSPNs as, among others, GreatSPN Editor projects\cite{Amparore14}.
Table~\ref{tab:experiments} gives some indications of the obtained sizes of the GSPNs for some DFT benchmarks from \cite{VJK17}.
All GSPN translations could be computed within a second.
As observed before, the GSPN size is linear in the size of the DFT.
\begin{table}[tb]
    \centering
    \scriptsize{
    \caption{Experimental evaluation of GSPN translations}
    \label{tab:experiments}
    \begin{tabular}{l|cccc|ccc}
        \toprule
        Benchmark           & \multicolumn{4}{c|}{DFT} & \multicolumn{3}{c}{GSPN} \\
                            & \#BE & \#Dyn. & \#Nodes & $\maxchildren$ & \#Places & \#Timed Trans. & \#Immed. Trans. \\
        \midrule
        HECS 5\_5\_2\_np    & 61 & 10 & 107 & 16 & 273 & 122 & 181 \\
        MCS 3\_3\_3\_dp\_x  & 46 & 21 &  80 &  7 & 246 &  92 & 163 \\
        RC 15\_15\_hc       & 69 & 33 & 103 & 34 & 376 & 138 & 240 \\
        \bottomrule
    \end{tabular}
    }
\end{table}%
\section{A Unifying DFT Semantics}
\label{sec:semantics}
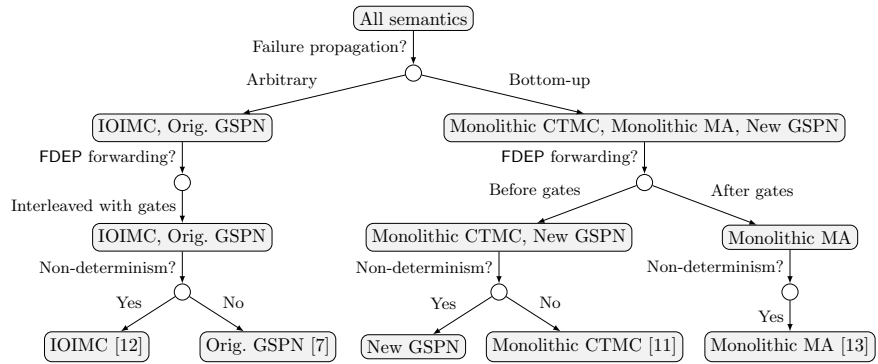
\begin{figure}[b]
\centering
\scalebox{0.7}{
\begin{tikzpicture}
  [
    treenode/.style         = {shape=rectangle, rounded corners, draw, align=center, fill=black!5},
    env/.style              = {treenode, font=\normalsize},
    dummy/.style            = {circle,draw},
    grow                    = down,
    sibling distance        = 4em,
    level distance          = 3.2em,
    edge from parent/.style = {draw, -latex},
    every node/.style       = {font=\footnotesize},
    level 2/.style={sibling distance=27em},
    level 4/.style={sibling distance=17em},
    level 6/.style={sibling distance=10em},
  ]
  \node [env] {All semantics}
    child { node [dummy] {}
        child { node [env] {IOIMC, Orig. GSPN}
            child { node [dummy] {}
                child { node [env] {IOIMC, Orig. GSPN}
                    child { node [dummy] {}
                        child { node [env] {IOIMC~\cite{Boudali2010}}
                            edge from parent node[above left] {Yes}
                        }
                        child { node [env] {Orig. GSPN~\cite{DBLP:journals/entcs/Raiteri05}}
                            edge from parent node[above right] {No}
                        }
                        edge from parent node[left] {Non-determinism?}
                    }
                    edge from parent node[left] {Interleaved with gates}
                }
                edge from parent node[left] {\FDEP forwarding?}
            }
            edge from parent node [above left] {Arbitrary}
        }
        child { node [env] {Monolithic CTMC, Monolithic MA, New GSPN}
            child { node [dummy] {}
                child { node [env] {Monolithic CTMC, New GSPN}
                    child { node [dummy] {}
                        child { node [env] {New GSPN}
                            edge from parent node[above left] {Yes}
                        }
                        child { node [env] {Monolithic CTMC~\cite{CSD00}}
                            edge from parent node[above right] {No}
                        }
                        edge from parent node[left] {Non-determinism?}
                    }
                    edge from parent node[above left] {Before gates}
                }
                child { node [env] {Monolithic MA}
                    child { node [dummy] {}
                        child { node [env] {Monolithic MA~\cite{VJK17}}
                            edge from parent node[left] {Yes}
                        }
                        edge from parent node[left] {Non-determinism?}
                    }
                    edge from parent node[above right] {After gates}
                }
                edge from parent node[left] {\FDEP forwarding?}
            }
            edge from parent node [above right] {Bottom-up}
        }
        edge from parent node [left] {Failure propagation?}
    }
    ;
\end{tikzpicture}%
}
\caption{Decision tree to compare five different DFT semantics}
\label{fig:decision_tree_semantics}
\end{figure}

The interpretation of DFTs is subject to various subtleties, as surveyed in~\cite{JGKS16}.
Varying interpretations have given rise to various DFT semantics in the literature.
The key aspects are summarised in Table~\ref{tab:semantic_differences} on page~\pageref{tab:semantic_differences}.
In the following, we focus on three key aspects --- failure propagation, \FDEP forwarding, and non-determinism --- and show that these suffice to differentiate \emph{all} five DFT semantics, see Fig.~\ref{fig:decision_tree_semantics}.
Note that we consider the interleaving semantics of nets.

We expose the subtle semantic differences by considering the three aspects using the translated GSPNs of some simple DFTs.
The simple DFTs contain structures which occur in industrial case-studies \cite{JGKS16}.
We vary two ingredients in our net semantics: \emph{instantiations of the priority variables $\prioVar$}, and the \emph{partitioning $\petriPartition$ of immediate transitions}.
The former constrain the ordering of transitions, while the latter control the treatment of non-determinism.
This highlights a key advantage of our net translation: all different DFT semantics from the literature can be captured by small changes in the GSPN.
In particular, the net structure itself stays the same for all semantics.
Each of the following subsections is devoted to one of the aspects: failure propagation, \FDEP forwarding, and non-determinism.
Afterwards, we summarise the differences in Table~\ref{tab:gspn_differences} on page~\pageref{tab:gspn_differences}.

\subsection{Failure propagation}
\label{sec:semantics_failure_propagation}

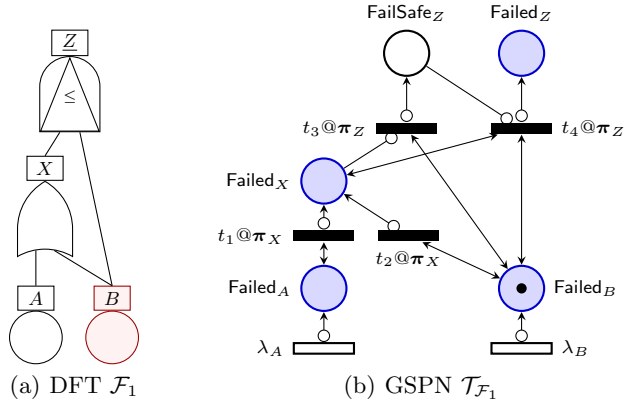
\begin{figure}[tb]
\centering
\subfigure[DFT $\DFT_1$]{
    \begin{tikzpicture}[scale=.6,text=black]
    \node[and2] (pand) {{\tiny\rotatebox{270}{$\leq$}}};
    \node[triangle,scale=1.62,yshift=-3.5,xscale=0.80] (triangle_b) at (pand) {};
	\node[labelbox] (and_label) at (pand.east) {$\underline{Z}$};

	\node[or2,below=1.7cm of pand.center, yshift=0.7cm] (or) {};
	\node[labelbox] (or_label) at (or.east) {$X$};

	\node[be,below=0.8cm of or.input 1] (A) {};
	\node[labelbox] (a_label) at (A.north) {$A$};
	\node[be,failed,right=0.3cm of A] (B) {};
	\node[labelbox,failed] (b_label) at (B.north) {$B$};

	\draw[-] (pand.input 1) -- (or_label.north);
	\draw[-] (pand.input 2) -- (b_label.north);
	\draw[-] (or.input 1) -- (a_label.north);
	\draw[-] (or.input 2) -- (b_label.north);
\end{tikzpicture}%
    \label{fig:failure_propagation_dft}
}
\hspace{0.5cm}
\subfigure[GSPN $\petriTemplate_{\DFT_1}$]{
    \begin{tikzpicture}[every label/.append style={font=\scriptsize}]
  	\node[Ttransition, label=left:{$\lambda_A$}] (Af) {};
  	\node[Ttransition, right=1.8cm of Af, label=right:{$\lambda_B$}] (Bf) {};
  	\node[iplace, above=0.4cm of Af, label=left:{$\Failed_A$}] (AF) {};
  	\node[iplace, tokens=1, above=0.4cm of Bf, label=right:{$\Failed_B$}] (BF) {};

  	\draw[o->] (Af) -- (AF);
  	\draw[o->] (Bf) -- (BF);

  	\node[iplace, above=0.8cm of AF, label=left:{$\Failed_X$}] (XF) {};

    \node[Itransition, above=0.3cm of AF, label=left:{$t_1$@$\prioVar_X$}] (Xf1) {};
    \node[Itransition, right=0.3cm of Xf1, label=below:{$t_2$@$\prioVar_X$}] (Xf2) {};

	\draw[<->] (AF) -- (Xf1);
  	\draw[<->] (BF) -- (Xf2);

  	\draw[o->] (Xf1) -- (XF);
  	\draw[o->] (Xf2) -- (XF);

    \node[iplace,label=90:{$\Failed_Z$}, above=2.5cm of BF] (v_failed) {};
    \node[place,label=90:{$\Failsafe_Z$}, left=0.9cm of v_failed] (v_failsafe) {};
    \node[Itransition, below=0.6cm of v_failsafe,label=left:{$t_3$@$\prioVar_Z$}] (Zt1) {};
    \node[Itransition, below=0.6cm of v_failed, label=right:{$t_4$@$\prioVar_Z$}] (Zt2) {};

    \draw[<->] (XF) -- (Zt2);
    \draw[<->] (BF) -- (Zt2);
    \draw[-o] (v_failsafe) -- (Zt2);
    \draw[<-o] (v_failed) -- (Zt2);
    \draw[<->] (BF) -- (Zt1);
    \draw[-o] (XF) -- (Zt1);
    \draw[<-o] (v_failsafe) -- (Zt1);
\end{tikzpicture}%
    \label{fig:failure_propagation_gspn}
}
\caption{Example for failure propagation}
\label{fig:failure_propagation}
\end{figure}

This aspect is concerned with the order in which failures propagate through the DFT.
Consider (a) the DFT $\DFT_1$ and (b) its GSPN $\petriTemplate_{\DFT_1}$ in Fig.~\ref{fig:failure_propagation} and suppose $B$ has failed, as indicated in red and the token in place $\Failed_B$
(the same example was used in the introduction).
The question is how $B$'s failure propagates through the DFT.
Considering a total ordering on failure propagations, there are two scenarios.
\emph{Is $B$'s failure first propagated to gate $X$, causing \PAND $Z$ to fail, or is $B$'s failure first propagated to gate $Z$, turning $Z$ fail-safe?}

The question reflects in net $\petriTemplate_{\DFT_1}$:
Consider the enabled transitions $t_2$ and $t_3$.
Firing $t_2$ places a token in $\Failed_X$ (and in $\Failed_B$) and models that $B$'s failure first propagates to $X$.
Next, firing $t_4$ places a token in $\Failed_Z$ and models that the failures of $B$ and $X$ propagate to $Z$.
Now consider first propagating $B$'s failure to $Z$.
This corresponds to firing $t_3$ and a token in $\Failsafe_Z$ modelling that $Z$ is fail-safe.
($B$'s failure can still be propagated to $X$, but $Z$ remains fail-safe as transition $t_4$ is disabled due to the token in $\Failsafe_Z$.)

The order of failure propagation is thus crucial as it may cause a gate to either fail or to be fail-safe.
Existing ways to treat failure propagation are: (1) allow for all possible orders, or (2) propagate failures in a bottom-up manner through the DFT.
The former is adopted in the IOIMC and the original GSPN semantics.
This amounts in $\petriTemplate_{\DFT_1}$ to give all transitions the same priority, e.g., $\prioVar_v = 1$ for all $v \in \DFT$.
Case (2) forces failures to propagate in a bottom-up manner, \ie a gate is not evaluated before all its children have been evaluated.
This principle is used by the other three semantics.
To model this, the priority of a gate $v$ must be lower than the priorities of its children, \ie
$\prioVar_v < \prioVar_{\child{v}{i}}, \forall i \in \{1, \dots, |\children(v)|\}$.
In $\petriTemplate_{\DFT_1}$, this yields $\prioVar_Z < \prioVar_X$, forcing firing $t_2$ before $t_3$, see Table~\ref{tab:gspn_differences}.

\subsection{\FDEP forwarding}
\label{sec:semantics_fdep_forwarding}

\begin{figure}[tb]
\centering
\subfigure[DFT $\DFT_2$ \cite{JGKS16}]{
    \begin{tikzpicture}[scale=.6,text=black]
    \node[and2] (pand) {{\tiny\rotatebox{270}{$\leq$}}};
    \node[triangle,scale=1.62,yshift=-3.5,xscale=0.80] (triangle) at (pand) {};
	\node[labelbox] (pand_label) at (pand.east) {$\underline{Z}$};

	\node[be,below=1.8cm of pand.center, xshift=-0.5cm] (A) {};
	\node[labelbox] (a_label) at (A.north) {$A$};
	\node[be,failed,below=1.8cm of pand.center, xshift=0.5cm] (B) {};
	\node[labelbox,failed] (b_label) at (B.north) {$B$};
	\node[fdep, right=1.0cm of pand.center] (fdep) {};
	\node[labelbox] (fdep_label) at (fdep.north) {$D$};

	\draw[-] (pand.input 1) -- (a_label.north);
	\draw[-] (pand.input 2) -- (b_label.north);
	\draw[-] (fdep.T) -- (b_label.north);
	\draw[-] (fdep.EA) -- (a_label.north);
\end{tikzpicture}%
    \label{fig:fdep_forwarding_dft}
}
\hspace{0.5cm}
\subfigure[GSPN $\petriTemplate_{\DFT_2}$]{
    \begin{tikzpicture}[every label/.append style={font=\scriptsize}]

  	\node[Ttransition, label=left:{$\lambda_A$}] (Af) {};
  	\node[Ttransition, right=1.0cm of Af, label=right:{$\lambda_B$}] (Bf) {};
  	\node[iplace, above=0.4cm of Af, label=left:{$\Failed_A$}] (AF) {};
  	\node[iplace, tokens=1, above=0.4cm of Bf, label=right:{$\Failed_B$}] (BF) {};

  	\draw[o->] (Af) -- (AF);
  	\draw[o->] (Bf) -- (BF);

    \node[iplace,label=90:{$\Failed_Z$}, above=1.4cm of BF] (v_failed) {};
    \node[place,label=90:{$\Failsafe_Z$}, above=1.4cm of AF] (v_failsafe) {};
    \node[Itransition, below=0.6cm of v_failsafe,label=above left:{$t_2$@$\prioVar_Z$}] (Tt1) {};
    \node[Itransition, below=0.6cm of v_failed, label=above right:{$t_3$@$\prioVar_Z$}] (Tt2) {};

    \node[Itransition, right=1.0cm of Tt2, label=above:{$t_1$@$\prioVar_D$}] (Ft1) {};

    \draw[<->] (AF) -- (Tt2);
    \draw[<->] (BF) -- (Tt2);
    \draw[-o] (v_failsafe) -- (Tt2);
    \draw[<-o] (v_failed) -- (Tt2);
    \draw[<->] (BF) -- (Tt1);
    \draw[-o] (AF) -- (Tt1);
    \draw[<-o] (v_failsafe) -- (Tt1);

    \draw[<-o] (AF) -- (Ft1);
    \draw[<->] (BF) -- (Ft1);
\end{tikzpicture}%
    \label{fig:fdep_forwarding_gspn}
}
\caption{Example for \FDEP forwarding}
\label{fig:fdep_forwarding}
\end{figure}
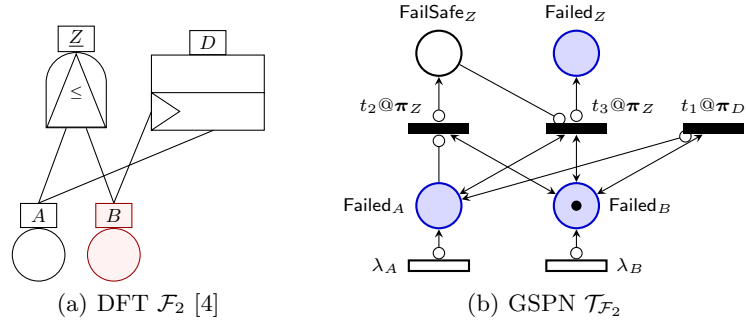

The second aspect concerns how \FDEP{}s forward failures in the DFT.
Consider (a) the DFT $\DFT_2$ and (b) its GSPN $\petriTemplate_{\DFT_2}$ in Fig.~\ref{fig:fdep_forwarding}.
Suppose $B$ fails.
The crucial question is --- similar to failure propagation --- when to propagate $B$'s failure via \FDEP $D$ to $A$.
\emph{Is $B$'s failure first propagated via $D$, causing $A$ and $Z$ to fail, or does $B$'s failure first cause $Z$ to become fail-safe before $A$ fails?}
The first scenario is possible as $Z$ is inclusive and $A$ and $B$ are interpreted to fail simultaneously.
In $\petriTemplate_{\DFT_2}$, the scenarios are reflected by letting either of the enabled transitions $t_1$ and $t_2$ fire first.
A similar scenario can be constructed with a $\PANDexcl$ and an $\FDEP$ from $A$ to $B$.

The order of evaluating \FDEP{}s is thus crucial (as above).
We distinguish three options: evaluating \FDEP{}s (1) before, (2) after, or (3) interleaved with failure propagation in gates.
The first two options evaluate \FDEP{}s either before or after all other gates, respectively.
In $\petriTemplate_{\DFT_2}$, these options require that all transitions of an \FDEP template get the (1) highest (or (2) lowest, respectively) priority, \ie
\[\forall f \in \DFT_{\FDEP}: \prioVar_f > \prioVar_v, \forall v \in \DFT \setminus \DFT_{\FDEP} \quad\text{(or, }\prioVar_f < \prioVar_v \text{ respectively).}\]
The monolithic CTMC and the new GSPN semantics\footnote{The new GSPN semantics needs further adaptions for downward \FDEP{}s, \cf Sect.~\ref{sec:semantics_further_issues}.} evaluate \FDEP{}s before gates, whereas the monolithic MA semantics evaluate them after gates.
In option (3), \FDEP{}s are evaluated interleaved with the other gates.
This option is used by the IOIMC and the original GSPN semantics.
In $\petriTemplate_{\DFT_2}$, interleaving corresponds to giving all transitions the same priority, e.g. $\prioVar_v = 1, \forall v \in \DFT$, see Table~\ref{tab:gspn_differences}.

\subsection{Non-determinism}
\label{sec:semantics_nondeterminism}

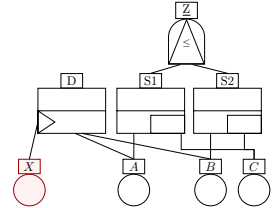
\begin{wrapfigure}[10]{R}{0.3\textwidth}
\centering
\scalebox{\dftscale}{
    \begin{tikzpicture}[scale=.6,text=black]
    \node[and2] (pand) {{\tiny\rotatebox{270}{$\leq$}}};
    \node[triangle,scale=1.62,yshift=-3.5,xscale=0.80] (triangle) at (pand) {};
	\node[labelbox] (pandl) at (pand.east) {\underline{Z}};
	\node[spare,below=of pand,fill=white!100!black, xshift=-1.2cm] (spare1) {};
	\node[spare,right=0.20cm of spare1,fill=white!100!black] (spare2) {};
	\node[labelbox] (s1) at (spare1.north) {S1};
	\node[labelbox] (s2) at (spare2.north) {S2};
    \node[fdep, left=1.0cm of spare1.center] (fdep) {};
	\node[labelbox] (fdep_label) at (fdep.north) {D};

	\node[be,node distance=0.9cm,  below=of spare1.P] (A) {};
	\node[labelbox] (al) at (A.north) {$A$};
	\node[be,node distance=0.9cm,  below=of spare2.P] (B) {};
	\node[labelbox] (bl) at (B.north) {$B$};
	\node[be,right=0.25cm of B] (C) {};
	\node[labelbox] (cl) at (C.north) {$C$};
	\node[be,failed,left=1.60cm of A] (X) {};
	\node[labelbox,failed] (xl) at (X.north) {$X$};

	\draw[-] (s1.north) -- (pand.input 1);
	\draw[-] (s2.north) -- (pand.input 2);
	\draw[-] (spare1.P) -- (al.north);
    \draw[-] (spare2.P) -- (bl.north);
    \draw[-] (spare1.SE) -| +(0,-0.6)  -| (cl.north);
    \draw[-] (spare2.SC) -| +(0,-0.6)  -| (cl.north);
	\draw[-] (fdep.T) -- (xl.north);
	\draw[-] (fdep.EA) -- (al.north);
	\draw[-] (fdep.EB) -- (bl.north);
\end{tikzpicture}%
}
\label{fig:nondeterminism_dft}
\caption{Example for non-determinism (DFT $\DFT_3$)}
\label{fig:nondeterminism}
\end{wrapfigure}

The third aspect is how to resolve non-determinism in DFTs.
Consider DFT $\DFT_3$ in Fig.~\ref{fig:nondeterminism} where \BE $X$ has failed and \FDEP $D$ forwards the failure to \BE{}s $A$ and $B$.
This renders $A$ and $B$ unavailable for \SPARE{}s $S1$ and $S2$.
\emph{The question is which one of the failed \SPARE{}s ($S1$ or $S2$) claims the spare component $C$?}
This phenomenon is known as a \emph{spare race}.
How the spare race is resolved is important: the outcome determines whether \PAND $Z$ fails or becomes fail-safe.

The spare race is represented in $\petriTemplate_{\DFT_3}$ (depicted in Fig.~\ref{fig:greatspn_example} in App.~\ref{sec:examples}) by a conflict between the claiming transitions of the nets of $S1$ and $S2$.
Depending on the previous semantic choices, the race is resolved in different ways.
For the monolithic MA semantics, the race is resolved by the order of the \FDEP forwarding.
For the new GSPN semantics, the race is resolved by the order in which the claim-transitions originating from $\templ{\SPARE}{S_1}$ and $\templ{\SPARE}{S_2}$ are handled.
In the IOIMC semantics, the winner of the race is determined by the order of interleaving.

For any semantics, the race is represented by a conflict between immediate transitions (with the same priority).
We resolve a conflict either by (1)~\emph{randomisation}, or (2)~\emph{non-determinism}.
We realise the randomisation by using weights, i.e., by equipping every immediate transition with the same weight like $\petriWeight(t) = 1, \forall t \in \petriTransitions$ and letting $\petriPartition= \petriImmediate$ contain all immediate transitions.
A conflict between enabled transitions is then resolved by means of a uniform distribution:
each enabled transition is equally probable.
This approach reflects the monolithic CTMC and the original GSPN semantics for DFTs.

Case~(2) takes non-determinism \emph{as is} and reflects the other three DFT semantics.
In this case, in $\petriTemplate_{\DFT_3}$ each immediate transition is a separate partition: $\petriPartition = \{\{t\} \sep t \in \petriImmediate\}$.
In many DFTs, the non-determinism is \emph{spurious} and its resolution does not affect standard measures such as reliability and availability.
The example $\DFT_3$ however yields significantly different analysis results depending on how non-determinism is resolved.

\begin{remark}
The semantics of GSPNs~\cite{DBLP:journals/tocs/MarsanCB84,Mar95} assigns a weight to every immediate transition.
These weights induce a probabilistic choice between conflicting immediate transitions.
If several immediate transitions are enabled, the probability of selecting one is determined by its weight relative to the sum of the weights of all enabled transitions, see Sect.~\ref{sec:gspns}.
Under this interpretation, the stochastic process underlying a confusion-free GSPNs is a CTMC.
In order to capture the possibility of non-deterministically resolving, e.g., spare races, we use a GSPN semantics~\cite{EHKZ13} where immediate transitions are partitioned. Transitions resolved in a random manner (by using weights) are in a single partition, transitions resolved non-deterministically constitute their own partition --- their weights are irrelevant.
For confusion-free GSPNs, our interpretation corresponds to \cite{DBLP:journals/tocs/MarsanCB84,Mar95} and yields a CTMC.
In general, however, the underlying process is an MA.
\end{remark}

The GSPN adaptations for the different DFT semantics are summarised in Table~\ref{tab:gspn_differences}.
The last two rows of the table concern \FDEP{}s that are triggered by gates (rather than BEs) and are discussed in detail below.
\begin{table}[tb]
    \scriptsize{
    \centering
    \caption{GSPN differences between supported semantics}
    \label{tab:gspn_differences}
    \begin{tabular}{@{}p{2.5cm}p{2.0cm}p{5.0cm}p{2.5cm}@{}}
        \toprule
        DFT semantics   & \multicolumn{2}{l}{GSPN priority variables}   & GSPN partitioning \\
        \midrule
        Monolithic CTMC & $\prioVar_v < \prioVar_{\child{v}{i}}$ & $\forall v \in \DFT, \forall i \in \{1, \dots, |\children(v)|\}$ & \multirow{2}{*}{$\{\petriImmediate\}$} \\
                        & $\prioVar_f > \prioVar_v$ & $\forall f \in \DFT_\FDEP, \forall v \not\in \DFT_\FDEP$  &\\ \midrule
        IOIMC           & $\prioVar_v = \prioVar_{v'}$ & $\forall v, v' \in \DFT$ & \multirow{1}{*}{$\{\{t\} \sep t \in \petriImmediate\}$} \\
        \midrule
        Monolithic MA   & $\prioVar_v < \prioVar_{\child{v}{i}}$ & $\forall v \in \DFT, \forall i \in \{1, \dots, |\children(v)|\}$ & \multirow{2}{*}{$\{\{t\} \sep t \in \petriImmediate\}$} \\
                        & $\prioVar_f < \prioVar_v$ & $\forall f \in \DFT_\FDEP, \forall v \not\in \DFT_\FDEP$  &\\
        \midrule
        Original GSPN   & $\prioVar_v = \prioVar_{v'}$ & $\forall v, v' \in \DFT$ & \multirow{1}{*}{$\{\petriImmediate\}$} \\
        \midrule
        New GSPN        & $\prioVar_v \leq \prioVar_{\child{v}{i}}$ & $\forall v \in \DFT_\AND \cup \DFT_\OR, \forall i \in \{1, \dots, |\children(v)|\}$ & \multirow{4}{*}{$\{\{t\} \sep t \in \petriImmediate\}$} \\
                        & $\prioVar_v < \prioVar_{\child{v}{i}}$ & $\forall v \not\in \DFT_\AND \cup \DFT_\OR, \forall i \in \{1, \dots, |\children(v)|\}$ & \\
                        & $\prioVar_f \geq \prioVar_{\child{f}{i}}$ & $\forall f \in \DFT_\FDEP, \forall i \in \{2, \dots, |\children(v)|\}$ & \\
                        & $\prioVar_f \leq \prioVar_{\child{f}{1}}$ & $\forall f \in \DFT_\FDEP$    &\\
        \bottomrule
        \end{tabular}
    }
\end{table}

\subsection{Allow \FDEP{}s triggered by gates}
\label{sec:semantics_further_issues}

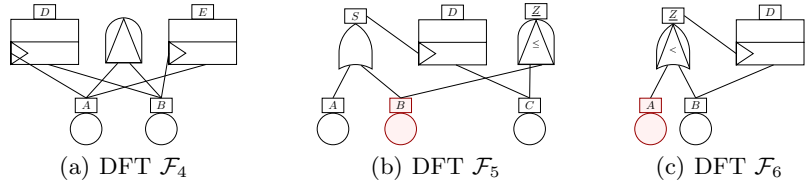
\begin{figure}[tb]
\centering
\subfigure[DFT $\DFT_4$]{
    \scalebox{\dftscale}{
    \begin{tikzpicture}[scale=.6,text=black]
	\node[and2] (and) {};
	\node[triangle,scale=1.62,yshift=-3.5,xscale=0.80] (triangle) at (and) {};
	\node[be, below=1.1cm of and.input 1, xshift=-0.7cm] (A) {};
	\node[labelbox] (a_label) at (A.north){$A$};
	\node[be, below=1.1cm of and.input 2, xshift=0.7cm] (B) {};
	\node[labelbox] (b_label) at (B.north){$B$};

	\node[fdep, left=1.0cm of and.center] (fdep1) {};
	\node[labelbox] (fdep1_label) at (fdep1.north) {$D$};
	\node[fdep, right=1.0cm of and.center] (fdep2) {};
	\node[labelbox] (fdep2_label) at (fdep2.north) {$E$};

	\draw[-] (and.input 1) -- (a_label.north);
	\draw[-] (and.input 2) -- (b_label.north);

	\draw[-] (fdep1.T) -- (a_label.north);
	\draw[-] (fdep1.EA) -- (b_label.north);
	\draw[-] (fdep2.T) -- (b_label.north);
	\draw[-] (fdep2.EA) -- (a_label.north);
\end{tikzpicture}%
    }
    \label{fig:feedback_loop}
}
\hspace{0.5cm}
\subfigure[DFT $\DFT_5$]{
    \scalebox{\dftscale}{
    \begin{tikzpicture}[scale=.6,text=black]
    \node[or2] (or) {};
	\node[labelbox] (or_label) at (or.east) {$S$};
	\node[fdep, right=1.4cm of or.west, yshift=0.5cm] (fdep) {};
	\node[labelbox] (fdep_label) at (fdep.north) {$D$};
    \node[and2, right=4cm of or.west] (pand) {{\tiny\rotatebox{270}{$\leq$}}};
    \node[triangle,scale=1.62,yshift=-3.5,xscale=0.80] (triangle) at (pand) {};
	\node[labelbox] (pand_label) at (pand.east) {$\underline{Z}$};

	\node[be,below=1.4cm of or.center, xshift=-0.5cm] (A) {};
	\node[labelbox] (a_label) at (A.north) {$A$};
	\node[be,failed,below=1.4cm of or.center, xshift=1cm] (B) {};
	\node[labelbox,failed] (b_label) at (B.north) {$B$};
	\node[be,right=2.5cm of B.center] (C) {};
	\node[labelbox] (c_label) at (C.north) {$C$};

	\draw[-] (or.input 1) -- (a_label.north);
	\draw[-] (or.input 2) -- (b_label.north);
	\draw[-] (pand.input 1) -- (c_label.north);
	\draw[-] (pand.input 2) -- (b_label.north);
	\draw[-] (fdep.T) -- (or_label.east);
	\draw[-] (fdep.EA) -- (c_label.north);
\end{tikzpicture}%
    }
    \label{fig:downward_fdep_valid}
}
\hspace{0.5cm}
\subfigure[DFT $\DFT_6$]{
    \scalebox{\dftscale}{
    \begin{tikzpicture}[scale=.6,text=black]
    \node[or2] (por) {{\tiny\rotatebox{270}{$<$}}};
    \node[btriangle,scale=1.61,yscale=0.915, xshift=-0.113cm] (triangle) at (por) {};
	\node[labelbox] (por_label) at (por.east) {$\underline{Z}$};
	\node[fdep, right=1.4cm of por.west, yshift=0.5cm] (fdep) {};
	\node[labelbox] (fdep_label) at (fdep.north) {$D$};

	\node[be,failed,below=1.4cm of por.center, xshift=-0.5cm] (A) {};
	\node[labelbox,failed] (a_label) at (A.north) {$A$};
	\node[be,below=1.4cm of por.center, xshift=0.5cm] (B) {};
	\node[labelbox] (b_label) at (B.north) {$B$};

	\draw[-] (por.input 1) -- (a_label.north);
	\draw[-] (por.input 2) -- (b_label.north);
	\draw[-] (fdep.T) -- (por_label.east);
	\draw[-] (fdep.EA) -- (b_label.north);
\end{tikzpicture}%
    \label{fig:downward_fdep_invalid}
    }
}
\caption{Examples for downward \FDEP forwarding}
\label{fig:downward_fdep}
\end{figure}

So far we assumed that \FDEP triggers are \BE{}s.
We now lift this restriction simplifying the presentation and discuss the options when \FDEP{}s can be triggered by a gate, see Fig.~\ref{fig:downward_fdep_valid} and~\ref{fig:downward_fdep_invalid}.
The row ``downward'' \FDEP{}s in Table~\ref{tab:syntax_overview} on page~\pageref{tab:syntax_overview} reflects this notion.
The challenge is to treat \emph{cyclic dependencies}.
Cyclic dependencies already occur at the level of \BE{}s, see Fig.~\ref{fig:feedback_loop}.
According to the monolithic CTMC and new GSPN semantics, \FDEP{}s forward failures immediately:
All \BE{}s that fail are marked failed before any gate is evaluated, naturally matching bottom-up propagation.
The effect is as-if the \BE{}s $A$ and $B$ failed simultaneously.
For the new GSPN semantics, we generalise this propagation, and support \FDEP{}s triggered by gates.
Consider $\DFT_5$ in Fig.~\ref{fig:downward_fdep_valid}:
The failure of $B$ indirectly (via $S$ and $D$) forwards to $C$.
If $Z$ is evaluated after the failure is forwarded to $C$, the interpretation is that $B$ and $C$ failed simultaneously and the \PAND fails, as intended.
To guarantee that $C$ is marked failed before $Z$ is evaluated, \emph{$S$ and $D$ require higher priorities than $Z$} in the net.
Consequently, all children of $Z$ are evaluated before $Z$ is evaluated.

Concretely, we generalise bottom-up propagation by refining the priorities:
First, we observe that only for dynamic gates, where the order in which children fail matters, the children need to be evaluated strictly before the parents.
For other gates, we may weaken the constraints on the priorities. A non-strict ordering suffices: $\forall v \in \DFT_{\AND} \cup \DFT_{\OR}: \prioVar_v \leq \prioVar_{\child{v}{i}}, \forall i \in \{1, \dots, |\children(v)|\}$.
Second, we mimic bottom-up propagation in \FDEP forwarding, meaning that dependent events require a priority not larger than their triggers.
Thus, we ensure for each \FDEP $f$, $\prioVar_f \leq \prioVar_{\child{f}{1}}$, and $\prioVar_f \geq \prioVar_{\child{f}{i}}$ for all children $i{\neq}1$.
Equal priorities are admitted. For \FDEP{}s, like for static gates, the status change is order-independent.

Some DFTs (with \FDEP{}s triggered by gates and cyclic forwarding) do not admit a valid priority-assignment.
We argue that the absence of a suitable priority assignment is natural; DFTs without valid priority assignment can model a paradox.
The DFT $\DFT_6$ in Fig.~\ref{fig:downward_fdep_invalid} illustrates this.
The new GSPN semantics induce the following constraints:
\[\prioVar_A < \prioVar_Z, \quad \prioVar_B < \prioVar_Z, \quad \prioVar_Z \leq \prioVar_D, \quad \mbox{and} \quad \prioVar_D \leq \prioVar_B.\]
The constraints imply $\prioVar_B < \prioVar_B$, which is unsatisfiable.
\BE $A$ has failed and the exclusive \POR $Z$ fails too.
(A detailed account of \POR-gates is given in App.~\ref{sec:extensions}.)
But then $B$ fails because of \FDEP $D$.
If we now assume $A$ and $B$ to fail simultaneously, the exclusive \POR cannot fail, as its left child $A$ did not fail strictly before $B$.
Then, $D$'s trigger would have never failed.
Thus, it is reasonable to exclude such DFTs and consider them ill-formed.

The IOIMC and the monolithic MA semantics support \FDEP{}s triggered by gates, but have different interpretations of simultaneity.
The monolithic CTMC semantics is in line with our interpretation, but the algorithm \cite{MCSB99} claimed to match this semantics produces deviating results for the DFTs in this sub-section.

\section{Conclusions and Future Work}
\label{sec:conclusion}
This paper presents a unifying GSPN semantics for Dynamic Fault Trees (DFTs).
The semantics is compositional, the GSPN for each gate is rather simple.
The most appealing aspect of the semantics is that design choices for DFT interpretations are concisely captured by changing only transition priorities and the partitioning of transitions.
Our semantics thus provides a framework for comparing DFT interpretations.
Future work consists of extending the framework to DFTs with repairs~\cite{DBLP:conf/icfem/GuckSS15,bobbio2004parametric} and to study unfoldings~\cite{DBLP:journals/acta/Engelfriet91} of the underlying nets.

%

%========================================================================================
% Bibliography
%========================================================================================
\bibliographystyle{splncs}
%\bibliography{references}
\bibliography{bibliography}
\clearpage
\pagebreak
\appendix
\section{Extensions}
\label{sec:extensions}
In this section, we showcase additional gates to live up to the claim that the presented framework is able to represent the various elements in DFTs.
We additionally concretise some (minor) semantic misconceptions and subtleties that were pointed out in \cite{JGKS16} by referring to the GSPN semantics.

\subsection{Additional gates}
\label{sec:additional_gates}
\subsubsection{Cold \BE{}.}
\label{sec:cold_be}
\begin{figure}[tb]
    \centering
    \begin{tikzpicture}[every label/.append style={font=\scriptsize}]
    \node[iplace,label=above:{$\Active_v$}] (v_active) {};
    \node[iplace,label=below:{$\Disabled_v$}, below=0.3cm of v_active] (v_enabled) {};
    \node[iplace,label=90:{$\Failed_v$}, right=1.9cm of v_active] (v_failed) {};
    \node[iplace,label=below:{$\Unavailable_v$}, below=0.3cm of v_failed] (v_unavailable) {};
    \node[ttransition, label=above:{fail-active}, label=below:{@$\prioVar_v$}, right=1cm of v_active, yshift=-0.5cm, label=60:{$\lambda$}] (v_tactive) {};

    \draw[<->] (v_active) -- ([yshift=2pt] v_tactive.west);
    \draw[o->] (v_tactive) -- (v_failed);
    \draw[->] (v_tactive) -- (v_unavailable);

    \draw[-o] (v_enabled) edge (v_tactive);
\end{tikzpicture}%
    \caption{GSPN template for cold \BE}
    \label{fig:cold_be_template}
\end{figure}
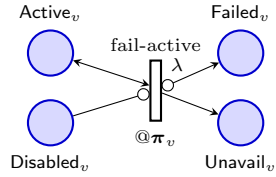

A cold \BE has a passive failure rate of zero, \ie $\mu = 0 \cdot \lambda$.
Thus, if a cold \BE is not active, it cannot fail.
The corresponding template is depicted in Fig.~\ref{fig:cold_be_template}.

\subsubsection{Voting gate.}
\label{sec:votgate}
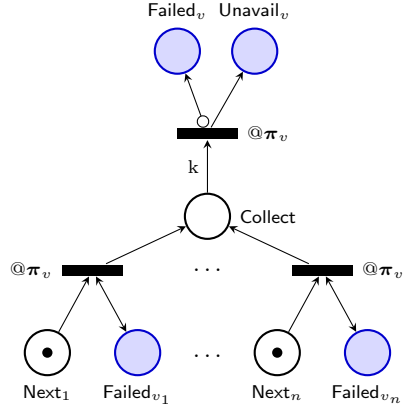
\begin{figure}[tb]
    \centering
    \begin{tikzpicture}[every label/.append style={font=\scriptsize}]
    \node[iplace,label=above:{$\Failed_v$}] (v_failed) {};
    \node[iplace,label=above:{$\Unavailable_v$}, right=0.4cm of v_failed] (v_unavailable) {};

    \node[Itransition, below=0.7cm of v_failed, xshift=0.4cm, label=right:{@$\prioVar_v$}] (tcollect) {};
    \node[place, below=0.7cm of tcollect, label=right:{$\Collect$}] (collect) {};

    \node[Itransition,dummy, below=0.3cm of collect] (dummy) {$\hdots$};
    \node[place,dummy, below=0.7cm of dummy] (dummy2) {$\hdots$};

    \node[Itransition, left=0.7cm of dummy, label=left:{@$\prioVar_v$}] (t1fail) {};
    \node[iplace, below=0.7cm of t1fail, xshift=0.6cm, label=below:{$\Failed_{\child{v}{1}}$}] (sv1_failed) {};
    \node[place, tokens=1, below=0.7cm of t1fail, xshift=-0.6cm, label=below:{$\Consider_1$}] (sv1_consider) {};

    \node[Itransition, right=0.7cm of dummy, label=right:{@$\prioVar_v$}] (tnfail) {};
    \node[iplace, below=0.7cm of tnfail, xshift=0.6cm, label=below:{$\Failed_{\child{v}{n}}$}] (svn_failed) {};
    \node[place, tokens=1, below=0.7cm of tnfail, xshift=-0.6cm, label=below:{$\Consider_n$}] (svn_consider) {};

    \draw[<->] (sv1_failed) -- (t1fail);
    \draw[->] (sv1_consider) -- (t1fail);
    \draw[->] (t1fail) -- (collect);
    \draw[<->] (svn_failed) -- (tnfail);
    \draw[->] (svn_consider) -- (tnfail);
    \draw[->] (tnfail) -- (collect);
    \draw[->] (collect) -- node[left, font=\scriptsize]{k} (tcollect);
    \draw[o->] (tcollect) -- (v_failed);
    \draw[->] (tcollect) -- (v_unavailable);
\end{tikzpicture}%
    \caption{GSPN template for a $\VOT{k}$-gate}
    \label{fig:vot_template}
\end{figure}

A $\VOT{k}$-gate fails if $k$ out of $n$ (with $k \leq n$) of its inputs have failed in arbitrary order.
This gate does not add any expressive power; it is equivalent to a combination of \AND- and \OR-gates.
As this however can result in an exponentially-sized DFT, it is convenient to include the $\VOT{k}$-gate as a first-class citizen.
Fig.~\ref{fig:vot_template} shows the GSPN template $\templ{\VOT{k}}{v}$ for a $\VOT{k}$-gate $v$ with $n$ inputs.
If child $\child{v}{i}$ fails, its transition fires and puts a token in the shared place $\Collect$.
As the token from place $\Consider_i$ is removed, $\child{v}{i}$'s transition is disabled afterwards.
This prevents the generation of multiple tokens in $\Collect$ by the same transition.
If $\Collect$ contains $k$ tokens, the corresponding transition can fire and places a token in $\Failed_v$ indicating that $\VOT{k}$ has failed.
It should be noted that the resulting net is $k$-bounded.

\subsubsection{\POR.}
Similar to the $\PAND$, two versions of the $\POR$ are considered: the inclusive (denoted $\leq$) and the exclusive (denoted $<$) variant.
\begin{figure}[tb]
\centering
\subfigure[$\PORincl$]{
    \begin{tikzpicture}[every label/.append style={font=\scriptsize}]
    \node[iplace,label=90:{$\Failed_v$}] (v_failed) {};
    \node[iplace,label=90:{$\Unavailable_v$}, right=0.6cm of v_failed] (v_unavailable) {};
    \node[place,label=90:{$\Failsafe$}, right=1.1cm of v_unavailable] (v_failsafe) {};

    \node[Itransition, below=0.5cm of v_failed, label=left:{@$\prioVar_v$}] (t_vfailed) {};

    \node[iplace,below=1.4cm of v_failed,label=below:{$\Failed_{\child{v}{1}}$}] (sv1_failed) {};
    \node[iplace,right=0.4cm of sv1_failed,label=below:{$\Failed_{\child{v}{2}}$}] (sv2_failed) {};
    \node[place,dummy,right=0.3cm of sv2_failed] (dummy) {$\hdots$};
    \node[iplace,below=1.4cm of v_failsafe,label=270:{$\Failed_{\child{v}{n}}$}] (svn_failed) {};

    \node[Itransition, above=0.75cm of sv2_failed, label=315:{@$\prioVar_v$}] (t_vfailsafe1) {};
    \node[Itransition, above=0.75cm of svn_failed, label=right:{@$\prioVar_v$}] (t_vfailsafen) {};
    \node[Itransition,dummy, above=0.75cm of dummy] {$\hdots$};

    \draw[<->] (sv1_failed) -- (t_vfailed);
    \draw[o->] (t_vfailed) -- (v_failed);
    \draw[->] (t_vfailed) -- (v_unavailable);
    \draw[o-]  (t_vfailed) -- (v_failsafe);

    \draw[<->] (sv2_failed) -- (t_vfailsafe1);
    \draw[<->] (svn_failed) -- (t_vfailsafen);
    \draw[o-]  (t_vfailsafe1) -- (sv1_failed.70);

    \draw[o-]  (t_vfailsafen) -- (sv1_failed.55);
    \draw[o->] (t_vfailsafe1) -- (v_failsafe);
    \draw[o->] (t_vfailsafen) -- (v_failsafe);
\end{tikzpicture}%
    \label{fig:porincl_template}
}
\hspace{0.5cm}
\subfigure[$\PORexcl$]{
    \begin{tikzpicture}[every label/.append style={font=\scriptsize}]
    \node[iplace,label=90:{$\Failed_v$}] (v_failed) {};
    \node[iplace,label=90:{$\Unavailable_v$}, right=0.6cm of v_failed] (v_unavailable) {};

    \node[Itransition, below=0.5cm of v_failed, label=left:{@$\prioVar_v$}] (t_vfailed) {};

    \node[iplace,below=1.4cm of v_failed,label=below:{$\Failed_{\child{v}{1}}$}] (sv1_failed) {};
    \node[iplace,right=0.4cm of sv1_failed,label=below:{$\Failed_{\child{v}{2}}$}] (sv2_failed) {};
    \node[place,dummy, right=0.3cm of sv2_failed] (dummy) {$\hdots$};
    \node[iplace,right=1.1cm of sv2_failed,label=270:{$\Failed_{\child{v}{n}}$}] (svn_failed) {};

    \node[place, above=1.4cm of svn_failed, label=90:{$\Failsafe$}] (v_failsafe) {};

    \draw[<->] (sv1_failed) -- (t_vfailed);
    \draw[o->] (t_vfailed) -- (v_failed);
    \draw[->] (t_vfailed) -- (v_unavailable);

    \draw[-o] (sv2_failed) -- (t_vfailed);
    \draw[-o] (svn_failed) -- (t_vfailed);
\end{tikzpicture}%
    \label{fig:porexcl_template}
}
\caption{GSPN templates for inclusive and exclusive \POR}
\label{fig:por_templates}
\end{figure}
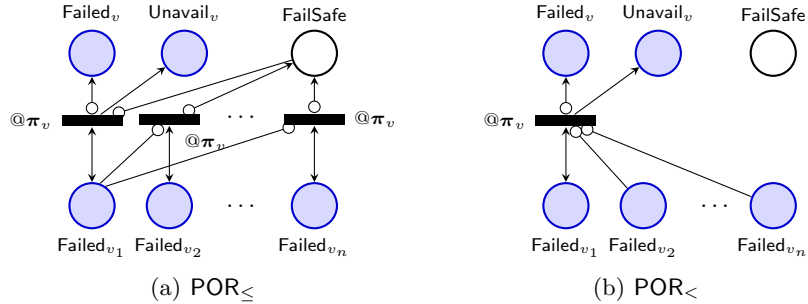

The inclusive $\PORincl$ fails if its leftmost child fails before or simultaneously with the other children.
Fig.~\ref{fig:porincl_template} depicts the template $\templ{\PORincl}{v}$ for $\PORincl$-gate $v$ with $n$ children.
The leftmost transition can fire if $\child{v}{1}$ has failed and the $\PORincl$ is not yet fail-safe.
The token put in $\Failed_v$ indicates that the failure condition of $\PORincl$ is fulfilled.
If a sibling of $\child{v}{1}$ fails before $\child{v}{1}$ does, the corresponding transition can fire and a token is put in $\Failsafe$.
If the leftmost child and another child fail simultaneously, only the leftmost transition is enabled, the $\PORincl$ fails and cannot become fail-safe.

The exclusive $\PORexcl$ fails if the leftmost child fails strictly before all its siblings.
For $\PORexcl$ $v$ with $n$ children, the template $\templ{\PORexcl}{v}$ is depicted in Fig.~\ref{fig:porexcl_template}.
The transition can only fire if $\Failed_{\child{v}{1}}$ contains a token but all other $\Failed_{\child{v}{i}}$ do not.
The $\Failsafe$ place is superfluous here.

\subsubsection{Probabilistic dependencies.}
\label{sec:pdep_template}
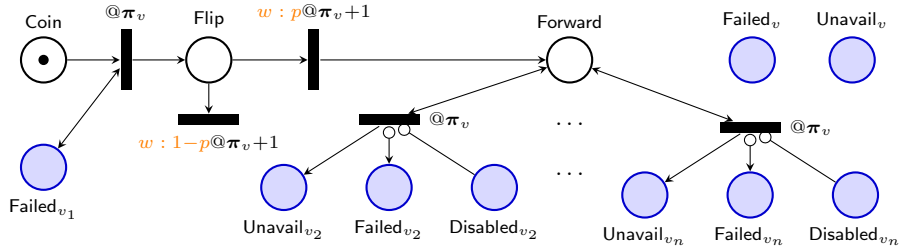
\begin{figure}[tb]
\begin{tikzpicture}[every label/.append style={font=\scriptsize}]

	\node[iplace, label=below:{$\Failed_{\child{v}{1}}$}] (sv1_failed) {};
	\node[place, above=0.8cm of sv1_failed, label=above:{$\Coin$}, tokens=1] (pdep_go) {};
	\node[itransition, right=0.7cm of pdep_go, label=above:{@$\prioVar_{v}$}] (tgo) {};
	\node[place, right=0.7cm of tgo, label=above:{$\Flip$}] (startgamble) {};
	\node[itransition, right=1.0cm of startgamble, label=above:{\color{orange}$w:p$\color{black}@$\prioVar_{v}{+}1$}] (tgamblesucc) {};
	\node[place, right=3.0cm of tgamblesucc, label=above:{$\Forward$}] (distr) {};
	\node[Itransition, below=0.4cm of startgamble, label=below:{\color{orange}$w:1{-}p$\color{black}@$\prioVar_{v}{+}1$}] (tgamblefail) {};

	\node[iplace,label=above:{$\Failed_v$}, right=1.8cm of distr] (v_failed) {};
	\node[iplace,label=above:{$\Unavailable_v$}, right=0.7cm of v_failed] (v_unavailable) {};

	\node[Itransition, below=0.4cm of distr, xshift=-2.4cm, label=right:{@$\prioVar_{v}$}] (tf1) {};
	\node[below=0.4cm of distr] (tf_dots) {$\hdots$};
	\node[below=0.4cm of tf_dots] (tf_dots2) {$\hdots$};
	\node[Itransition, below=0.5cm of distr, xshift=2.4cm, label=right:{@$\prioVar_{v}$}] (tfn) {};

	\node[iplace, below=0.5cm of tf1, label=below:{$\Failed_{\child{v}{2}}$}] (sv2_failed) {};
	\node[iplace, below=0.5cm of tf1, xshift=-1.4cm, label=below:{$\Unavailable_{\child{v}{2}}$}] (sv2_unavail) {};
	\node[iplace, below=0.5cm of tf1, xshift=1.4cm, label=below:{$\Disabled_{\child{v}{2}}$}] (sv2_enabled) {};
	\node[iplace, below=0.5cm of tfn, label=below:{$\Failed_{\child{v}{n}}$}] (svn_failed) {};
	\node[iplace, below=0.5cm of tfn, xshift=-1.4cm, label=below:{$\Unavailable_{\child{v}{n}}$}] (svn_unavail) {};
	\node[iplace, below=0.5cm of tfn, xshift=1.4cm, label=below:{$\Disabled_{\child{v}{n}}$}] (svn_enabled) {};

	\draw[<->] (sv1_failed) -- (tgo);
	\draw[->] (pdep_go) -- (tgo);
	\draw[->] (tgo) -- (startgamble);
	\draw[->] (startgamble) -- (tgamblefail);
	\draw[->] (startgamble) -- (tgamblesucc);
	\draw[->] (tgamblesucc) -- (distr);

	\draw[<->] (distr) -- (tf1);
	\draw[<->] (distr) -- (tfn);
	\draw[-o] (sv2_enabled) -- (tf1);
	\draw[-o] (svn_enabled) -- (tfn);
	\draw[<-o] (sv2_failed) -- (tf1);
	\draw[<-o] (svn_failed) -- (tfn);
	\draw[<-] (sv2_unavail) -- (tf1);
	\draw[<-] (svn_unavail) -- (tfn);
\end{tikzpicture}%
\caption{GSPN template for $\PDEP_p$}
\label{fig:pdep_template}
\end{figure}%
The template for \FDEP{}s was treated in Sect.~\ref{sec:templates}.
Recap that $\FDEP = \PDEP_p$ for $p=1$.
We now consider the general $\PDEP_p$.
Fig.~\ref{fig:pdep_template} depicts the template $\templ{\PDEP_p}{v}$ for a $\PDEP_p$-gate $v$ with $n$ children.
Once the trigger of $\PDEP{p}$ fails, the failure is propagated to the children with probability $p$.
With probability $1{-}p$ no propagation happens, as ensured by the weights.
If a token is placed in $\Failed_{\child{v}{1}}$, the leftmost transition can fire and moves the token from $\Coin$ to $\Flip$.
Next, the token is either moved from $\Flip$ to $\Forward$ with probability $p$ or is removed from $\Flip$ with probability $1{-}p$.
In the latter case no transitions are enabled anymore and the failure propagation is stopped.
If a token is placed in $\Forward$, the failure propagation to the dependent children takes places just as in the \FDEP.

The auxiliary $\mathsf{Flip}$ place ensures that only the two incident transitions are enabled.
As only these two transitions are enabled, we can ensure that probability to move the token to $\mathsf{Forward}$ is indeed $p$,
 without the usage of more complex partitions in the GSPN definition.

\subsubsection{Restrictors.}
\label{sec:seq_template} 
The common restrictor is the \SEQ.
It allows its children only to fail from left to right.
Thus, there is at most one child that is allowed to fail (the \emph{current} child).
Its right sibling is the \emph{next} child.
Initially, the leftmost child is the current child.
If the current child fails, the \SEQ lifts its restriction to the next child.
Contrarily to what is sometimes claimed in the literatures, \SEQ{}s cannot be  modelled by \SPARE{}s in general.

We only consider restrictions over \BE{s} as \SEQ{}s with gates as children raise several semantic complications \cite{JGKS16}.
Furthermore, to the best of our knowledge, \SEQ{}s over gates are used only to model a \SPARE or a \MUTEX.

A modular translation of \SEQ implements a counter for each \BE (the $\Disabled$ place), initialised with the number of \SEQ{}s which potentially prevent the \BE from failing.
Every \SEQ which grants its concession decreases the counter.
If the counter becomes zero, the \BE is free to fail.
The GSPN is now bounded with the maximal number of \SEQ{}s restricting one \BE, \ie $\textsf{max}_{b \in \DFT_{\BE}} |\{v \in \DFT_{\SEQ} \sep \exists i: \child{v}{i} = b\}|$.
As for failure forwarding and failure forwarding, the selection of priorities for transitions from \templ{\PDEP_p}{v} changes the semantics. 
In particular, it is interesting whether tokens are removed from the \Disabled{} places before or after failure forwarding. 
We refrain from a in-depth discussion of this semantic issue.

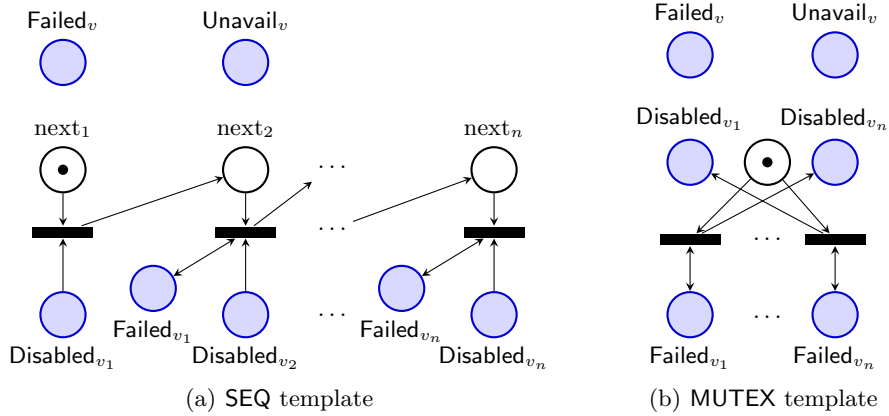
\begin{figure}[tb]
\subfigure[$\SEQ$ template]{
    \begin{tikzpicture}
	\node[iplace, label=below:{$\Disabled_{\child{v}{1}}$}] (d1) {};
	\node[place, label=above:{$\text{next}_1$}, above=1.3cm of d1, tokens=1] (n1) {};
	\node[Itransition, above=0.7cm of d1] (t1) {};
	\node[iplace, label=below:{$\Disabled_{\child{v}{2}}$}, right=1.8cm of d1] (d2) {};
	\node[place, label=above:{$\text{next}_2$}, above=1.3cm of d2] (n2) {};
	\node[iplace, left=0.6cm of d2, yshift=0.35cm, label=below:{$\Failed_{\child{v}{1}}$}] (f1) {};
	\node[Itransition, above=0.7cm of d2] (t2) {};

	\node[place,dummy, right=0.5cm of d2] (dummy) {$\hdots$};
	\node[place,dummy, above=1.3cm of dummy] (ddummy) {$\hdots$};
	\node[Itransition,dummy, above=0.7cm of dummy] (tdummy) {$\hdots$};

	\node[iplace, label=below:{$\Disabled_{\child{v}{n}}$}, right=1.5cm of dummy] (dn) {};
	\node[place, label=above:{$\text{next}_n$}, above=1.3cm of dn] (nn) {};
	\node[iplace, left=0.6cm of dn, yshift=0.35cm, label=below:{$\Failed_{\child{v}{n}}$}] (fn) {};
	\node[Itransition, above=0.7cm of dn] (tn) {};

	\node[iplace,label=above:{$\Failed_v$}, above=0.8cm of n1] (v_failed) {};
	\node[iplace,label=above:{$\Unavailable_v$}, right=1.8cm of v_failed] (v_unavailable) {};

	\draw[->] (n1) -- (t1);
	\draw[->] (d1) -- (t1);
	\draw[->] (t1) -- (n2);

	\draw[<->] (f1) -- (t2);
	\draw[->] (n2) -- (t2);
	\draw[->] (d2) -- (t2);
	\draw[->] (t2) -- (ddummy);

	\draw[->] (tdummy) -- (nn);

	\draw[<->] (fn) -- (tn);
	\draw[->] (nn) -- (tn);
	\draw[->] (dn) -- (tn);
\end{tikzpicture}%
    \label{fig:seq_template}
}
\hspace{0.5cm}
\subfigure[$\MUTEX$ template]{
    \begin{tikzpicture}
	\node[iplace, label=above:{$\Disabled_{\child{v}{1}}$}] (d1) {};
	\node[iplace, label=above:{$\Disabled_{\child{v}{n}}$}, right=1.3cm of d1] (dn) {};
	\node[iplace, label=below:{$\Failed_{\child{v}{1}}$}, below=1.4cm of d1] (f1) {};
	\node[iplace, label=below:{$\Failed_{\child{v}{n}}$}, below=1.4cm of dn] (fn) {};
	\node[place, tokens=1, right=0.4cm of d1] (c) {};
	\node[iplace,dummy, below=1.4cm of c] (dummy) {$\hdots$};

	\node[Itransition, above=0.6cm of f1] (t1) {};
	\node[Itransition, above=0.6cm of fn] (tn) {};
	\node[Itransition,dummy, above=0.6cm of dummy] (tdummy) {$\hdots$};

	\node[iplace,label=above:{$\Failed_v$}, above=0.8cm of d1] (v_failed) {};
	\node[iplace,label=above:{$\Unavailable_v$}, right=1.3cm of v_failed] (v_unavailable) {};

	\draw[<->] (f1) -- (t1);
	\draw[<->] (fn) -- (tn);

	\draw[->] (t1) -- (dn);
	\draw[->] (tn) -- (d1);

	\draw[<-] (t1) -- (c);
	\draw[<-] (tn) -- (c);
\end{tikzpicture}%
    \label{fig:mutex_template}
}
\caption{Restrictor templates}
\label{fig:restrictor_templates}
\end{figure}

\paragraph{Can \SEQ{}s be used to model mutual exclusion?}
As we only allow restrictions over \BE{s},  mutual exclusion of two \BE{s} is no longer syntactic sugar (via a construction from \cite{JGKS16}):
A \SEQ{} never disables a \BE once it is free to fail, contrary to the concept of mutual exclusion.
Therefore, we add a dedicated template (Fig.~\ref{fig:mutex_template}) for mutual exclusion. The template uses the well-known Petri net construction.

\begin{remark}[Can we lift the syntactic restriction on restrictors?]
Allowing \SEQ{s} to have gates as children would allow more constructions that describe which failure can occur.
To realise support for restrictors over gates, three approaches can be used:
(1)~\emph{Actively} preventing failures which would lead to a disabled gate failing.
Active prevention could be implemented in either the GSPN or via a static analysis.
While the former is compositional, it makes the Petri net complex to understand.
(2)~\emph{Passively} preventing failures of disabled gates.
 If such a failure occurs, a rollback to the marking which initially caused the disabled gate to fail has to be initialised.
While such a rollback is simple in a monolithic and explicit state based approach (as in \cite{VJK17}), it is significantly harder to implement symbolically (in the GSPN).
(3)~\emph{Ignore} the failure of \SEQ{}s while marking states where the \SEQ is failed. Then, a suitable analysis technique needs to handle the occurrence of such states.
\end{remark}

\subsection{Claiming variants}
\subsubsection{Arbitrary claiming order.}
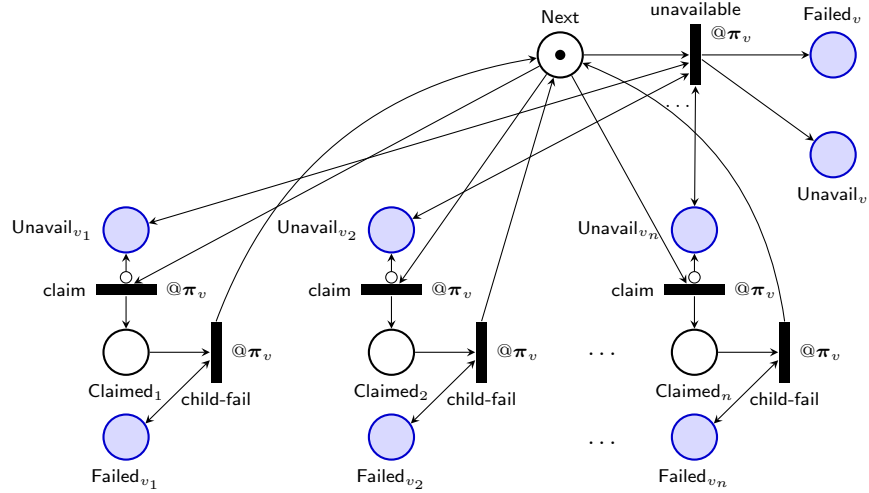
\begin{figure}[tb]
    \centering
    \begin{tikzpicture}[every label/.append style={font=\scriptsize}]
	\node[place, label=above:{$\Consider$}, tokens=1] (consider) {};
	\node[itransition, right=1.4cm of consider, label=above:{$\tunavailable$}, label=35:{@$\prioVar_v$}, label=below:{$\hdots\phantom{123}$}] (tclaimfail) {};

    % Child 1
	\node[place,dummy, below=1cm of consider, xshift=-6.5cm] (dummy1) {};
	\node[iplace, right=0.1cm of dummy1, yshift=-0.7cm, label=left:{$\Unavailable_{\child{v}{1}}$}] (unavailable1) {};
	\node[place, below=1cm of unavailable1, label=below:{$\Claimed_1$}] (claimed1) {};
	\node[iplace, below=0.5cm of claimed1, label=below:{$\Failed_{\child{v}{1}}$}] (failed1) {};

	\node[itransition, right=0.8cm of claimed1, label=below:{$\tchildfail$}, label=right:{@$\prioVar_v$}] (tfailedclaimed1) {};
	\node[Itransition, below=0.4cm of unavailable1, label=left:{$\tclaim$}, label=right:{@$\prioVar_v$}] (tclaimsuccess1) {};

	\node[place,dummy, right=2.9cm of dummy1] (dummy2) {};

	\draw[->] (consider) -- ([xshift=3pt] tclaimsuccess1.north);
	\draw[<-o] (unavailable1) -- (tclaimsuccess1.north);
	\draw[->] (tclaimsuccess1) -- (claimed1);

	\draw[->] (claimed1) -- (tfailedclaimed1.west);
	\draw[<->] (failed1.35) -- ([yshift=-3pt] tfailedclaimed1.west);
	\draw[->] (tfailedclaimed1.north) edge[bend left] (consider);

    % Child 2
	\node[iplace, right=0.1cm of dummy2, yshift=-0.7cm, label=left:{$\Unavailable_{\child{v}{2}}$}] (unavailable2) {};
	\node[place, below=1cm of unavailable2, label=below:{$\Claimed_2$}] (claimed2) {};
	\node[iplace, below=0.5cm of claimed2, label=below:{$\Failed_{\child{v}{2}}$}] (failed2) {};

	\node[itransition, right=0.8cm of claimed2, label=below:{$\tchildfail$}, label=right:{@$\prioVar_v$}] (tfailedclaimed2) {};
	\node[Itransition, below=0.4cm of unavailable2, label=left:{$\tclaim$}, label=right:{@$\prioVar_v$}] (tclaimsuccess2) {};

	\node[place,dummy, right=2.9cm of dummy2] (considerdummy) {};

	\draw[->] (consider) -- ([xshift=3pt] tclaimsuccess2.north);
	\draw[<-o] (unavailable2) -- (tclaimsuccess2.north);
	\draw[->] (tclaimsuccess2) -- (claimed2);

	\draw[->] (claimed2) -- (tfailedclaimed2.west);
	\draw[<->] (failed2.35) -- ([yshift=-3pt] tfailedclaimed2.west);
	\draw[->] (tfailedclaimed2.north) -- (consider);

    % Child dummy
	\node[place,dummy, below=1.7cm of considerdummy] (claimeddummy) {$\hdots$};
	\node[place,dummy, below=0.5cm of claimeddummy] (faileddummy) {$\hdots$};

    % Child n
	\node[iplace, right=0.6cm of considerdummy, yshift=-0.7cm, label=left:{$\Unavailable_{\child{v}{n}}$}] (unavailablen) {};
	\node[place, below=1cm of unavailablen, label=below:{$\Claimed_n$}] (claimedn) {};
	\node[iplace, below=0.5cm of claimedn, label=below:{$\Failed_{\child{v}{n}}$}] (failedn) {};

	\node[itransition, right=0.8cm of claimedn, label=below:{$\tchildfail$}, label=right:{@$\prioVar_v$}] (tfailedclaimedn) {};
	\node[Itransition, below=0.4cm of unavailablen, label=left:{$\tclaim$}, label=right:{@$\prioVar_v$}] (tclaimsuccessn) {};

	\node[iplace, label=above:{$\Failed_v$}, right=3cm of consider] (v_failed) {};

%	\draw[->] (considerdummy) -- (tclaimfailn.west);
%	\draw[<->] (unavailablen) -- ([yshift=-3pt] tclaimfailn.west);
%	\draw[->] (tclaimfailn) -- (v_failed);

	\draw[->] (consider) -- ([xshift=-3pt] tclaimsuccessn.north);
	\draw[<-o] (unavailablen) -- (tclaimsuccessn.north);
	\draw[->] (tclaimsuccessn) -- (claimedn);

	\draw[->] (claimedn) -- (tfailedclaimedn.west);
	\draw[<->] (failedn.35) -- ([yshift=-3pt] tfailedclaimedn.west);
	\draw[->] (tfailedclaimedn.north) edge[bend right] (consider);

	\node[iplace,label=below:{$\Unavailable_v$}, below=0.7cm of v_failed] (v_unavailable) {};

    \draw[<->] (unavailable1) -- ([yshift=-3pt] tclaimfail.west);
    \draw[<->] (unavailable2) -- ([yshift=-7pt] tclaimfail.west);
    \draw[<->] (unavailablen) -- (tclaimfail.south);
    \draw[->] (consider) -- (tclaimfail);
    \draw[->] (tclaimfail) -- (v_failed);
    \draw[->] (tclaimfail) -- (v_unavailable);

\end{tikzpicture}%
    \caption{GSPN template for a \SPARE with arbitrary claiming order}
    \label{fig:spare_arbitrary_claiming}
\end{figure}

The \SPARE{}s considered so far try to claim the children in order from left to right.
We now adapt the \SPARE template to allow claiming in arbitrary order.
The template is depicted in Fig.~\ref{fig:spare_arbitrary_claiming}.

If a token is in $\Consider$, all available spare components can be claimed.
This is reflected by the $\tclaim$ transitions which are enabled.
The choice which child is claimed is resolved non-deterministically.
After a child is claimed, there is no token in $\Consider$ anymore.
Thus, no other child can be claimed.
If the used child fails, the token is put back in $\Consider$ and another child can be claimed.
If all spare components are unavailable, the transition $\tunavailable$ can fire and places a token in $\Failed_v$ marking the failure of the \SPARE.

\subsubsection{Non-exclusive claiming.}
So far, shared spare components are claimed exclusively by one \SPARE.
It is interesting to lift this restriction and allow non-exclusive claiming.
That means that a spare component $c$ can be claimed by multiple \SPARE{}s at the same time.
If $c$ fails, all \SPARE{}s which use $c$ have to claim a new child.
We leave an extension which allows non-exclusive claiming as future work.

\subsection{Nested \SPARE semantics}
\label{sec:nested_spares}
In Sect.~\ref{sec:dfts}, we introduced \SPARE{}s with \emph{early claiming}.
There are variants for these semantics \cite{JGKS16}, which differ when the \SPARE is not activated: In particular, inactive \SPARE{}s might be prevented to claim.

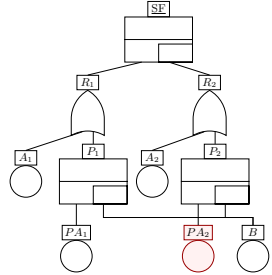
\begin{figure}[tb]
\centering
\scalebox{\dftscale}{
\begin{tikzpicture}
    \node[spare] (System) {};
    \node[labelbox] (System_label) at (System.north) {\underline{SF}};

    \node[or2, below=1.2cm of System, anchor=south, yshift=1.2cm] (R1) {};
    \node[labelbox] (R1_label) at (R1.east) {$R_1$};
    \node[or2, below=1.2cm of System, anchor=south, yshift=-1.5cm] (R2) {};
    \node[labelbox] (R2_label) at (R2.east) {$R_2$};

    \node[spare, below=0.6cm of R1.input 2] (P1) {};
    \node[labelbox] (P1_label) at (P1.north) {$P_1$};
    \node[be, below=0.8cm of P1.P] (PA1) {};
    \node[labelbox] (PA1_label) at (PA1.north) {$PA_1$};
    \node[be, left=1.1cm of P1, anchor=west] (A1) {};
    \node[labelbox] (A1_label) at (A1.north) {$A_1$};

    \node[spare, below=0.6cm of R2.input 2] (P2) {};
    \node[labelbox] (P2_label) at (P2.north) {$P_2$};
    \node[be,failed, below=0.8cm of P2.P] (PA2) {};
    \node[labelbox,failed] (PA2_label) at (PA2.north) {$PA_2$};
    \node[be, left=1.0cm of P2, anchor=west] (A2) {};
    \node[labelbox] (A2_label) at (A2.north) {$A_2$};

    \node[be, right=0.5cm of PA2] (B) {};
    \node[labelbox] (B_label) at (B.north) {$B$};

    %\node[be, right=0.5cm of PA1] (PS) {};
    %\node[labelbox] (PS_label) at (PS.north) {PS};

    \draw[-] (P2.P) -- (PA2_label.north);
    \draw[-] (P2.SB) -| +(0,-0.3) -| (B_label.north);

    \draw[-] (P1.P) -- (PA1_label.north);
    \draw[-] (P1.SB) -| +(0,-0.3) -| (B_label.north);

    \draw[-] (System.P) -- (R1_label.north);
    \draw[-] (System.SB) -- (R2_label.north);

    \draw[-] (R2.input 1) -- (A2_label.north);
    \draw[-] (R2.input 2) -- (P2_label.north);

    \draw[-] (R1.input 1) -- (A1_label.north);
    \draw[-] (R1.input 2) -- (P1_label.north);
\end{tikzpicture}%
}
\caption{DFT with nested \SPARE{}s}
\label{fig:nested_spare}
\end{figure}

\begin{example}(based on \cite{JGKS16})
\label{ex:nested_spare_late_claiming}
The DFT depicted in Fig.~\ref{fig:nested_spare} describes a communication system consisting of two radios $R_1$ and $R_2$, where $R_2$ is the back-up.
Each radio consists of an antenna ($A_1$ and $A_2$, respectively) and a power unit ($P_1$ and $P_2$, respectively).
Both power units have their own power adaptor ($PA_1$ and $PA_2$, respectively).
Every power unit can use the spare battery ($B$).
Consider the failure of $PA_2$.
Under early claiming, the power unit $P_2$ directly claims battery $B$ which then cannot be claimed anymore by $P_1$.
Under late claiming, $P_2$ does not claim $B$ yet.
Instead, it will only claim $B$ once $R_1$ failed and $R_2$ has subsequently been activated.
\end{example}
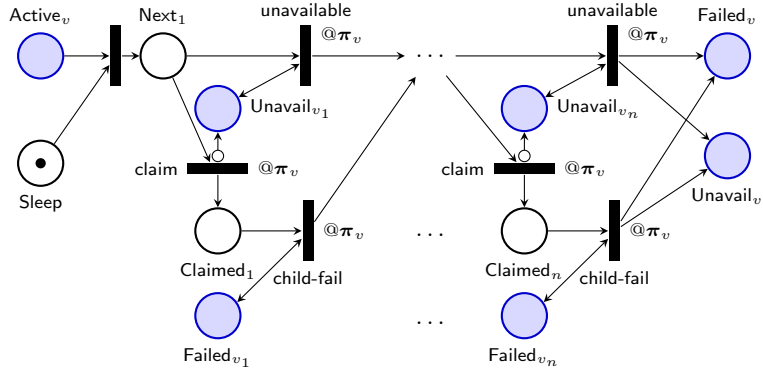
\begin{figure}[tb]
\centering
\begin{tikzpicture}[every label/.append style={font=\scriptsize}]
    % Child 1
	\node[place, label=above:{$\Consider_1$}] (consider1) {};
	\node[iplace, right=0.1cm of consider1, yshift=-0.7cm, label=right:{$\Unavailable_{\child{v}{1}}$}] (unavailable1) {};
	\node[place, below=1cm of unavailable1, label=below:{$\Claimed_1$}] (claimed1) {};
	\node[iplace, below=0.5cm of claimed1, label=below:{$\Failed_{\child{v}{1}}$}] (failed1) {};

	\node[itransition, right=0.8cm of claimed1, label=below:{$\tchildfail$}, label=right:{@$\prioVar_v$}] (tfailedclaimed1) {};
	\node[Itransition, below=0.4cm of unavailable1, label=left:{$\tclaim$}, label=right:{@$\prioVar_v$}] (tclaimsuccess1) {};
	\node[itransition, right=1.5cm of consider1, label=above:{$\tunavailable$}, label=35:{@$\prioVar_v$}] (tclaimfail1) {};

	\node[place,dummy, right=2.9cm of consider1] (considerdummy) {$\hdots$};

	\draw[->] (consider1) -- (tclaimfail1.west);
	\draw[<->] (unavailable1) -- ([yshift=-3pt] tclaimfail1.west);
	\draw[->] (tclaimfail1) -- (considerdummy);

	\draw[->] (consider1) -- ([xshift=-3pt] tclaimsuccess1.north);
	\draw[<-o] (unavailable1) -- (tclaimsuccess1.north);
	\draw[->] (tclaimsuccess1) -- (claimed1);

	\draw[->] (claimed1) -- (tfailedclaimed1.west);
	\draw[<->] (failed1.35) -- ([yshift=-3pt] tfailedclaimed1.west);
	\draw[->] (tfailedclaimed1) -- (considerdummy);

    % Child dummy
	\node[place,dummy, below=1.7cm of considerdummy] (claimeddummy) {$\hdots$};
	\node[place,dummy, below=0.5cm of claimeddummy] (faileddummy) {$\hdots$};

    % Child n
	\node[iplace, right=0.6cm of considerdummy, yshift=-0.7cm, label=right:{$\Unavailable_{\child{v}{n}}$}] (unavailablen) {};
	\node[place, below=1cm of unavailablen, label=below:{$\Claimed_n$}] (claimedn) {};
	\node[iplace, below=0.5cm of claimedn, label=below:{$\Failed_{\child{v}{n}}$}] (failedn) {};

	\node[itransition, right=0.8cm of claimedn, label=below:{$\tchildfail$}, label=right:{@$\prioVar_v$}] (tfailedclaimedn) {};
	\node[Itransition, below=0.4cm of unavailablen, label=left:{$\tclaim$}, label=right:{@$\prioVar_v$}] (tclaimsuccessn) {};
	\node[itransition, right=2.0cm of considerdummy, label=above:{$\tunavailable$}, label=35:{@$\prioVar_v$}] (tclaimfailn) {};

	\node[iplace, label=above:{$\Failed_v$}, right=3.3cm of considerdummy] (v_failed) {};

	\draw[->] (considerdummy) -- (tclaimfailn.west);
	\draw[<->] (unavailablen) -- ([yshift=-3pt] tclaimfailn.west);
	\draw[->] (tclaimfailn) -- (v_failed);

	\draw[->] (considerdummy) -- ([xshift=-3pt] tclaimsuccessn.north);
	\draw[<-o] (unavailablen) -- (tclaimsuccessn.north);
	\draw[->] (tclaimsuccessn) -- (claimedn);

	\draw[->] (claimedn) -- (tfailedclaimedn.west);
	\draw[<->] (failedn.35) -- ([yshift=-3pt] tfailedclaimedn.west);
	\draw[->] (tfailedclaimedn) -- (v_failed);

	\node[iplace,label=below:{$\Unavailable_v$}, below=0.7cm of v_failed] (v_unavailable) {};

	\draw[->] (tfailedclaimedn) -- (v_unavailable);
	\draw[->] (tclaimfailn) -- (v_unavailable);

    % Late claiming
	\node[iplace, left=1cm of consider1, label=above:{$\Active_v$}] (active) {};
	\node[place, below=0.8cm of active,  label=below:{$\mathsf{Sleep}$},tokens=1] (holder) {};
	\node[itransition, right=0.6cm of active] (tstart) {};
	\draw[->] (active) -- (tstart);
	\draw[->] (holder) -- (tstart);
	\draw[->] (tstart) -- (consider1);
\end{tikzpicture}%
\caption{Part of \SPARE template for \emph{late claiming/late failing}}
\label{fig:late_claiming}
\end{figure}

Fig.~\ref{fig:late_claiming} depicts part of the template for \SPARE{}s with late claiming.
The changes are minimal: $\Consider_1$ no longer initially gets a token, instead, the token is only placed there upon activation.
Late claiming semantics raises some questions:

\paragraph{Can an operational, inactive \SPARE have zero operational children?}
First, observe that with early claiming, a \SPARE with zero operational children will have failed to claim a new used child, and thus it will fail.
In particular, \SPARE{}s only fail after a used child fails.
With late claiming, an inactive \SPARE has no used child.
Thus, without adapting the template, a \SPARE can have zero operational children.
We refer to this as \emph{late failing}.
\emph{Early failing} circumvents this situation by failing as soon as a \SPARE has zero operational children.
\begin{example}
We continue with Example~\ref{ex:nested_spare_late_claiming}.
Using late failing, $P_2$ fails only if it fails to claim upon activation of $R_2$.
Using early failing, $P_2$ fails---regardless of $R_2$ being activated or not---whenever $B$ failed or was claimed by $P_1$.
\end{example}

\begin{figure}[tb]
\centering
\begin{tikzpicture}[every label/.append style={font=\scriptsize}]
    % Child 1
	\node[place, label=above:{$\Consider_1$}] (consider1) {};
	\node[iplace, right=0.1cm of consider1, yshift=-0.7cm, label=right:{$\Unavailable_{\child{v}{1}}$}] (unavailable1) {};
	\node[place, below=1cm of unavailable1, label=below:{$\Claimed_1$}] (claimed1) {};
	\node[iplace, below=0.5cm of claimed1, label=below:{$\Failed_{\child{v}{1}}$}] (failed1) {};

	\node[itransition, right=0.8cm of claimed1, label=below:{$\tchildfail$}, label=right:{@$\prioVar_v$}] (tfailedclaimed1) {};
	\node[Itransition, below=0.4cm of unavailable1, label=left:{$\tclaim$}, label=right:{@$\prioVar_v$}] (tclaimsuccess1) {};
	\node[itransition, right=1.5cm of consider1, label=above:{$\tunavailable$}, label=35:{@$\prioVar_v$}] (tclaimfail1) {};

	\node[place,dummy, right=2.9cm of consider1] (considerdummy) {$\hdots$};

	\draw[->] (consider1) -- (tclaimfail1.west);
	\draw[<->] (unavailable1) -- ([yshift=-3pt] tclaimfail1.west);
	\draw[->] (tclaimfail1) -- (considerdummy);

	\draw[->] (consider1) -- ([xshift=-3pt] tclaimsuccess1.north);
	\draw[<-o] (unavailable1) -- (tclaimsuccess1.north);
	\draw[->] (tclaimsuccess1) -- (claimed1);

	\draw[->] (claimed1) -- (tfailedclaimed1.west);
	\draw[<->] (failed1.35) -- ([yshift=-3pt] tfailedclaimed1.west);
	\draw[->] (tfailedclaimed1) -- (considerdummy);

    % Child dummy
	\node[place,dummy, below=1.7cm of considerdummy] (claimeddummy) {$\hdots$};
	\node[place,dummy, below=0.5cm of claimeddummy] (faileddummy) {$\hdots$};

    % Child n
	\node[iplace, right=0.6cm of considerdummy, yshift=-0.7cm, label=right:{$\Unavailable_{\child{v}{n}}$}] (unavailablen) {};
	\node[place, below=1cm of unavailablen, label=below:{$\Claimed_n$}] (claimedn) {};
	\node[iplace, below=0.5cm of claimedn, label=below:{$\Failed_{\child{v}{n}}$}] (failedn) {};

	\node[itransition, right=0.8cm of claimedn, label=below:{$\tchildfail$}, label=right:{@$\prioVar_v$}] (tfailedclaimedn) {};
	\node[Itransition, below=0.4cm of unavailablen, label=left:{$\tclaim$}, label=right:{@$\prioVar_v$}] (tclaimsuccessn) {};
	\node[itransition, right=2.0cm of considerdummy, label=above:{$\tunavailable$}, label=35:{@$\prioVar_v$}] (tclaimfailn) {};

	\node[iplace, label=above:{$\Failed_v$}, right=3.3cm of considerdummy] (v_failed) {};

	\draw[->] (considerdummy) -- (tclaimfailn.west);
	\draw[<->] (unavailablen) -- ([yshift=-3pt] tclaimfailn.west);
	\draw[->] (tclaimfailn) -- (v_failed);

	\draw[->] (considerdummy) -- ([xshift=-3pt] tclaimsuccessn.north);
	\draw[<-o] (unavailablen) -- (tclaimsuccessn.north);
	\draw[->] (tclaimsuccessn) -- (claimedn);

	\draw[->] (claimedn) -- (tfailedclaimedn.west);
	\draw[<->] (failedn.35) -- ([yshift=-3pt] tfailedclaimedn.west);
	\draw[->] (tfailedclaimedn) -- (v_failed);

	\node[iplace,label=below:{$\Unavailable_v$}, below=0.7cm of v_failed] (v_unavailable) {};

	\draw[->] (tfailedclaimedn) -- (v_unavailable);
	\draw[->] (tclaimfailn) -- (v_unavailable);

    % Late claiming
	\node[iplace, left=1cm of consider1, label=above:{$\Active_v$}] (active) {};
	\node[place, below=0.8cm of active,  label=below:{$\mathsf{Sleep}$},tokens=1] (holder) {};
	\node[itransition, right=0.6cm of active] (tstart) {};
	\draw[->] (active) -- (tstart);
	\draw[->] (holder) -- (tstart);
	\draw[->] (tstart) -- (consider1);

    % Early failing
	\node[itransition, below=3.5cm of v_failed] (earlyfail) {};
	\node[place,dummy, left=0.6cm of earlyfail, yshift=0.15cm] (dummyearlyfail) {$\vdots$};
	\draw[o->] (earlyfail) edge[bend right] (v_failed);
	\draw[->] (failed1) edge[bend right=15] (earlyfail);
	\draw[->] (failedn) -- (earlyfail);
\end{tikzpicture}%
\caption{Part of \SPARE template for \emph{late claiming/early failing}}
\label{fig:late_claiming_early_failing}
\end{figure}
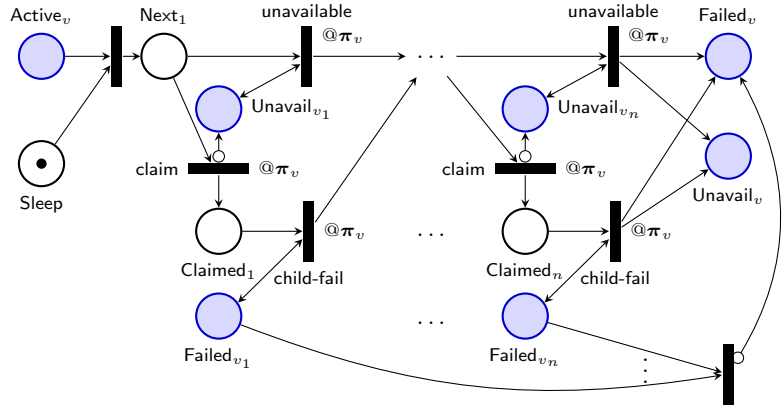
In the template (Fig.~\ref{fig:late_claiming_early_failing}), early failing is realised by adding a transition which places a token in $\Failed_v$ once all children have failed.
Although the additional part is similar to an \AND, early failing cannot be mimicked by adding an \OR and an \AND to the DFT, as such constructions are typically syntactically disallowed.

\paragraph{When to activate spare components?}
In the early-claiming semantics, claiming affected activation, but not the other way around.
In particular, as claiming might fail, firing  $\mathsf{activate}$ may indirectly cause a token to be placed in $\Failed_v$.
\begin{example}
Consider again Fig.~\ref{fig:nested_spare} where $PA_2$ failed.
Let subsequently $PA_1$ and $B$ fail.
Under late claiming, $P_2$ does not claim $B$ (as it is not yet active) when $PA_2$ fails.
$P_1$ thus claims $B$.
Under early failing, $P_2$ now fails as none of its children is available anymore.
It thus fails before $B$ does.
Under late failing, $P_1$ fails once $B$ fails, and $R_2$ is activated.
Now $P_2$ is activated.
As it cannot claim any child, $P_2$ fails after $B$.
\end{example}

The moment when we update the activation now matters, as it affects claiming, and for claiming the order matters.
Thus, we need to consider the priorities in activation propagation.
Again, a variety of options is available, in particular in relation to the priorities used for claiming, failure forwarding and propagation. We refrain from an in-depth analysis of these variants and stress that nested spares and late claiming are ill-supported by existing semantics.

\subsection{Template adaptions}
\label{sec:extension_adaptions}

\subsubsection{Adaption to ensure $1$-boundedness of $\Unavailable$.}
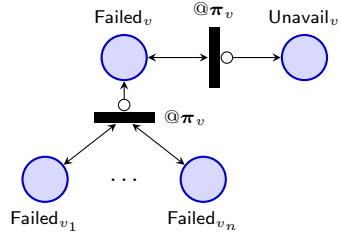
\begin{figure}[tb]
    \centering
    \begin{tikzpicture}[every label/.append style={font=\scriptsize}]
    \node[iplace,label=above:{$\Failed_v$}] (v_failed) {};

    \node[Itransition, below=0.4cm of v_failed, label=right:{@$\prioVar_v$}] (v_tfail) {};
    \node[place,dummy, below=0.4cm of v_tfail] (dummy) {$\hdots$};
    \node[iplace, left=0.4cm of dummy, label=below:{$\Failed_{\child{v}{1}}$}] (sv1_failed) {};
    \node[iplace, right=0.4cm of dummy, label=below:{$\Failed_{\child{v}{n}}$}] (svn_failed) {};

    \node[itransition, right=0.8cm of v_failed, label=above:{@$\prioVar_v$}] (tunavail) {};
    \node[iplace,right=0.8cm of tunavail, label=above:{$\Unavailable_v$}] (v_unavailable) {};

    \draw[<->] (sv1_failed) -- (v_tfail);
    \draw[<->] (svn_failed) -- (v_tfail);
    \draw[o->] (v_tfail) -- (v_failed);

    \draw[<->] (v_failed) -- (tunavail);
    \draw[o->] (tunavail) -- (v_unavailable);
\end{tikzpicture}%
    \caption{GSPN template adaptions for \AND}
    \label{fig:template_adaption_unavail}
\end{figure}

In Sect.~\ref{sec:properties} we claimed that the GSPN templates can be easily adapted to ensure $1$-boundedness of $\Unavailable_v$.
We now present these adaptions exemplary for \AND in Fig.~\ref{fig:template_adaption_unavail}.
This adaption can be made for all gates.

A token can only be placed in $\Unavailable_v$ if there was none before and $\Failed_v$ does contain a token.
Moreover, in the $\templ{\SPARE}{v}$ the transition $\tclaim$ can only place a token in $\Unavailable_{\child{v}{i}}$ if there was none before.
Thus, $\petriTemplate_\DFT$ constructed with the presented adaptions ensures $1$-boundedness of the GSPN.
However, the templates for the $\VOT{k}$ gate and $\SEQ$ violate the $1$-bound as explained in App.~\ref{sec:additional_gates}.

\section{Proofs}
\label{sec:proofs}
Let $\petriTemplate_\DFT$ be the obtained GSPN for a conventional DFT $\DFT$.
We do not consider the extensions of App.~\ref{sec:extensions}.
We denote the maximal number of children with $\maxchildren = \max_{v \in \DFT} |\children(v)|$.
\begin{theorem}
    The GSPN $\petriTemplate_\DFT$ has at most $6 \cdot |V| \cdot \maxchildren + 2$ places, $2 \cdot |\DFT_{\BE}|$ timed transitions and at most $6 \cdot |V| \cdot \maxchildren$ immediate transitions.
\end{theorem}
\noindent\emph{Proof sketch.}

\begin{compactitem}
\item The number of interface places $\interfaces_\DFT$ is bounded by $|\interfaces_\DFT| \leq 4 \cdot |V|$.
    The number of auxiliary places is bounded by $2 \cdot |V| \cdot \maxchildren$, plus two for the initial template.
    The exact number of auxiliary places is given in the second column of Table~\ref{tab:gspn_size}, where $n$ denotes the number of children $\sizechildren$.
\item The number of timed transitions is given by $|\petriTimed| = 2 \cdot |\DFT_{\BE}|$ as for each \BE $v$ the GSPN $\templ{\BE}{v}$ has $2$ timed transitions.
\item The number of immediate transitions is bounded by $|\petriImmediate| \leq 4 \cdot |\DFT_{\SPARE}| \cdot \maxchildren + 2 \cdot |\DFT \setminus \{\DFT_{\SPARE}\}| \cdot \maxchildren + 2$.
    The precise number for each gate is given in the third column of Table~\ref{tab:gspn_size}.
    Additionally each gate has $\sizechildren$ immediate transitions resulting from the activation template.
\end{compactitem}
    \begin{table}[tb]
        \centering
        \caption{Sizes for GSPN templates}
        \label{tab:gspn_size}
        \begin{tabular}{l|ccc}
            \toprule
            Gate type              & \# Aux. places & \# Transitions & \# Arcs \\
            \midrule
            $\templ{\BE}{v}$       & $0$   & $2$ timed  & $11$ \\
            $\templ{\AND}{v}$      & $0$   & $1$ immed. & $2n+3$ \\
            $\templ{\OR}{v}$       & $0$   & $n$ immed. & $5n$ \\
            $\templ{\PANDincl}{v}$ & $1$   & $n$ immed. & $5(n-1)+6$ \\
            $\templ{\PANDexcl}{v}$ & $n-1$ & $n$ immed. & $5(n-1)+6$ \\
            $\templ{\SPARE}{v}$    & $2n$  & $3n$ immed.& $12n+2+6n$ \\
            $\templ{\FDEP}{v}$     & $0$ & $n-1$ immed. & $6(n-1)$ \\
            $\initMarkTempl$       & $2$ & $2$ immed.   & $4+|\{e_1, \dots, e_k\}|$ \\
            \midrule
            activation gates       & $0$    & $n$ immed.& $4n$ \\
            activation \SPARE      & $0$    & $n$ immed.& $6n$ \\
            \bottomrule
            \end{tabular}
        \end{table}

The only places in $\petriTemplate_\DFT$ that initially possess a token are $\Consider_1$ in $\templ{\SPARE}{v}$ and $\Init$ in $\initMarkTempl$.
Furthermore, $I(t)(p) \leq 1$, $O(t)(p) \leq 1$, and $H(t)(p) \leq 1$ for all $t \in \petriTransitions$ and $p \in \petriPlaces$, \ie all transitions consider at most one token per place.

\begin{theorem}
\label{thm:tokens_not_removed}
    Let marking $m$ and place $p \in \{\Failed_v, \Active_v, \Unavailable_v \sep v \in \DFT\}$ with $m(p) \geq 1$. Then for all $t \in \petriTransitions$, $\fire(m,t)(p) \geq 1$.
    \end{theorem}
\noindent\emph{Proof sketch.}
    Let $\petriPlaces'$ be the set of places where for all transitions $t \in \petriTransitions$ and $p \in \petriPlaces'$ it holds $I(t)(p)=1 \Rightarrow O(t)(p)=1$, \ie every token which is removed by a transition is put back.
    We give $\petriPlaces'$ for all templates:
    \begin{itemize}
        \item In $\templ{\BE}{v}$, $\templ{\AND}{v}$, $\templ{\OR}{v}$, $\templ{\PANDincl}{v}$, $\templ{\FDEP}{v}$ and all activation templates, we have $\petriPlaces' = \petriPlaces$.
        \item In $\templ{\PANDexcl}{v}$, it holds $\petriPlaces' = \petriPlaces \setminus \{X_1, \dots, X_{n-1}\}$.
        \item In $\templ{\SPARE}{v}$, it holds $\petriPlaces' = \petriPlaces \setminus \bigcup_{1\leq i \leq n}\{\Consider_i, \Claimed_i\}$.
        \item In $\initMarkTempl$, it holds $\petriPlaces' = \petriPlaces \setminus \{\Init, \Evidence\}$.
    \end{itemize}
    As for all gates $\petriPlaces' \supseteq \{\Failed_v, \Active_v, \Unavailable_v \sep v \in \DFT\}$ the claim holds for all interface places $\interfaces_\DFT \setminus \{\Disabled_v \sep v \in \DFT_\BE\}$.

\begin{theorem}
\label{thm:transitions_fire_once}
    Each transition $t \in \petriTransitions$ can fire at most once.
\end{theorem}
\noindent\emph{Proof sketch.}
We define a mapping $\mapping\colon \petriTransitions \rightarrow \petriPlaces$ such that $\mapping(t)$ denotes a place which both
\begin{compactitem}
\item  prevents firing $t$ multiple times;
formally, we ensure $\mapping(t) = p$ with $O(t)(p)>0$ and $H(t)(p)>0$, and
\item and tokens placed in $p$ are never removed; formally for all $t \in \petriTransitions$, $\fire(m,t)(p) \geq 1$.
\end{compactitem}
Then, if $t$ fires, it places a token in $p$ and as $H(t)(p)>0$ an immediate refiring of $t$ is prohibited.
Furthermore, as tokens are not removed for $p \in \mapping$, the transition can never refire.

For some templates, choosing $\marking$ is trivial:
    \begin{itemize}
        \item For any transition $t \in \templ{\BE}{v}$, $\templ{\AND}{v}$, and $\templ{\OR}{v}$ we set $\mapping(t) = \Failed_v$.
        \item For any transition $t \in \templ{\PANDincl}{v}$ we set $\mapping(t) \in \{ \Failsafe,\Failed_v \}$.
        \item For any transition $t \in \templ{\FDEP}{v}$ we set $\mapping(t) \in \{\Failed_{\child{v}{2}}, \dots \Failed_{\child{v}{n}}\}$.
        \item For any transition governing the activation, $\mapping(t) \in \{\Active_{\child{v}{1}}, \dots, \Active_{\child{v}{n}}\}$.
       \end{itemize}
       
For the following templates, we observe the following (and set $\mapping$ accordingly)       
       
       \begin{itemize}
        \item In $\templ{\PANDexcl}{v}$ the transitions can only fire from left to right moving the token from $X_1$ to $X_{n-1}$ and $\Failed_v$.
        If a transition removes the token from $X_{i-1}$ and places it in $X_i$, $\Failed_{\child{v}{i}}$ contains a token (and will forever contain it).
        Then the inhibitor arc prevents a refiring of the transition, which places a token in $X_{i-1}$.
        If a token is in $\Failed_v$ the rightmost transition is disabled forever.
        \item In $\templ{\SPARE}{v}$ the token in $\Consider_i$ either directly moves to $\Consider_{i+1}$ (or to $\Failed_v$ in the end) or to $\Claimed_i$.
        In both cases, $\tclaim$ and $\tunavailable$ are disabled afterwards.
        If a token is in $\Consider_i$ and transition $\tchildfail$ fires the token moves to $\Consider_{i+1}$ and $\tchildfail$ is disabled afterwards.
        \item In $\initMarkTempl$ the first transition removes the token $\Init$ and afterwards this transition is disabled.
        The next transition is then enabled as $\Evidence$ contains a token.
        After this transition fires, $\Evidence$ does not contain a token anymore, therefore disabling the transition.
    \end{itemize}
    $\mapping$ ensures that each transition is only fired at most once.

\begin{corollary}
    $\petriTemplate_\DFT$ contains no time-traps.
\end{corollary}
\noindent\emph{Proof sketch.}
    By Theorem~\ref{thm:transitions_fire_once}, each transition $t \in \petriTransitions$ is fired at most once.
    Thus, the marking graph of $\petriTemplate_\DFT$ cannot contain a cycle.

\begin{theorem}
    $\petriTemplate_\DFT$ is $2$-bounded.
\end{theorem}
\noindent\emph{Proof sketch.}
   All places except $\Unavailable_v$ are $1$-bounded.
    A transition can only place an additional token in $\Unavailable_v$ if the corresponding $\Failed_v$ did not contain a token before.
    For all places $p \in \mapping$ with $\mapping$ as before, a new token is only placed in $p$ if there was no token before.
    This is ensured by the inhibitor arcs as explained before.
    We consider all templates in detail:
    \begin{itemize}
        \item In $\templ{\BE}{v}$, $\templ{\AND}{v}$, $\templ{\OR}{v}$ and $\templ{\PANDincl}{v}$ at most one transition can be fired and places a token in $p \in \mapping$.
        Thus all places $p \neq \Unavailable_v$ are $1$-bounded.
        \item In $\templ{\PANDexcl}{v}$ the transitions  only fire from left to right, while moving the token from $X_1$ to $X_{n-1}$ and finally placing it in $\Failed_v$ if there was none before.
        \item In $\templ{\FDEP}{v}$ all transitions can only place a token in $\Failed_{\child{v}{2}}, \dots, \Failed_{\child{v}{n}}$ if there was none before.
        \item In the activation templates a token can only be placed in $\Active_{\child{v}{1}}, \dots, \Active_{\child{v}{n}}$ if there was none before.
        \item In $\templ{\SPARE}{v}$ the token moves from left to right through the auxiliary places, until finally a token is placed in $\Failed_v$.
        There is no inhibitor arc for $\Failed_v$ but due to the single moving token, only one token is placed in $\Failed_v$.
        Furthermore, a token is placed in $\Unavailable_{\child{v}{i}}$ only if there was none before.
        \item In $\initMarkTempl$ the first transition puts a token in $\Active_{top}$.
        As these transitions have the highest priorities, the interface tokens have not received a token before.
        The same argument holds for the tokens placed in $\Failed_{e_1}, \dots, \Failed_{e_n}$.
    \end{itemize}

\section{Example}
\label{sec:examples}
\begin{figure}[b]
\centering
\includegraphics[width=\textwidth]{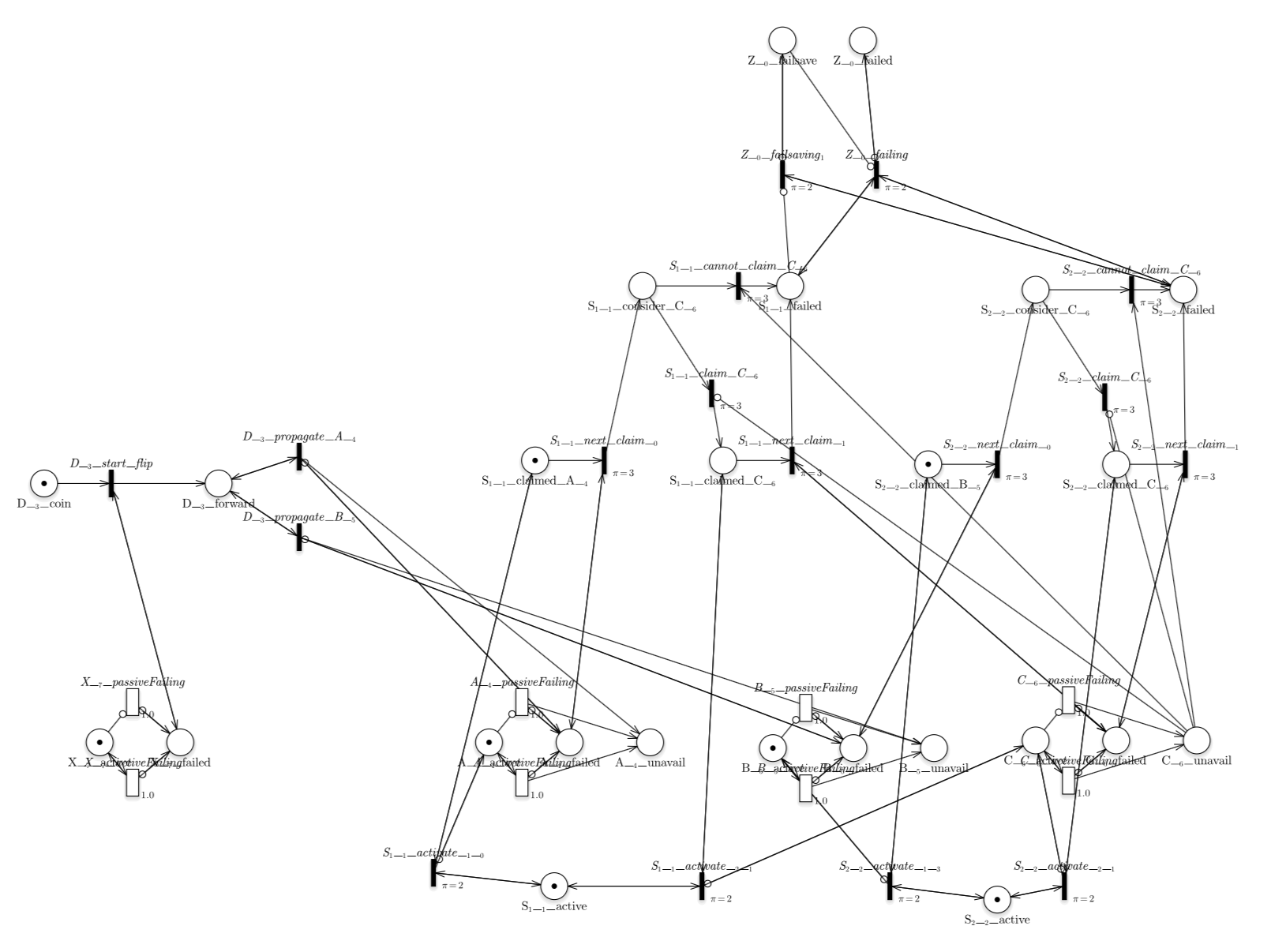}
\caption{GSPN $\petriTemplate_{\DFT_3}$ from GreatSPN Editor}
\label{fig:greatspn_example}
\end{figure}

We give an example of a larger GSPN.
Recall the DFT $\DFT_3$ from Fig.~\ref{fig:nondeterminism} on page~\pageref{fig:nondeterminism}.
Fig.~\ref{fig:greatspn_example} depicts the corresponding GSPN $\petriTemplate_{\DFT_3}$ exported from the GreatSPN Editor~\cite{Amparore14}.

\end{document}